\documentclass[twocolumn,tighten,times]{aastex631}

\usepackage{amsmath}
\newcommand{\avg}[1]{\left\langle{#1}\right\rangle}

\iftrack
\renewcommand{\added}[1]{{\bf #1}}
\renewcommand{\deleted}[1]{}
\renewcommand{\replaced}[2]{{\bf #2}}
\fi
\shorttitle{Line Broadening Models for LIM}
\shortauthors{Chung et al.}
\graphicspath{{./}{figures/}}

\begin{document}

\title{A Model of Spectral Line Broadening in Signal Forecasts for Line-intensity Mapping Experiments}

\correspondingauthor{Dongwoo T.~Chung}
\email{dongwooc@cita.utoronto.ca}

\author[0000-0003-2618-6504]{Dongwoo T.~Chung}
\affiliation{Canadian Institute for Theoretical Astrophysics, University of Toronto, 60 St. George Street, Toronto, ON M5S 3H8, Canada}
\affiliation{Dunlap Institute for Astronomy and Astrophysics, University of Toronto, 50 St. George Street, Toronto, ON M5S 3H4, Canada}

\author[0000-0001-8382-5275]{Patrick C.~Breysse}
\affiliation{Center for Cosmology and Particle Physics, Department of Physics, New York University, 726 Broadway, New York, NY, 10003, USA}

\author[0000-0003-3420-7766]{H\aa vard T.~Ihle}
\affiliation{Institute of Theoretical Astrophysics, University of Oslo, P.O. Box 1029 Blindern, N-0315 Oslo, Norway}

\author[0000-0002-8800-5740]{Hamsa Padmanabhan}
\affiliation{D\'epartement de Physique Théorique, Universite de Genève, 24 Quai Ernest-Ansermet, CH-1211 Genève 4, Switzerland}

\author[0000-0003-0209-4816]{Marta B.~Silva}
\affiliation{Institute of Theoretical Astrophysics, University of Oslo, P.O. Box 1029 Blindern, N-0315 Oslo, Norway}

\author[0000-0003-2358-9949]{J.~Richard Bond}
\affiliation{Canadian Institute for Theoretical Astrophysics, University of Toronto, 60 St. George Street, Toronto, ON M5S 3H8, Canada}

\author{Jowita Borowska}
\affiliation{Institute of Theoretical Astrophysics, University of Oslo, P.O. Box 1029 Blindern, N-0315 Oslo, Norway}

\author{Kieran A.~Cleary}
\affiliation{California Institute of Technology, 1200 E. California Blvd., Pasadena, CA 91125, USA}

\author[0000-0003-2332-5281]{Hans Kristian Eriksen}
\affiliation{Institute of Theoretical Astrophysics, University of Oslo, P.O. Box 1029 Blindern, N-0315 Oslo, Norway}

\author{Marie Kristine Foss}
\affiliation{Institute of Theoretical Astrophysics, University of Oslo, P.O. Box 1029 Blindern, N-0315 Oslo, Norway}

\author{Joshua Ott Gundersen}
\affiliation{Department of Physics, University of Miami, 1320 Campo Sano Avenue, Coral Gables, FL 33146, USA}

\author[0000-0001-5211-1958]{Laura C.~Keating}
\affiliation{Leibniz-Institut f{\"u}r Astrophysik Potsdam (AIP), An der Sternwarte 16, D-14482 Potsdam, Germany}

\author{Jonas Gahr Sturtzel Lunde}
\affiliation{Institute of Theoretical Astrophysics, University of Oslo, P.O. Box 1029 Blindern, N-0315 Oslo, Norway}

\author{Liju Philip}
\affiliation{Jet Propulsion Laboratory, California Institute of Technology, 4800 Oak Grove Drive, Pasadena, CA 91109, USA}

\author{Nils-Ole Stutzer}
\affiliation{Institute of Theoretical Astrophysics, University of Oslo, P.O. Box 1029 Blindern, N-0315 Oslo, Norway}

\author{Marco P.~Viero}
\affiliation{California Institute of Technology, 1200 E. California Blvd., Pasadena, CA 91125, USA}

\author[0000-0002-5437-6121]{Duncan J.~Watts}
\affiliation{Institute of Theoretical Astrophysics, University of Oslo, P.O. Box 1029 Blindern, N-0315 Oslo, Norway}

\author[0000-0003-3821-7275]{Ingunn Kathrine Wehus}
\affiliation{Institute of Theoretical Astrophysics, University of Oslo, P.O. Box 1029 Blindern, N-0315 Oslo, Norway}

\collaboration{18}{(COMAP Collaboration)}



\begin{abstract}

Line-intensity mapping observations will find fluctuations of integrated line emission are attenuated by varying degrees at small scales due to the width of the line emission profiles. This attenuation may significantly impact estimates of astrophysical or cosmological quantities derived from measurements. We consider a theoretical treatment of the effect of line broadening on both the clustering and shot-noise components of the power spectrum of a generic line-intensity power spectrum using a halo model. We then consider possible simplifications to allow easier application in analysis, particularly in the context of inferences that require numerous, repeated, fast computations of model line-intensity signals across a large parameter space. For the CO Mapping Array Project (COMAP) and the CO(1--0) line-intensity field at $z\sim3$ serving as our primary case study, we expect a $\sim10$\% attenuation of the spherically averaged power spectrum on average at relevant scales of $k\approx0.2$--0.3\,Mpc$^{-1}$, compared to $\sim25$\% for the interferometric Millimetre-wave Intensity Mapping Experiment (mmIME) targeting shot noise from CO lines at $z\sim1$--5 at scales of $k\gtrsim1$\,Mpc$^{-1}$. We also consider the nature and amplitude of errors introduced by simplified treatments of line broadening, and find that while an approximation using a single effective velocity scale is sufficient for spherically-averaged power spectra, a more careful treatment is necessary when considering other statistics such as higher multipoles of the anisotropic power spectrum or the voxel intensity distribution.

\end{abstract}



\section{Introduction}
Line-intensity mapping (LIM) or intensity mapping (IM) is the study of the aggregate emission in a given spectral line across large cosmological volumes. As previous overviews of the field by~\cite{Kovetz2017} and~\cite{Kovetz2019} (and references therein) have discussed, such observations will allow cosmological and astrophysical inferences in understudied redshift ranges where targeted galaxy surveys are difficult to undertake over large sky areas. In particular, LIM should enable astrophysical inferences about the faint end of luminosity functions at high redshift, which will be less challenging to survey through integrated line emission than in isolated targeted observations.

Interest in surveying reionization topology and large-scale structure through 21 cm IM~\citep{Madau97,Chang08} led the initial scientific work in LIM. But the field has since evolved to include other lines such as carbon monoxide (CO) and ionized carbon ([\ion{C}{2}]), and the past decade has seen a great abundance of literature around models of the LIM signal to be expected from such lines~\citep{Lidz11,Pullen13,Mashian15,Yue15,Li16,LidzTaylor16,Breysse17,Padmanabhan2018a,Dumitru19,Padmanabhan2019,BreysseAlexandroff19,Ihle19,MoradinezhadKeating19,Sun19,cii_um}. Many of these works highlight specific aspects not necessarily heavily emphasized in previous literature that carry implications for signal expectations, including astrophysical or cosmological effects like photodissociation of CO in high-redshift low-metallicity environments~\citep{Mashian15} and scale-dependent corrections to the tracer bias of CO emission in relation to the underlying matter density~\citep{MoradinezhadKeating19}.

However, we have not seen detailed, explicit models of the effect of line broadening\footnote{We use the term `line broadening' in this work rather than describe the effect as a finger-of-God (FoG) effect, which traditionally refers to suppression of clustering at small scales due to the pairwise velocity dispersion of galaxies. In this work, we deal with suppression of line-intensity shot noise as well as clustering, and the smearing of a continuous temperature field rather than positions of individual galaxies.}---the fact that spectral line emission from each source is not confined to a single exact frequency, but rather extends over a finite line width---in the context of CO or [\ion{C}{2}] line-intensity mapping. As the effect will be to reduce the amplitude of spectral line-intensity fluctuations, interpretation of line-intensity power spectrum measurements should ideally account for it to avoid biased recovery of astrophysical or cosmological quantities. That said, the impact of line broadening will vary based on the line surveyed and the scales of interest.

The 21 cm IM literature does model the velocity dispersion of neutral hydrogen separately from that of matter, with~\cite{VN18} computing the dispersion as a function of halo mass and~\cite{SarkarBharadwaj19} building a careful model of the 21 cm line profile as a function of the host dark matter halo's properties. However, 21 cm experiments measuring baryon acoustic oscillations at $z\gtrsim2$ will target larger scales ($k\sim0.1$\,Mpc$^{-1}$) where there is minimal suppression of the power spectrum from such small-scale corrections.

In other contexts, the effect of line broadening is not the dominant source of the suppression of observable fluctuations at small scales. The upcoming generation of [\ion{C}{2}] intensity mapping experiments observing at 200--400 GHz have resolving power of $R\sim100$--300~\citep{EoR-Spec,TIME,CONCERTO}. This would effectively correspond to a spectral width at 300 GHz of around 1--3 GHz, or $\sim10^3$\,km\,s$^{-1}$ channels, versus the $\sim10^2$\,km\,s$^{-1}$ line widths seen in high-redshift [\ion{C}{2}] sources~\citep{Capak15,Pentericci16}. Therefore, line broadening should be subdominant to the limited frequency resolution of these experiments.

But we suggest that line broadening cannot be neglected in every LIM context, and in particular must be considered explicitly for CO intensity mapping. Interferometric experiments such as the CO Power Spectrum Survey~\citep[COPSS;][]{COPSS} and the Millimetre-wave Intensity Mapping Experiment \citep[mmIME;][]{mmIME-ACA} specifically target small-scale CO intensity fluctuations. Even a single-dish experiment like the CO Mapping Array Project~\citep[COMAP;][]{Ihle19}, which chiefly targets large-scale fluctuations, cannot entirely ignore line broadening due to the effect it will have on the voxel intensity distribution (VID).

Yet the vast majority of signal and sensitivity forecasts essentially treat line emitters as point sources along the line of sight, with the implicit assumption that the effect of line broadening is subdominant. When we do see models of line profiles of CO in previous LIM literature, it is typically as a single number describing a width essentially expected of all line profiles in the observation. For instance,~\cite{MoradinezhadKeating19} model a single intrinsic line width at each redshift, set to $1+z$ times ${0.001c\approx300\ \mathrm{km\,s}^{-1}}$ without justification. (The resulting attenuation is also applied only in the context of the clustering contribution to the line-intensity power spectrum.) More recently, in presenting observational results from the previously mentioned mmIME, \cite{mmIME-ACA} estimate attenuation for specific line widths that are reasonable expectations based on high-redshift observations and models. However, the work does not explicitly model a line width prescription across the distribution of CO emitters, or apply a correction to the measured power spectrum based on the estimated attenuation.

Therefore, in this work, we set out to devise a model for line broadening suitable for intensity mapping in any spectral line, but with an emphasis on CO intensity mapping where the effect is particularly relevant in the short- to medium-term future. Using our model, we aim to answer these questions:
\begin{enumerate}
\item What is the level of signal attenuation\footnote{When we refer to the `signal' and its attenuation in this work, we refer to the power spectrum rather than the line-intensity field. Line widths do not reduce the mean line intensity, only the fluctuations about it.} that we can expect for experiments like COMAP and mmIME due to line broadening?
\item Is it sufficient to describe the effect of line broadening using a single parameter (as has been done by previous works), such as an effective global line width?
\item If not, how does this simplification fail?
\end{enumerate}

We have structured the paper as follows. In~\autoref{sec:theory}, we outline the theoretical formalism for the anisotropic power spectrum and multipoles with and without line broadening, given an analytic halo model of the line luminosity and line width. We define such a model in~\autoref{sec:linemodel} for CO(1--0) emission at $z\sim3$ as targeted by COMAP, which will allow us to quantify the effect of line broadening in this case. In~\autoref{sec:approx} we consider simplified treatments of line broadening for practical use in analysis, including a prescription for a single effective line width rather than a mass-dependent broadening. Then in~\autoref{sec:precheck} we validate our effective line width prescription in the context of signals targeted by COMAP and mmIME, but using numerical calculations based on our analytic formalism rather than using N-body simulations. We do use lightcones from N-body simulations in~\autoref{sec:sim}, where we calculate the power spectrum and VID with and without line broadening as might be observed by COMAP. We discuss all of our calculations and their implications for LIM experiments in~\autoref{sec:discussion}, before summarising and concluding in~\autoref{sec:conclusion}.

Where necessary, unless otherwise specified, we assume base-10 logarithms, and a $\Lambda$CDM cosmology with parameters $\Omega_m = 0.286$, $\Omega_\Lambda = 0.714$, $\Omega_b =0.047$, $H_0=100h$\,km\,s$^{-1}$\,Mpc$^{-1}$ with $h=0.7$, $\sigma_8 =0.82$, and $n_s =0.96$. The cosmology of choice matches the one assumed for the N-body simulation used in~\autoref{sec:sim}, and is broadly consistent with nine-year \emph{WMAP} results~\citep{WMAP9}. Distances carry an implicit $h^{-1}$ dependence throughout, which propagates through masses (all based on virial halo masses, proportional to $h^{-1}$) and volume densities ($\propto h^3$).

\section{Theoretical Formalism: Anisotropic Power Spectrum}
\label{sec:theory}
The present work will use lightcones from a N-body cosmological simulation to directly numerically calculate various statistics of the line-intensity field with line broadening taken into account. However, to understand the eventual results better, we will first outline the theoretical calculation of the line-intensity power spectrum using a halo model of line emission---first as in existing literature~\citep[e.g.:][]{Lidz11,LidzTaylor16,BreysseAlexandroff19,MoradinezhadKeating19,anisotropies,Bernal19b}, without line broadening but with other leading anisotropies, and then incorporating line broadening into both clustering and shot-noise components.

We will not extend theoretical treatment of line broadening to the VID as the high complexity around the calculation of the VID (as outlined by, e.g.,~\citealt{Breysse17}) makes it far more straightforward to find directly in numerical simulations, and a rigorous theoretical treatment will not add much to our understanding of the results. The numerical simulations alone still allow us to understand qualitative aspects of the effect of line broadening on the VID later in this work.

\subsection{Redshift-space Power Spectrum without Line Broadening}
We begin with the line-intensity power spectrum in real space, for sources at some fixed redshift, each associated with a dark matter halo. In our halo model at this redshift $z$, a dark matter halo with halo mass $M_h$ is associated with line luminosity $L(M_h)$, and the distribution of halo masses is given by a halo mass function $dn/dM_h$ describing number density per mass bin. Then the clustering component of the line-intensity power spectrum, associated with the large-scale structure formed by the underlying halo population, is found by scaling the matter power spectrum $P_m(k)$ by the bias $b$ with which the line emission traces the large-scale structure (i.e., the proportionality between line-intensity contrast and matter density contrast) and then by the cosmic average line brightness temperature $\avg{T}$:
\begin{equation}
    P_\text{clust}(k) = \avg{T}^2b^2P_m(k).
\end{equation}
We find $\avg{T}$ by integrating luminosity density across all $M_h$ and multiplying by an appropriate conversion factor $C_{LT}$:
\begin{equation}\avg{T}=\underbrace{\left(\frac{c^3(1+z)^2}{8\pi k_B\nu_\text{rest}^3H(z)}\right)}_{C_{LT}}\int dM_h\,\frac{dn}{dM_h}\,L(M_h).\label{eq:Tline}\end{equation}
As for the line bias $b$, if a halo mass bin at $M_h$ traces matter density contrast with halo bias $b(M_h)$, then we can average $b(M_h)$ weighted by luminosity density at each $M_h$:
\begin{equation}b=\frac{\int dM_h\,dn/dM_h\,L(M_h)b(M_h)}{\int dM_h\,dn/dM_h\,L(M_h)}.\label{eq:bline}\end{equation}

Meanwhile, a shot-noise component to the power spectrum describes scale-independent fluctuations arising from the fact that line emitters are discrete objects, which subjects a measurement of line-intensity fluctuations to Poisson statistics. The shot noise is described by the average squared line-luminosity density:
\begin{equation}P_\text{shot}=C_{LT}^2\int dM_h\,\frac{dn}{dM_h}\,L^2(M_h).\label{eq:Pshot_dL}\end{equation}
If we prescribe log-normal scatter of $\sigma_L$ (in units of dex) around the average $L(M_h)$ relation, this modifies $P_\text{shot}$:
\begin{equation}P_\text{shot}(\sigma_L) = \exp{(\sigma_L^2\ln^2{10})}P_\text{shot}(\sigma_L=0).\end{equation}
For brevity, we will consider this additional factor implicit in most expressions below. The present work will never consider a mass-dependent $\sigma_L$, so the factor will not change with consideration of line broadening.

The shot noise variance is independent of the clustering variance, and so the components add linearly to give the total line-intensity power spectrum in real space:
\begin{equation}P(k) = P_\text{clust}(k)+P_\text{shot}.\end{equation}

Using~\cite{anisotropies} as our primary reference, we can consider two leading effects (omitting the finger-of-God effect as it was shown to be small, and as it will likely be subdominant to line broadening). Both effects, as well as the effect from line broadening to be considered below, preserve the angular isotropy in the real-space signal. Therefore the full three-dimensional power spectrum, while strictly speaking a function of the 3D wavevector $\mathbf{k}$, will still depend effectively only on $k=|\mathbf{k}|$ and $\mu=\hat{\mathbf{k}}\cdot\hat{\mathbf{z}}$, the latter being the cosine of the angle between $\mathbf{k}$ and the line of sight described by the unit vector $\hat{\mathbf{z}}$.

The first leading effect is the Kaiser effect, due to large-scale coherent halo migration into matter overdensities, causing redshift-space distortions. This modifies only the clustering component,
\begin{equation}P_\text{clust}^r(k,\mu)=\left(1+\frac{\Omega_m(z)^{0.55}}{b}\mu^2\right)^2P_\text{clust}(k).\end{equation}
The second leading effect is instrumental resolution in both angular and line-of-sight directions. Suppose the signal is subject to a Gaussian beam profile with standard deviation of $\sigma_\perp$ in comoving space, and also to a Gaussian spectral profile approximating the instrumental frequency resolution with standard deviation of $\sigma_\parallel$ also in comoving space. Here, if the angular beam profile has standard deviation $\sigma_\text{beam}$ in units of radians, then $\sigma_\perp$ is simply $\sigma_\text{beam}$ times the comoving distance $R(z)$ to the emission redshift $z$.

\replaced{The line-intensity field is modified by convolution in comoving space with the a}{These a}ngular and line-of-sight Gaussian profiles\added{ modify the line-intensity field in comoving space by convolution}. We multiply the Fourier transform of the line-intensity field, $\tilde{T}(k)$, by the appropriate Fourier transforms to yield
\begin{align}\tilde{T}_\text{conv}(k,\mu)&=\tilde{T}(k)\times\nonumber\\&\qquad\exp{[-k^2\sigma_\perp^2(1-\mu^2)/2]}\exp{(-k^2\sigma_\parallel^2\mu^2/2)},\end{align}
and the power spectrum is modified as the squared Fourier transform would be. Putting this together with the Kaiser effect, the overall modification of the real-space power spectrum results in the redshift-space power spectrum,
\begin{align}P_\text{conv}(k,\mu)&=\nonumber\\&\left[\avg{T}^2b^2\left(1+\frac{\Omega_m(z)^{0.55}}{b}\mu^2\right)^2P_m(k)+P_\text{shot}\right]\notag\\*&\hspace{1cm}\times\exp{[-k^2\sigma_\perp^2(1-\mu^2)-k^2\sigma_\parallel^2\mu^2]}.\label{eq:Pconv}\end{align}

The spherically-averaged power spectrum that a survey can actually measure corresponds to the monopole $P_{\ell=0}(k)$ from the multipole expansion of $P(k,\mu)$ in Legendre polynomials in $\mu$,
\begin{equation}P_\ell(k)=\frac{2\ell+1}{2}\int_{-1}^1d\mu\,P(k,\mu)\mathcal{L}_\ell(\mu),\label{eq:legendre}\end{equation}
where $\mathcal{L}_\ell$ denotes the Legendre polynomial of order $\ell$. Thus,
\begin{equation}
    P_\text{conv}(k,\mu)=\sum_\ell P_\ell(k)\mathcal{L}_\ell(\mu).
\end{equation}
The quadrupole power spectrum $P_{\ell=2}(k)$ then describes the leading anisotropies, as $P_\text{conv}(k,\mu)$ is even in $\mu$ and thus the dipole $P_{\ell=1}(k)=0$. We use the notations $P_0(k)$ and $P_2(k)$ throughout the remainder of this work to refer to the monopole $P_{\ell=0}(k)$ and quadrupole $P_{\ell=2}(k)$.

If line broadening is a subdominant effect compared to the spectral response of the instrument, then we can simply use the above. However, we will proceed under the assumption that the instrument's native frequency resolution is subdominant compared to the line profile size, so that we discard the above \replaced{$\sigma_\parallel$ that is independent of mass}{mass-independent $\sigma_\parallel$} and instead consider a mass-dependent $\sigma_v(M)$ that introduces additional complications.

\subsection{Redshift-space Power Spectrum with Line Broadening}
We now introduce line broadening to consider its effect on the signal. As with the bias, line luminosity, and number density, the line width depends on halo mass. Suppose that the line profile of emission from a halo of mass $M_h$ is Gaussian, with full width at half maximum (FWHM) given by $v(M_h)$ in units of physical velocity. Then the standard deviation of the corresponding line-of-sight Gaussian profile in comoving space is
\begin{equation}
    \sigma_v(M_h)=\frac{(1+z)}{H(z)}\frac{v(M_h)}{2\sqrt{2\ln{2}}}.\label{eq:veltosig}
\end{equation}

Line broadening is a small-scale effect and therefore by and large it suffices to consider its effect on the shot-noise component only. The attenuation is mass-dependent and will thus differ for each mass bin contributing to the total shot noise. Therefore, instead of multiplying the total $P_\text{shot}$ by the squared Fourier transform of the Gaussian as in the previous section, we need to apply the attenuation with the appropriate $\sigma_v(M_h)$ to the integrand of~\autoref{eq:Pshot_dL}. Including the angular beam profile,
\begin{align}
    P_\text{shot,v}(k,\mu)&=C_{LT}^2\int dM_h\,\frac{dn}{dM_h}\,L^2(M_h)\times\nonumber\\&\hspace{1.1cm}\exp{\left[-k^2\sigma_\perp^2(1-\mu^2)-k^2\sigma_v^2(M_h)\mu^2\right]}.\label{eq:Pshotv}
\end{align}

The shot noise is scale-independent while $P_m(k)\sim k^{-3}$ at high $k$, so shot noise typically dominates at scales where attenuation of the power spectrum from line broadening is non-negligible. However, for completeness we do also consider the effect on the clustering component.

Recall that in calculating $P_\text{clust}(k)$, we model the scaling of line-intensity fluctuations from matter density contrast, where the former is around an average temperature $\avg{T}$ and traces the latter with some linear bias $b$:
\begin{equation}\Delta T(\mathbf{x})=\avg{T}b\,\delta_m(\mathbf{x}).\end{equation}
Then the Fourier transforms are scaled the same way, and since we take fluctuations in matter to be isotropic in real space, it suffices to describe the Fourier modes simply with $k$ rather than $\mathbf{k}$:
\begin{equation}\tilde{T}(k)=\avg{T}b\,\delta_m(k).\end{equation}
(We simply write $\tilde{T}$ rather than $\Delta\tilde{T}$ as we do not consider the Fourier transform at $k=0$ that would correspond to the mean value $\avg{T}$.)

Substituting~\autoref{eq:Tline} and~\autoref{eq:bline} and simplifying, we can express the scaling between matter density contrast and line-intensity contrast as
\begin{equation}\tilde{T}(k)=C_{LT}\delta_m(k)\int dM_h\,\frac{dn}{dM_h}\,L(M_h)b(M_h).\end{equation}
So while we typically scale $P_m(k)$ to $P_\text{clust}(k)$ by evaluating two integrals in mass, it is only really necessary to evaluate just one (i.e., we write $\avg{T}b$ but really mean $\avg{Tb}$). In fact, the way each mass bin contributes to the total signal is clearer if we move $\delta_m(k)$ behind the integral:
\begin{equation}\tilde{T}(k)=C_{LT}\int dM_h\,\frac{dn}{dM_h}\,L(M_h)b(M_h)\delta_m(k).\end{equation}
So for each halo mass bin $(M_h,M_h+dM_h)$, we scale the matter fluctuation by the halo bias and luminosity corresponding to that bin, and then this is integrated weighted by number density to give the total line-intensity fluctuation.

Since $P_m(k)\propto\delta_m^2(k)$ and $P_\text{clust}(k)\propto\tilde{T}^2(k)$ by the same proportionality---the comoving volume being studied, independent of $M_h$---we can rewrite the above as
\begin{equation}\sqrt{P_\text{clust}(k)}=C_{LT}\int dM_h\,\frac{dn}{dM_h}\,L(M_h)b(M_h)\sqrt{P_m(k)}.\end{equation}
We should note at this point that this is a very casual derivation, and in particular the way in which we treat $\sqrt{P(k)}$ as equivalent to $|\tilde{T}(k)|$ even for contributions from individual mass bins is not strictly acceptable in general but reasonable in this context.\footnote{Appendix A2 of~\cite{BreysseAlexandroff19} derives $P_\text{clust}$ somewhat more rigorously as an integral of $\avg{\tilde{T}(k,L_1)\tilde{T}(k,L_2)}\,dL_1\,dL_2$ over source luminosities $L_1$ and $L_2$. This derivation makes clear that to arrive at the familiar form of $P_\text{clust}(k)$, we must assume source luminosities are uncorrelated at the scales where we calculate $P_\text{clust}(k)$.}

When considering the total line-intensity fluctuation (before shot noise, which we consider to be independent from these large-scale fluctuations) as the integral (or sum) of contributions from different mass bins, the Fourier-space factors from the Kaiser effect, beam profile, and line profiles should apply to each of those contributions:
\begin{align}&\frac{\sqrt{P_{\text{clust},v}(k,\mu)}}{C_{LT}}=\nonumber\\&\int dM_h\,\frac{dn}{dM_h}\,L(M_h)b(M_h)\left(1+\frac{\Omega_m(z)^{0.55}}{b(M_h)}\mu^2\right)\times\nonumber\\&\quad\exp{\left[-\frac{k^2\sigma_\perp^2(1-\mu^2)+k^2\sigma_v^2(M_h)\mu^2}{2}\right]}\sqrt{P_m(k)}.\end{align}
The bias and $\sigma_\parallel$ are mass-dependent in our model. Simplifying this a little bit,
\begin{align}P_{\text{clust},v}(k,\mu)&=C_{LT}^2P_m(k)\times\nonumber\\&\hspace{-2mm}\left\{\int dM_h\,\frac{dn}{dM_h}\,L(M_h)[b(M_h)+\Omega_m(z)^{0.55}\mu^2]\right.\nonumber\\&\quad\left.\exp{\left[-\frac{k^2\sigma_\perp^2(1-\mu^2)+k^2\sigma_v^2(M_h)\mu^2}{2}\right]}\right\}^2.\label{eq:Pclustv}\end{align}
We note with interest that while the Gaussian with exponent proportional to $\sigma_v^2$ will be weighted by $L^2$ in the shot-noise attenuation, here the corresponding Gaussian is weighted by $L$. So fainter emitters will hold more influence than with the shot-noise component, as is expected for the clustering component. If $\sigma_v$ increases with $M_h$ as is our general expectation, this means we will see less attenuation of the clustering component at high $k$ than of the shot-noise component. That said, as noted above, $P_m(k)\sim k^{-3}$ at high $k$, so while it may be theoretically possible that the lower attenuation would allow the observable clustering component to become dominant again over the observable shot-noise component at very high $k$, it would require (at least in the context of star-formation lines like CO) extremely contrived circumstances for it to occur at a scale relevant to a real-world survey.

In sum, using~\autoref{eq:Pshotv} and~\autoref{eq:Pclustv}, we can calculate the total $P(k,\mu)$---and in turn $P_0(k)$ and $P_2(k)$---incorporating mass-dependent line broadening. As long as we can formulate $v(M_h)$ and thus $\sigma_v(M_h)$, we will be able to both examine the full calculation outlined here and compare this to various approximations.

Note that for our line models, we will ultimately make the assumption that the line FWHM is rotation-dominated, which requires accounting for random orientations of line emitters relative to the observer's line of sight. We outline the appropriate corrections in~\autoref{sec:incli}, but apart from alterations in functional form, the corrections do not add significant qualitative understanding.
\section{Example Line Model: CO(1--0) Emission at Redshift 3}
\label{sec:linemodel}
It is not possible to estimate the impact of the above effect on a line-intensity signal without specifying an exact model for the line-intensity signal. Specifically, we will lay out $L(M_h)$ and $v(M_h)$ for CO(1--0) at $z\sim3$, both of which we need to calculate a line-intensity power spectrum subject to line broadening.

The CO molecule emits in a series of rotational lines, resulting from transitions between the quantized rotational energy states of the CO molecule. The line associated with the transition between rotational quantum numbers $J$ and $J-1$ has a rest frequency of approximately $J\times115.27$ GHz. Since diatomic hydrogen has no dipole moment due to symmetry and thus has no rotational transitions of its own, the CO lines are the primary way to trace molecular gas within and outside our galaxy. Emission in higher-$J$ CO lines requires more energetic environments where the molecular gas can be excited to higher energy states in the first place, so targeting lower-$J$ CO lines at high redshift gives more weight to emission from the cool molecular gas that fuels star formation.

Throughout the remainder of this work, we will make frequent reference to the COMAP Pathfinder (often simply written as COMAP), which targets the CO(1--0) line at $z\sim3$. The instrument is based at the Owens Valley Radio Observatory, with the receiver operating across observing frequencies of 26--34 GHz and the 10 m telescope providing an angular resolution of $4.5'$. The initial survey with this receiver will span three patches of 4 deg$^2$ each, with observations planned for five years.

This work considers only predictions of the signal and of line broadening, and will not make claims about the sensitivity of COMAP, such that other parameters of the survey should not be relevant here. However, we note that the scales of interest for COMAP correspond to $k\sim0.1$--0.5\,Mpc$^{-1}$, much lower than the $k\gtrsim1$\,Mpc$^{-1}$ range probed by interferometric experiments like COPSS or mmIME. As a result, COMAP chiefly targets the clustering component whereas COPSS and mmIME chiefly target the shot noise. We leave detailed contemplation of COMAP sensitivities to future work (Chung et al., in prep.).
\subsection{Halo Mass--Line Luminosity Relation}
\label{sec:LMmodel}
For the average $L(M_h)$ relation, we will use the double power-law form from~\cite{anisotropies}, which is similar to the functional form of~\cite{Padmanabhan2018a} but omits redshift evolution and has a somewhat different parametrization. We also initially model the CO luminosity in velocity-integrated observer units:
\begin{equation}
    \frac{L_\text{CO}'(M_h)}{\text{K\,km\,s}^{-1}\text{ pc}^2} = \frac{C}{(M_h/M_1)^A+(M_h/M_1)^B},
\end{equation}
with $A<B$ to distinguish the otherwise equivalent power-law slope parameters. We also prescribe log-normal scatter of $\sigma_L$ (in units of dex) around the average relation. It is straightforward to then convert CO(1--0) luminosity from the above observer units into intrinsic units of $L_\odot$:
\begin{equation}
    \frac{L_\text{CO}}{L_\odot}=4.9\times10^{-5}\left(\frac{L_\text{CO}'}{\text{K\,km\,s}^{-1}\text{ pc}^2}\right).\label{eq:Lprime_to_L}
\end{equation}

While~\cite{anisotropies} fixed parameter values to broadly match the fiducial model of~\cite{Li16}, we will not do the same in this work for two reasons. First, we want to show that whatever approach we have to calculating line broadening works not just for a specific point in parameter space, but for a range of parameters that one might realistically consider in analysis. Second, we want to update the priors and assumptions behind~\cite{Li16} to reflect high-redshift CO(1--0) observations from the past half-decade. In particular, the CO Luminosity Density at High-$z$ (COLDz) survey~\citep{COLDz,COLDzLF} used the Very Large Array (VLA) to search for CO line candidates at $z=2.0$--2.9 and obtained constraints on the CO(1--0) luminosity function at $z\approx2.4$, while the previously mentioned COPSS made a tentative detection of CO(1--0) shot noise at $z\sim3$. These are not the only recent CO observations of note at high redshift, but other surveys like ASPECS~\citep{ASPECS-LP}, PHIBBS2~\citep{PHIBBS2}, and the previously mentioned mmIME observe higher-$J$ CO lines and translate resulting constraints to CO(1--0) constraints using specific assumptions about CO line excitation, where a great deal of uncertainty (and possible variance) exists at high redshift.

A forthcoming paper (Chung et al., in prep.) will explain the derivation of our fiducial model in greater detail, as part of analysis of the first round of COMAP data. However, briefly speaking, we combine the priors on the relation between CO luminosity and star-formation rate from~\cite{Li16} with the best-fit values and 68\% intervals for the parameters of the star-formation rate model of~\cite{Behroozi19}\footnote{Note that instead of the official Data Release 1, we use the Early Data Release best-fit model; the changes between the two versions are small enough that the differences in all relevant best-fit parameter values for the~\cite{Behroozi19} model at $z\approx2.4$ are subdominant to their uncertainties and to the uncertainties in the other parts of our CO model.} to derive empirical priors at $z\approx2.4$:
\begin{align}A&= -1.66\pm2.33,\\B&= 0.04\pm1.26,\\\log{C}&= 10.25\pm5.29,\\\log{(M_1/M_\odot)}&= 12.41\pm1.77.\end{align}
We also set an initial prior of $\sigma_L=0.4\pm0.2$ (dex). The best estimate comes from the 0.37 dex total scatter in the~\cite{Li16} fiducial model, but we assume a somewhat broader prior on $\sigma_L$ compared to~\cite{Li16}. As with the previous models of~\cite{Li16} and~\cite{anisotropies}, we also set a minimum halo mass of $10^{10}\,M_\odot$ for line emission, and set $L(M_h<10^{10}\,M_\odot)=0$.

We then use the luminosity function constraints from COLDz to formulate a likelihood function, and run a Markov chain Monte Carlo (MCMC) inference using \texttt{emcee}~\citep{emcee} to obtain a posterior distribution from the above priors and our COLDz-based likelihood. (We also ran a MCMC simulation incorporating the COPSS result into the likelihood, but did not find the posterior changed significantly.) Unlike the rest of this work, this MCMC procedure uses a snapshot from the BolshoiP simulation~\citep{MultiDark} to simulate CO emitters, and thus uses the cosmology from~\cite{Planck15}, all to match~\cite{Behroozi19} (from whose data release we source the halo catalogue). (Specifically, we use the snapshot closest to the COLDz central redshift of $z\approx2.4$.)

\begin{figure*}
\centering\includegraphics[width=0.48\linewidth]{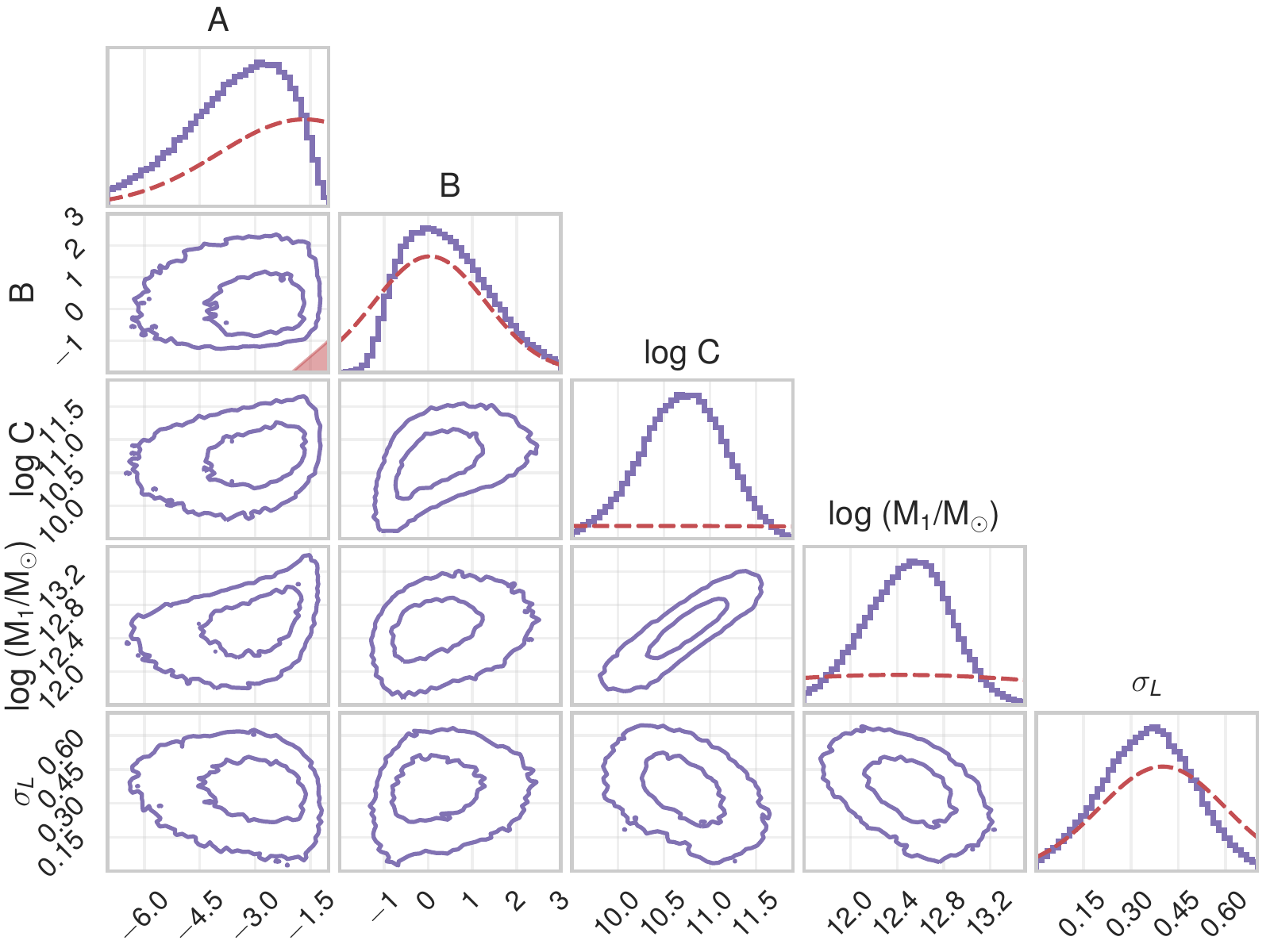}\quad\includegraphics[width=0.48\linewidth]{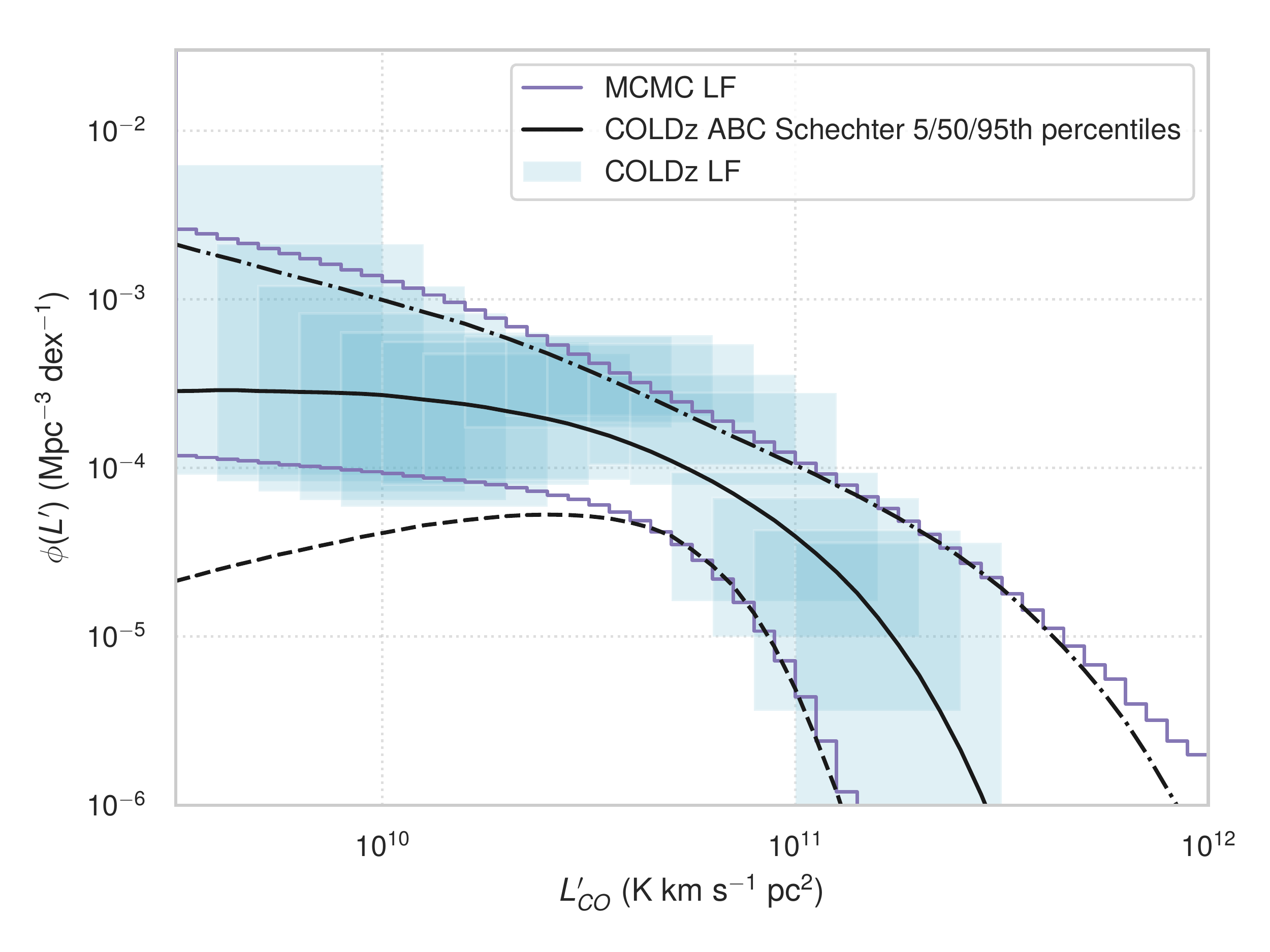}

\includegraphics[width=0.48\linewidth]{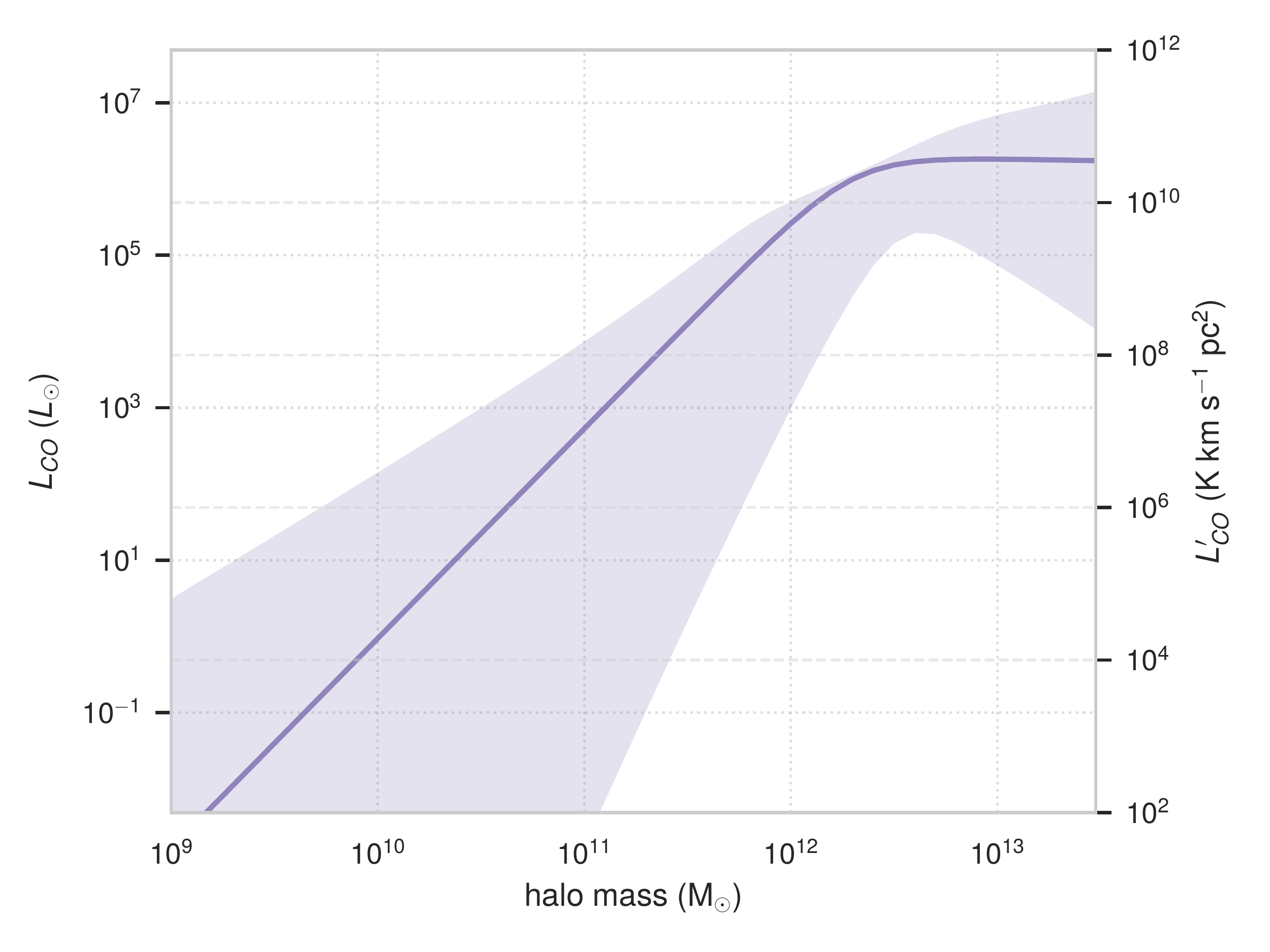}\quad \includegraphics[width=0.48\linewidth]{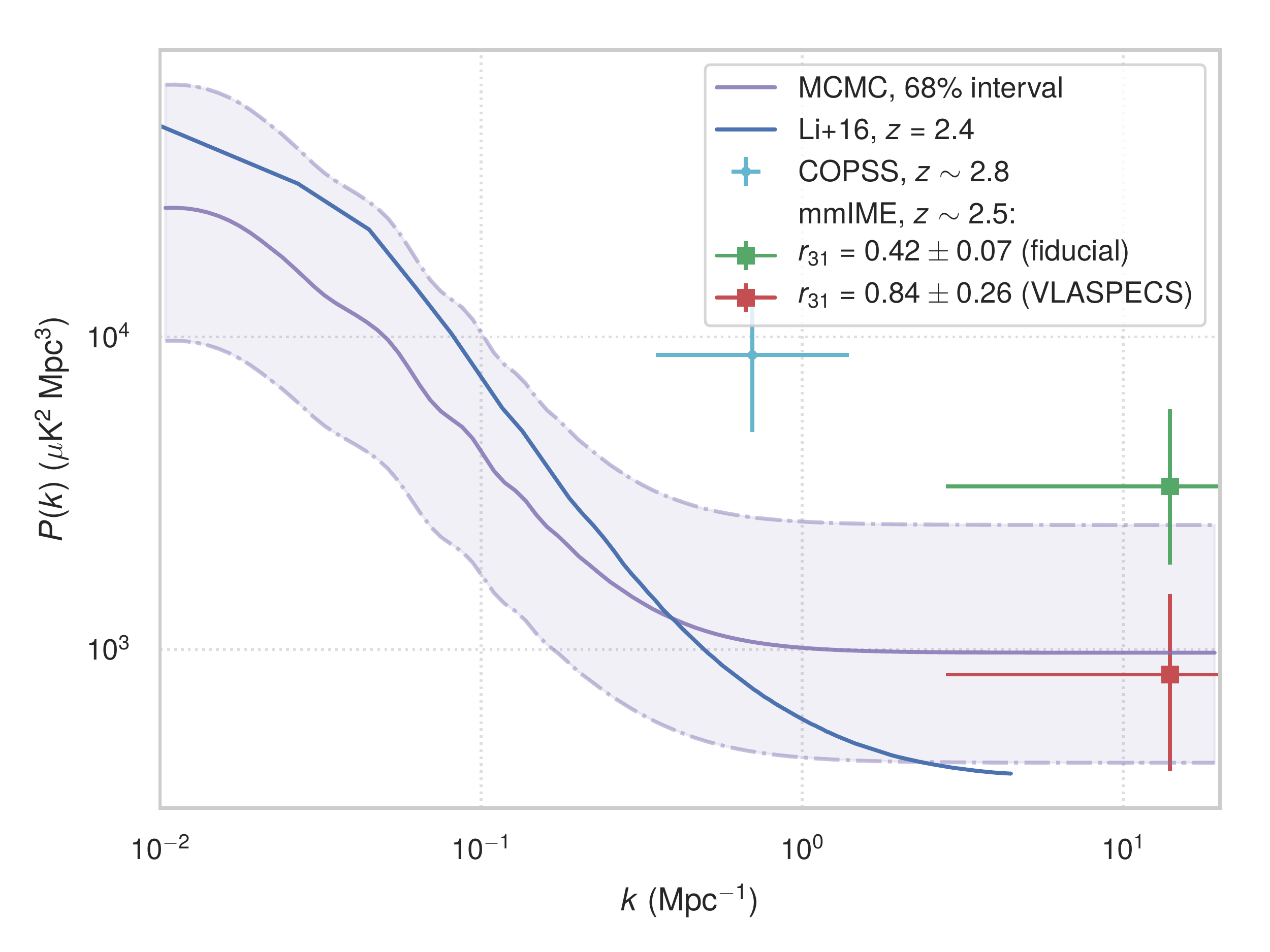}
\caption{Illustrations of the fiducial model ensemble for CO(1--0) emission at $z\sim3$. \emph{Upper left:} The model parameter posterior distribution from the MCMC combining our priors (red dashed lines in marginalized posterior plots) with a likelihood based on the COLDz ABC constraints. \emph{Upper right:} The luminosity function posterior distribution calculated from the MCMC. We show 90\% intervals for the MCMC (purple), the COLDz direct constraints (cyan shaded rectangles), and the COLDz ABC constraints on a Schechter luminosity function (black dashed, solid, and dash-dotted showing 5\%, 50\%, and 95\% percentiles). \emph{Lower left:} The 90\% interval for $L(M_h)$ from the MCMC, with the relation given by the parameter values of Equations~\ref{eq:realparami}--\ref{eq:realparamf} overplotted. \emph{Lower right:} Predictions from the MCMC for the CO(1--0) $P(k)$. We compare the 68\% interval from the MCMC (purple shaded area) to the fiducial model of~\protect\cite{Li16} at $z=2.4$ (blue) and to the direct measurement from COPSS (cyan), as well as estimates based on mmIME data plus either the fiducial $r_{31}$ used by~\protect\cite{mmIME-ACA} (green) or the~\protect\cite{VLASPECS} value based on VLASPECS follow-up (red).}
\label{fig:test}
\end{figure*}

Our MCMC output is a fiducial sample of models that we can use to validate models of line broadening across a reasonable range of $L(M_h)$ parameter values. The posterior distribution is shown in~\autoref{fig:test}, as are the empirically derived MCMC priors for comparison in the marginalized posterior plots. The COLDz data significantly constrain the characteristic mass and luminosity scales $M_1$ and $C$ beyond our priors, as well as put meaningful limits on the power-law slopes $A$ and $B$. We strongly favour a super-linear $L(M_h)$ relation at the faint end, with the power-law break point determined by $\log{C}\in(9.9,11.4)$ and $\log{(M_1/M_\odot)}\in(11.8,13.1)$ (both 90\% marginalized intervals).

We also show the 90\% intervals for the CO luminosity function and $L(M_h)$ relation. Alongside the luminosity function interval, we also show the COLDz results from~\cite{COLDzLF}, both as direct constraints on the luminosity function and as an approximate Bayesian computation (ABC) of constraints on a Schechter function description of the luminosity function. (Note that our MCMC likelihood function was based on the latter.) Our MCMC interval matches both quite well, although our interval favours a steeper faint-end luminosity function compared to the COLDz ABC constraints.

Although the inference is done at $z\approx2.4$, we apply the parameters without change to the COMAP central redshift of $z\approx2.8$ as we do not expect much evolution in cosmic star-formation activity between these redshifts. Any evolution would be subdominant to model uncertainties, so a sample of models that would be likely realistic for $z\approx2.4$ would largely be equally likely realistic for $z\approx2.8$ given our low level of information about high-redshift CO(1--0) emission. We will also revert from the~\cite{Planck15} cosmology to the fiducial cosmology for this work without altering our model values, as again uncertainties in cosmology are far subdominant to model uncertainties.

For reference, when we tune our $L(M_h)$ parameters to make the model $P(k)$ and luminosity function at $z\approx2.8$ approximately match the median $P(k)$ and luminosity function from our MCMC posterior distribution (again, found originally at $z\approx2.4$), the resulting representative set of parameters is as follows:
\begin{align}
A &= -2.75,\label{eq:realparami}\\
B &= 0.05,\\
\log{C} &= 10.6,\\
\log{(M_1/M_\odot)} &= 12.3,\\
\sigma_L &= 0.42.\label{eq:realparamf}
\end{align}
We show the relation based on these parameters alongside the 90\% MCMC interval for $L(M_h)$. The above parameters result in a $L(M_h)$ relation slightly more optimistic than the median (reflected mostly by the relatively low value of $M_1$), which is a counter-reaction to the slight downward shift of the halo mass function when moving from $z\approx2.4$ to $z\approx2.8$. However, the figure shows that the resulting $L(M_h)$ from the above parameter values still falls well within the 90\% sample interval. This should demonstrate that any shift in our best estimate between these two redshifts is subdominant compared to the model uncertainties involved.

We also compare our model distribution of the real-space $P(k)$ to the fiducial model of~\cite{Li16} previously used for COMAP forecasts, as well as the current extent of CO intensity mapping measurements from COPSS and mmIME. In particular, the mmIME $z\sim2.5$ estimate is converted from an estimate of CO(3--2) shot noise, and~\cite{mmIME-ACA} use a line-luminosity ratio $r_{31}=L'_\text{CO(3--2)}/L'_\text{CO(1--0)}$ from~\cite{Daddi15} of $0.42\pm0.07$. However, this is an average based on three near-IR selected `normal' star-forming galaxies at $z=1.5$. Meanwhile, follow-up of CO(3--2) detections from ASPECS~\citep{ASPECS-LP} by the VLA-ALMA SPECtroscopic Survey (VLASPECS) in the Hubble Ultra-Deep Field~\citep{VLASPECS} resulted in three robust CO(1--0) detections for which the line ratios were found to be closer to 0.8--1.1, the best overall estimate being $r_{31}=0.84\pm0.26$. Therefore, we show the mmIME result both with the fiducial $r_{31}$ from~\cite{Daddi15} used by~\cite{mmIME-ACA} and with the higher $r_{31}$ value from~\cite{VLASPECS}, to illustrate the level of uncertainty around the conversion from CO(3--2) to CO(1--0).

Our new model tends to predict higher shot noise and a dimmer clustering signal compared to~\cite{Li16}, and is also in some tension with the COPSS result. However, our predictions are very consistent with the mmIME results, particularly if the applicable $r_{31}$ value is higher than the value used by~\cite{mmIME-ACA}. That said, we do caution that neither of the experimental results include corrections for line broadening, which is the very effect we are setting out to describe.

\subsection{Halo Mass--Line FWHM Relation}
\label{sec:LWmodel}
Various classes of galaxies demonstrate well-measured correlations between galaxy luminosity and velocity scales---between luminosity and rotation velocity in disk galaxies as first identified by~\cite{TullyFisher77}, or between luminosity and velocity dispersion in elliptical galaxies as first identified by~\cite{FaberJackson76}. However, measuring such correlations requires fine observations across many galaxies of line profiles as well as morphology and inclination, which is easier in the local Universe than at $z\gtrsim2$.

Because we have comparatively limited information about the CO(1--0) line profiles of high-redshift galaxies, we will define our $v(M_h)$ model in two steps. We first examine the line FWHM in observations to consider $v(M_h)$ for sources around or brighter than the knee of the luminosity function---which broadly speaking should correspond to the characteristic mass and luminosity scales of our double power-law $L(M_h)$---and then consider how best to extrapolate to lower $M_h$ where we have no information.
\subsubsection{Average Line FWHM at Characteristic Mass}
\label{sec:LWchara}
While current constraints on the double power-law slopes are limited, observations do sufficiently probe the knee of the $z\gtrsim2$ CO luminosity function for information about the characteristic mass and luminosity scales at the double power-law break point, as we have already noted. Since observations identify individual line candidates with associated integrated line fluxes and line FWHM values, we can use these to get an approximate correspondence between CO luminosity and line FWHM, and in turn between halo mass and line FWHM, at these characteristic mass/luminosity scales.

To be clear, this is not a correlation we expect to be statistically very strong---the connection between halo properties and molecular gas dynamics is highly indirect, and we expect CO line FWHM for a given host halo mass to be highly variable. However, even an idea of the \emph{average} CO line FWHM at a characteristic halo virial mass would help us model line broadening in a way acceptable at the $P(k)$ and VID level, where many emitters are sampled simultaneously and thus the variability may not be as relevant as in the context of scanning for individual line candidates.

As when formulating the MCMC likelihood used in our $L(M_h)$ model above, we only consider CO-selected observations of CO(1--0) and no observations of higher-$J$ CO lines. Of the surveys mentioned in~\autoref{sec:LMmodel}, the COLDz survey found four secure line candidates at $z\sim2$--3~\citep{COLDz}, and~\cite{VLASPECS} reported three secure CO(1--0) line detections from VLASPECS as previously discussed. \autoref{tab:linewidths} lists these seven lines and their properties.

For the sources from~\cite{COLDz}, we convert the integrated line flux $S\,\Delta\nu$ to the observed line luminosity $L'$, via this relation (as found in, e.g.,~\citealt{SDR92}):

\begin{equation}
    L' = \frac{c^2}{2k}(S\,\Delta\nu)\frac{D_L^2}{\nu_{\text{obs}}^2(1+z)^3}.
\end{equation}
In a flat universe as described by our fiducial cosmology, the luminosity distance is simply the comoving distance $R(z)$ divided by the scale factor; since $\nu_{\text{obs}}=\nu_{\text{rest}}/(1+z)$ is the rest-frame line frequency multiplied by the scale factor,
\begin{equation}
    L' = \frac{c^2}{2k}(S\,\Delta\nu)\frac{(1+z)R^2(z)}{\nu_{\text{rest}}^2}.
\end{equation}
With $\nu_{\text{rest}} = 115.26$ GHz for CO(1--0), this equation becomes
\begin{equation}
    \frac{L'_{\text{CO(1--0)}}}{\text{K\,km\,s}^{-1}\text{ pc}^2} = 2.45\times{10}^{3}(1+z)\left(\frac{R(z)}{\text{Mpc}}\right)^2\frac{S\,\Delta\nu}{\text{Jy\,km\,s}^{-1}}.
\end{equation}

\begin{deluxetable*}{lcccl}
\tablecaption{Properties of compiled secure CO(1--0) line detections at $z\sim2$--3.\label{tab:linewidths}}
\tablehead{
\colhead{ID} & \colhead{Redshift} & \colhead{Line FWHM} & \colhead{$L'_{\text{CO(1--0)}}$} & \colhead{Reference} \\ & & \colhead{(km\,s$^{-1}$)} & \colhead{(${10}^{10}$ K\,km\,s$^{-1}$ pc$^2$)}}
\startdata
COLDz.COS.1&2.6675&$430\pm\phantom{0}80$&$\phantom{0}3.68\pm0.10$&\cite{COLDz}\\
COLDz.COS.2&2.4771&$830\pm130$&$\phantom{0}3.83\pm0.09$&\cite{COLDz}\\
COLDz.COS.3&1.9692&$240\pm \phantom{0}50$&$\phantom{0}7.27\pm1.97$&\cite{COLDz}\\
COLDz.GN.3 &2.4877&$580\pm120$&$10.08\pm3.56$&\cite{COLDz}\\
ASPECS-LP.9mm.1&2.5437&$447\pm110$&$\phantom{0}3.22\pm0.68$&\cite{VLASPECS}\\
ASPECS-LP.9mm.2&2.6976&$201\pm\phantom{0} 47$&$\phantom{0}1.32\pm0.19$&\cite{VLASPECS}\\
ASPECS-LP.9mm.3&2.6956&$560\pm230$&$\phantom{0}3.14\pm0.75$&\cite{VLASPECS}\enddata
\tablecomments{\protect\cite{COLDz} provide $S\,\Delta\nu$ rather than $L'_{\text{CO(1--0)}}$; the main text explains how we calculate the latter from the former. \protect\cite{VLASPECS} provide the $L'_{\text{CO(1--0)}}$ values shown above, and use a fiducial cosmology sufficiently similar to ours that we do not re-calculate luminosities.}
\end{deluxetable*}

\begin{figure}[htpb!]
    \centering
    \includegraphics[width=0.96\linewidth]{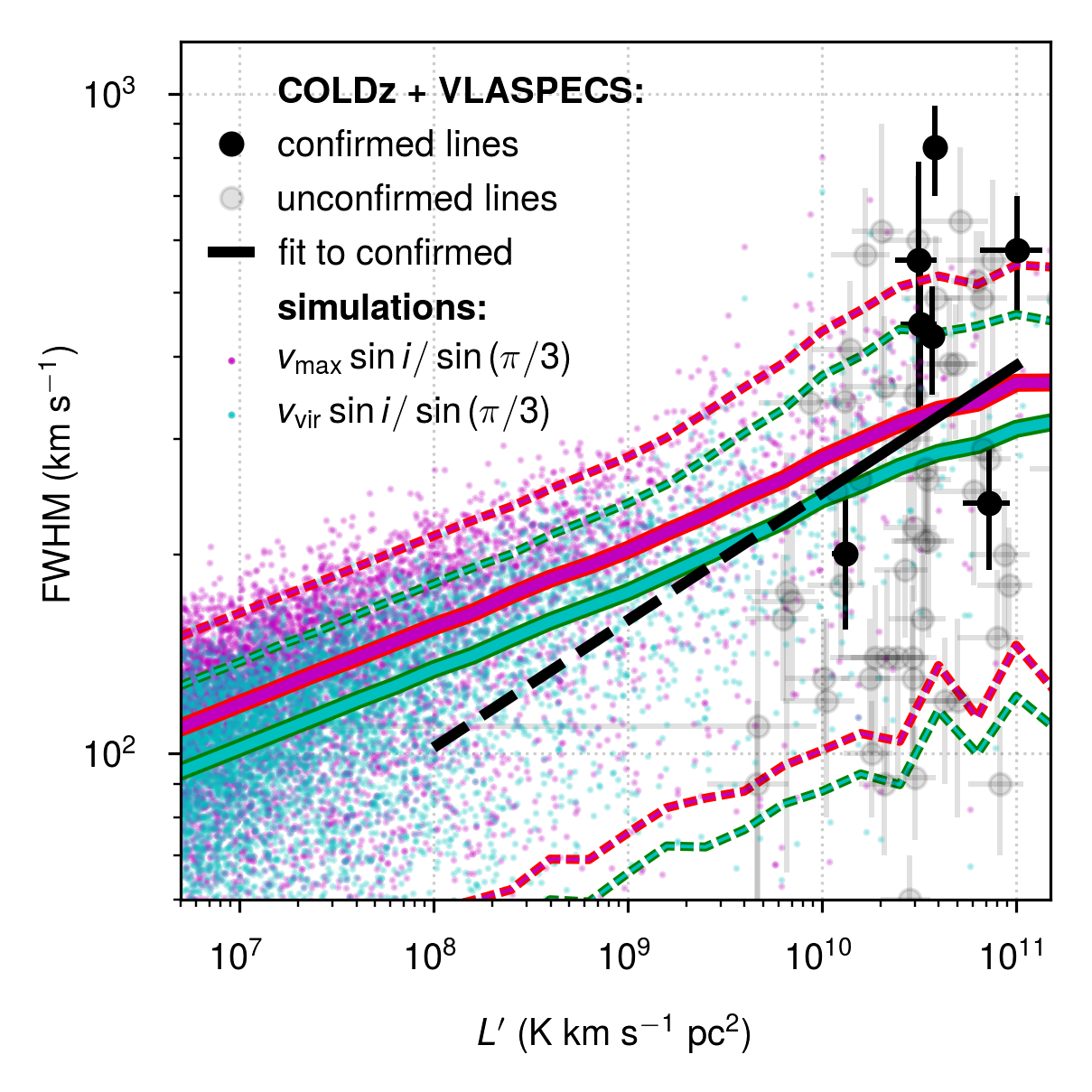}
    \caption{Plot of line luminosity $L'$ against line FWHM for the seven sources (black points with error bars) described in~\autoref{tab:linewidths}, along with the power-law fit (black solid line) described in the main text. We also show unconfirmed line candidates (grey points with error bars) from~\protect\cite{COLDz} and~\protect\cite{VLASPECS}, with no attempt made to correct for any magnification. We show the inclination-adjusted FWHM estimated from $v_\text{vir}$ (cyan) and $v_\text{max}$ (magenta) for a random subset (to avoid overcrowding the plot) of the halo catalogue of a sample lightcone from \replaced{our simulations (to be detailed in~\autoref{sec:sim})}{simulations detailed in~\autoref{sec:sim}}, with $L'$ calculated based on the fiducial model with representative values from Equations~\ref{eq:realparami}--\ref{eq:realparamf}, again assuming no magnification.\added{ Simulated FWHM for a given $L'$ varies due to random inclinations and scatter in $L(M_h)$, but average $L(M_h)$ and $v(M_h)$ relations are fixed.} We also show 90\% intervals for both velocities (cyan and magenta dashed lines) at each $L'$, to more clearly show the difference in the scaling of velocity with $L'$ versus our na\"{i}vely extrapolated power-law fit (black dashed).}
    \label{fig:linewidths}
\end{figure}

A fit across all seven line candidates to a power-law model, such that
\begin{equation}
    \log{\frac{\text{FWHM}}{\text{km\,s}^{-1}}}=\delta_0+\delta_1\log{\frac{L'}{10^{10}\text{ K\,km\,s}^{-1}\text{ pc}^2}},
    \label{eq:vLpowerlaw}
\end{equation}
finds $\delta_0=2.395\pm0.208$ and $\delta_1=0.193\pm0.302$. We show the fit graphically in~\autoref{fig:linewidths}. While the correlation is not statistically significant, it is broadly consistent with the `spherical' and `disk' model $L'$--FWHM relations of~\cite{ASPECS-LPISM}, which are examples given for comparison with estimated CO(1--0) luminosities based on higher-$J$ CO lines observed by ASPECS, and correspond to $\delta_0=2.14$ or 2.56 (for `spherical' and `disk' respectively) and $\delta_1=0.5$ (for both). Our fit is also consistent with correlations found from targeted observations of sub-millimetre galaxies. \cite{Harris12}, for instance, find a fit equivalent to $\delta_0=2.33$ and $\delta_1=0.59$ based on CO(1--0) observations only, and \cite{Goto15} find a fit equivalent to $\delta_0=2.52$ and $\delta_1=0.24$ also incorporating higher-$J$ observations.

Note that both~\cite{Harris12} and~\cite{Goto15} find an intrinsic luminosity by dividing $L'$ by the magnification $\mu$ in the presence of a gravitational lens where known, but such information is not available for the line candidates from~\cite{COLDz} or~\cite{VLASPECS}. Our assumption in formulating the $L(M_h)$ model has been that $\mu=1$, and if we did not believe this then we would replace $L'$ with $L'/\mu$ in the right-hand side of~\autoref{eq:Lprime_to_L}. We will assume that $\mu\approx1$ for the sources in~\autoref{tab:linewidths}, although we caution that three of the seven sources do show at least marginally resolved spatial extension.

Note also the lack of information on the inclination angle $i$ of the CO emitter's axis of rotation (with respect to the observer's line of sight), which would scale the line profile by $\sin{i}$. The assumption that sources are randomly oriented corresponds to a uniform distribution of $\cos{i}$, such that the observed line FWHM is scaled down from the intrinsic rotation-dominated line FWHM by a median multiplier of $\sqrt{3}/2\approx0.866$, and by no less than a multiplier of $1/2$ in $\approx86.6$\% of cases. Therefore, we will not consider any explicit corrections to the average $v(M_h)$ based on $i$, given the lack of information in high-redshift observations.

We will in fact incorporate random $i$ in the detailed simulations presented in~\autoref{sec:sim}, as shown in~\autoref{fig:linewidths}. But reflecting the absence of any corrections for $i$ in the above work, we will use a correction of $\sin{i}/0.866=\sin{i}/\sin{(\pi/3)}$ rather than $\sin{i}$ by itself, so that the correction is relative to the median adjustment for inclination. The implicit assumption here is that all of our emitters are disc-like or at least primarily rotationally supported; we will address this issue further in~\autoref{sec:rot_vs_disp}.

We also note that there are more unconfirmed line candidates from both~\cite{COLDz} and~\cite{VLASPECS}, which we partially show in grey in~\autoref{fig:linewidths}. Given that the $z\sim3$ CO(1--0) identification presumed for all of the unconfirmed line candidates is not definitive and other potential biases exist in flux or luminosity recovery, we will not present a fit using both secure and unconfirmed line candidates. Designing a procedure to infer a $L'$--FWHM relation while accounting for source fidelity (as COLDz or ASPECS would in inferring the luminosity function) is beyond the scope of this paper.

In any event, our own fit suggests that on average, assuming negligible magnification for our secure line candidates, CO emitters with $M_h=M_1$ and thus \replaced{$L'$}{$\log{L'}$} in a fiducial 90\% confidence interval\footnote{This interval does not account for log-normal scatter in $L(M_h)$, but that scatter tends to be modest with typical values of $\sigma_L\approx0.3$--0.4 dex in model space. The upper left part of~\autoref{fig:test} suggests that values of $\sigma_L$ above 0.5 dex would be considered unusual. Note also that when we apply log-normal scatter in this work, we do so while preserving the linear mean $L(M_h)$.} of ${\log{(C/2)}\in(9.6,11.1)}$ should have line FWHM values in the range of 210--400~km~s$^{-1}$, with the log-space midpoint at around 290\,km\,s$^{-1}$.

How does this compare to the virial velocity expected from a halo with virial mass in our 90\% confidence interval of $\log{M}\in(11.8,13.1)$? This mass range corresponds to a virial velocity range of approximately 200--540\,km\,s$^{-1}$, with the log-space midpoint at 330\,km\,s$^{-1}$. Therefore, at the double power-law break point, we find that the FWHM of a CO line profile approximately lines up with the virial velocity of its host dark matter halo.

The virial velocity $v_{\text{vir}}$ and maximum circular velocity $v_{\text{max}}$ of a halo are typically within a factor of order unity of each other. We suppose that the maximum circular velocity calculated by the halo finder, given that it reflects halo dynamics, may best reflect dynamics associated with the hosted molecular gas. If both velocities are available, $v_{\text{max}}$ takes precedence as the FWHM for our simulated CO line profiles.

That the maximum circular velocity of the overall mass profile should be equal to the full width of the CO line profile, as opposed to the velocity dispersion (smaller than the FWHM by a factor of $2\sqrt{2\ln{2}}\approx2.355$), is not unreasonable. While the velocity dispersion of atomic hydrogen is usually similar to that of matter (chiefly dark matter) for the $M_h>10^{10}\,M_\odot$ halo population that we consider (see, e.g., Figure 13 of~\citealt{VN18}), the distribution of molecular gas in a galaxy is considerably more compact even than that of atomic gas, let alone matter in general. As a result of this compactness, CO profiles take on a shape closer to a Gaussian\footnote{Both~\cite{COLDz} and~\cite{VLASPECS} use Gaussian fits to obtain the line FWHM values that inform our $v(M_h)$ model.} than the double-horned profile typical of 21 cm line emission, and are also likely to trace velocity widths at $\sim2.5\times$ smaller radii than 21 cm profiles, resulting in line FWHM values smaller by up to a similar factor~\citep{deBlokFabian14}.

Before we move to prescribe the line FWHM for all halo masses, note that our model of the galaxy--halo connection does not specify a non-trivial halo occupation distribution (HOD). While we make the very simple assumption of one CO emitter per halo in this work, explicitly assigning multiple CO emitters to a high-mass halo would somewhat push down the simulated values of $L'$ for each emitter, allowing us to better explain some of the higher-FWHM line candidates from COLDz and VLASPECS shown in~\autoref{fig:linewidths}. While we consider a HOD model for CO emitters to be beyond the scope of this paper, it could be an important consideration for future work due to these sorts of effects on the contribution of high-mass halos to the simulated CO signal.

\subsubsection{Average Line FWHM at All Halo Masses}
The above work makes a case---if a highly tentative one---that at the double power-law break point specifically, the FWHM of the CO emitter line profile is approximately equal to the host halo maximum circular velocity (or virial velocity, largely similarly). However, this still raises the question of how the CO line FWHM scales on average away from this characteristic scale.

If we were to extrapolate the fit between FWHM and $L'$ to lower $L'$, it would suggest that the line FWHM scales approximately as $L'^{1/5}$. Our priors are consistent with fairly sharp $L(M_h)$ scalings at the faint end like $L'\sim M_\mathrm{vir}^{2.5}$, which suggests that the line FWHM would scale almost as the square root of the host halo mass. By contrast, the circular velocity approximately scales as $M_\mathrm{vir}^{1/3}$ on average, which is a fairly weak scaling. \autoref{fig:linewidths} illustrates how circular velocity (virial or maximum) scales more weakly with $L'$ than the extrapolated FWHM fit. 

However, we will adopt the more conservative prescription that even at lower halo mass, the CO line profile FWHM is equal (again, \emph{on average}, specifically for the median inclination angle of $\pi/3$) to the halo maximum circular velocity. This reflects not necessarily high confidence in this prescription \emph{per se}, but rather the low amount of information we have about high-redshift CO. Our simulations include emitters with luminosities several orders of magnitude below the double power-law break point; therefore, strictly in principle, extrapolation of a fit across points that barely span one order of magnitude in $L'$ is unsafe in comparison to the assumption that rotation of molecular gas in a galaxy will scale with host halo mass in roughly the same way as the host halo's rotation.

Furthermore, for a sufficiently faint CO emitter (or light host halo), the velocity dispersion of the CO gas itself will begin to dominate over rotational dynamics, weakening any scaling of FWHM with host halo mass. Our assumed minimum luminous host halo mass is $M_\text{vir}=10^{10}\,M_\odot$, corresponding to a virial velocity of approximately 50\,km\,s$^{-1}$---roughly equal to the expected gas velocity dispersion for high-redshift galaxies~\citep{deBlokFabian14}.

\subsubsection{Additional Notes on Inclination and Scatter}
\label{sec:rot_vs_disp}
We noted above that we will assume CO emitters are randomly oriented and that this reduces the observed line FWHM by the sine of the inclination angle relative to the FWHM for an edge-on emitter. Introducing this random inclination angle for all emitters raises the question of whether we ought to assume CO emitters at $z\sim3$ are rotation-dominated, as such corrections for inclination should not apply to dispersion-dominated galaxies where random motions of molecular gas rather than galactic rotation contribute to most of the observed line profile.

So far we have made use of data from direct observations of CO(1--0) lines, but it is challenging to study kinematics across large numbers of such sources given the spatial resolution required and the long wavelength. Even looking to slightly higher-$J$ lines, blind line searches like ASPECS---which perhaps provide some of the larger samples of CO-selected galaxies in the literature---are not well-suited for kinematic studies as sources are spatially only marginally resolved. (That said, some CO-selected ASPECS sources do show clear rotation-dominated velocity gradients---see Appendix D of \citealt{ASPECS-LPISM}.)

As this cosmic epoch shows much greater star-formation activity compared to $z\sim0$, with a much more dominant interstellar gas component for fuel, there is a good reason to suspect that the fraction of rotation-dominated galaxies is much smaller than in the local universe. Targeted studies of ionized gas kinematics across large numbers of star-forming galaxies at $z\gtrsim1$ suggest that indeed at high redshift, rotation-dominated galaxies are far from an overwhelming majority. For example,~\cite{Wisnioski19} find a steadily declining fraction of rotation-dominated galaxies from 91\% at $z\sim1$ to 70\% at $z\sim2$, and~\cite{Turner17} find an even lower fraction of $34\pm8$\% at $z\sim3.5$.

It is safe to assume that similar physical considerations apply for molecular gas, so a na\"{i}ve prescription would be to model a rotation-dominated inclination-adjusted line profile for half of our CO emitters, and model a dispersion-dominated inclination-independent line profile for the other half. But it is not immediately clear that this is the correct tack to take. Depending on the exact criteria for deciding that a galaxy is `rotation-dominated', the circular rotation may still be a significant even if not dominant source of support for its dynamical mass, and thus a significant contributor to the line profile width.

Furthermore, while the works discussed above find a less-than-overwhelming majority of galaxies to be rotation-dominated, ground-based near-infrared observations of ionized gas are often susceptible to spatial resolution effects that mask lower rotation velocities, leading to smaller rotation-dominated systems being classified as dispersion-dominated. The work of~\cite{Newman13} provides a striking illustration of the effect with a sample of $z\sim2.2$ galaxies, where of 34 galaxies observed with adaptive optics, seeing-limited data would lead to 41\% being classified as dispersion-dominated, but using higher-resolution adaptive optics data would drop this fraction to 6--9\%. Although kinematics studies like~\cite{Turner17} and~\cite{Wisnioski19} will always attempt to correct for beam smearing, observational classification of galaxies as rotation- or dispersion-dominated is still far richer with complexities than the apparent simple binary would suggest at first glance.

Overall, there are not enough observational data---certainly not directly in CO lines---to advise against the assumption that a majority of CO emitters at $z\sim3$ are rotationally supported, at least somewhat disc-like systems. So for the remainder of this work, when we apply inclination corrections, we will apply them to all simulated emitters.

Note that the random inclination angles assigned to each emitter will be the only source of random scatter in $v(M_h)$. While for $L(M_h)$ we introduce random log-normal scatter on top of the average relation described by $\sigma_L$, we will not take similar steps for $v(M_h)$, or at least not explicitly through a parameter analogous to $\sigma_L$. The already-specified scatter in $L(M_h)$ actually accounts for much of the observed variation in FWHM given $L'$ for high-redshift CO emitters. Variations due to random inclination are smaller and skewed, but sufficient to account for any remaining variation (as possibly seen when including unconfirmed line candidates in~\autoref{fig:linewidths}). Beyond these two sources of scatter, we find insufficient information to support any empirically motivated non-specific scatter (log-normal or otherwise) in FWHM for fixed halo mass and redshift in the way we do for $L(M_h)$.
\section{Possible Simplifications}
\label{sec:approx}
Taking all of the above into account, with infinite computing time and space available to us, we would simulate line broadening in lightcones from cosmological N-body simulations by simulating a Gaussian line profile for each individual halo, taking the halo $v_\text{max}$ as the line width. Given the typical halo count ($\sim10^6$) in a simulated COMAP survey volume, this is infeasible. A more efficient approach would be binning in halo mass and applying Gaussian filters based on an average $v_\text{max}$ to the CO map generated from each mass bin. However, for a meaningful number of mass bins ($\sim100$), even this is too computationally expensive to be incorporated into a MCMC step that would complete within a reasonable amount of time. Furthermore, if we were to follow~\cite{Ihle19} and use approximate N-body simulations provided by the peak-patch method~\citep{mPP}, which do not in their current state provide halo properties like $v_\text{max}$, we would have to rely on calculations based on halo mass.

Therefore, we will consider some approaches that will use $v_\text{vir}(M_h)$ and make further simplifications. One approach is to use a single Gaussian filter with an effective velocity scale $v_\text{eff}$ to describe the broadening of the total CO line-intensity cube. This comes at the cost of some accuracy, but would bring significantly improved computational speed in any contexts where the approximation is applicable. The goal of~\autoref{sec:selwdesign} is to obtain a prescription for $v_\text{eff}$ that results in the same $P(k)$ attenuation as a simulation with halo mass bins broadened by $v_\text{max}$, to within $\sim10$\% up to $k=0.7$\,Mpc$^{-1}$ (beyond which attenuation of $P(k)$ due to the COMAP angular beam will exceed 50\%, even without any line-of-sight smearing). We then consider an alternate approach in~\autoref{sec:2tier} that still uses multiple bins in $v_\text{vir}$ but designs these bins more carefully to reduce the number of bins required and thus reduce computational burden.

\subsection{Use of a Single Effective Line Width}
\label{sec:selwdesign}
A description of line broadening using a single effective line width is extremely desirable as long as it achieves a reasonable accuracy. If we can design such a $v_\text{eff}$, we could treat the corresponding $\sigma_{\parallel,\text{eff}}$ the same way as the mass-independent $\sigma_\parallel$ in~\autoref{eq:Pconv}. This would in turn drastically simplify the computational work involved compared to the full calculation described in~\autoref{eq:Pshotv} and~\autoref{eq:Pclustv}. In a simulation using dark matter halo catalogues, the only necessary step to incorporate line broadening would be to apply a single Gaussian filter along the line of sight with its profile given by $v_\text{eff}$, as opposed to creating dozens of mass or velocity bins with individual profile widths.

To design $v_\text{eff}$, we focus on the shot-noise component of the power spectrum. Not only is attenuation stronger at higher wavenumbers---where the shot noise dominates $P(k)$---but also the emitters that contribute most to shot noise will tend to see greater line broadening.

Concentrating on shot noise attenuation allows us to state the problem in mathematical terms. A Gaussian line profile with FWHM of $v_\text{FWHM}$ in velocity units can be translated into a Gaussian profile in comoving space, with the standard deviation given by the same relation as~\autoref{eq:veltosig},
\begin{equation}
    \sigma_\parallel(v_\text{FWHM})=\frac{(1+z)}{H(z)}\frac{v_\text{FWHM}}{2\sqrt{2\ln{2}}}.\label{eq:vfwhmtosig}
\end{equation}
The virial velocity $v_\text{vir}$ is quite close to $v_\text{max}$---a visual inspection of~\autoref{fig:linewidths} suggests they fall within 20\% of each other. But unlike $v_\text{max}$, $v_\text{vir}$ is explicitly a function of the halo virial mass $M_h$ and redshift $z$:
\begin{align}
    v_\text{vir}&=\sqrt{\frac{GM_h}{r_\text{vir}}}=\left(\frac{\Delta_c}{2}\right)^{1/6}\left[GM_hH(z)\right]^{1/3}\\&\approx35\text{\,km\,s}^{-1}\left(\frac{\Delta_c}{200}\right)^{1/6}\times\nonumber\\&\qquad\left(\frac{M_h}{10^{10}\,M_\odot}\frac{H(z)}{100\text{\,km\,s}^{-1}\text{\,Mpc}^{-1}}\right)^{1/3},
    \end{align}
where $\Delta_c$ is the spherical overdensity relative to the critical density $\rho_\text{crit}$, and thus relates the virial mass and radius:
\begin{equation}
    r_\text{vir} = \left(\frac{3M_h}{4\pi\Delta_c\rho_\text{crit}}\right)^{1/3} = \left(\frac{2 GM_h}{\Delta_cH^2(z)}\right)^{1/3}.
\end{equation}
We use the definition of $\Delta_c$ and thus virial mass from~\cite{BryanNorman98}, which yields $\Delta_c\approx180$ for our cosmology and redshift. For reference, $H(z)$ at the central COMAP redshift is approximately $290\text{\,km\,s}^{-1}\text{\,Mpc}^{-1}$.

We can convert $v_\text{vir}(M_h)$ into the comoving $\sigma_v(M_h)$ expected for a CO emitter with host halo mass $M_h$ using~\autoref{eq:veltosig}. We would then feed this into~\autoref{eq:Pshotv} and integrate in $\mu$ to carry out a full calculation of the shot noise with line broadening.

Compare to using a single mass-independent velocity scale $v_\text{eff}$, with corresponding $\sigma_{\parallel,\text{eff}}=\sigma_\parallel(v_\text{eff})$. Based on Equation 22 of~\cite{anisotropies}, considering only the line-of-sight attenuation, the observed shot noise is the true $P_\text{shot}$ multiplied by
\begin{equation}
    S_0 = \frac{\pi^{1/2}\operatorname{erf}{(k\sigma_{\parallel,\text{eff}})}}{2k\sigma_{\parallel,\text{eff}}}.\label{eq:S0}
\end{equation}
Then we want to identify $v_\text{eff}$ such that
\begin{align}&\frac{\pi^{1/2}\operatorname{erf}{(k\sigma_{\parallel,\text{eff}})}}{2k\sigma_{\parallel,\text{eff}}}\nonumber\\&=\frac{\int_0^1d\mu\int dM_h\,dn/dM_h\,L^2(M_h)\exp{[-k^2\mu^2\sigma_v^2(M_h)]}}{\int dM_h\,dn/dM_h\,L^2(M_h)}.\label{eq:veff_tofit}\end{align}
We intentionally ignore the angular resolution of the survey so as to devise $v_\text{eff}$ as a well-defined function of the $L(M_h)$ and $v(M_h)$ models alone. We will see that this affects the fidelity of this approximation when $\sigma_\perp$ is non-negligible.

We can clearly solve~\autoref{eq:veff_tofit} exactly for $\sigma_{\parallel,\text{eff}}$ at each $k$, at least numerically. However, we want to find a single velocity scale $v_\text{erf}$ that \emph{approximately} satisfies~\autoref{eq:veff_tofit} across all $k$. In other words, for all $k$, $\sigma_{\parallel,\text{erf}}=\sigma_\parallel(v_\text{erf})$ satisfies
\begin{equation}\frac{\operatorname{erf}{(k\sigma_{\parallel,\text{erf}})}}{k\sigma_{\parallel,\text{erf}}}\approx\frac{\avg{L^2\text{erf}{\left[k\sigma_v(M_h)\right]}/[k\sigma_v(M_h)]}}{\avg{L^2}},\label{eq:veff_tofit2}\end{equation}
where $\avg{x}\equiv\int dM_h\,(dn/dM_h)\,x$. So for any given set of model parameters fully determining $L(M_h)$, we can calculate the shot noise transfer function across $k$ on the right-hand side and solve numerically for the $\sigma_{\parallel,\text{erf}}$ corresponding to that parameter set.

Rather than repeatedly solving for $v_\text{erf}$ (which comes with issues of reliability and computational cost), we want a closed-form quantity that we can calculate from the analytic halo model. Since the shot noise is the second moment of the luminosity function, a reasonable starting point for a representative velocity scale is the $L^2$-weighted average $v_\text{vir}$. Again using $\avg{x}\equiv\int dM_h\,(dn/dM_h)\,x$, we might write
\begin{equation}
    v_\text{eff}\stackrel{?}{=}\frac{\avg{L^2v_\text{vir}}}{\avg{L^2}}.
    \label{eq:veff_failed}
\end{equation}
However, note that $\operatorname{erf}{x}\to1$ as $x\to\infty$, such that for very high $k$ we actually expect a more reasonable estimate to be the inverse of the $L^2$-weighted average $v^{-1}(M)$,
\begin{equation}
    v_\text{eff}\stackrel{?}{=}\frac{\avg{L^2}}{\avg{L^2v_\text{vir}^{-1}}}.
    \label{eq:veff_failed2}
\end{equation}
Ultimately, however, we will find that the best estimate for $v_\text{erf}$ is the average of these two estimates,
\begin{equation}
    v_\text{eff}\equiv\frac{1}{2}\left(\frac{\avg{L^2v_\text{vir}}}{\avg{L^2}}+\frac{\avg{L^2}}{\avg{L^2v_\text{vir}^{-1}}}\right).
    \label{eq:veff_actual}
\end{equation}
We can then use~\autoref{eq:vfwhmtosig} to convert this to a $\sigma_\parallel$ value, with which we can then use~\autoref{eq:Pconv} to calculate the power spectrum analytically or numerically, or set the appropriate scale for the Gaussian filter to apply to a mock CO cube.

We give a very loose theoretical justification for our choice in the context of a simplified model in~\autoref{sec:ansatzcheck}, but our main justification will be based on explicit numerical calculations comparing $v_\text{erf}$, $v_\text{eff}$, and our other ans\"{a}tze across a broad range of model parameters in our fiducial distribution.

Note that the above calculation of $v_\text{eff}$ does not account for the effect of randomly distributed inclination angles. The effect can be expressed using special functions but the functions involved are more complex without necessarily yielding an improved qualitative understanding. Therefore, while we discuss explicit analytic derivations in~\autoref{sec:incli}, here we will simply show the required adjustment, which is a slight decrease in the high-$k$ ansatz:
\begin{equation}
    v_\text{eff}\equiv\frac{1}{2}\left(\frac{\avg{L^2v_\text{vir}}}{\avg{L^2}}+\frac{4}{\pi\sqrt{3}}\frac{\avg{L^2}}{\avg{L^2v_\text{vir}^{-1}}}\right).
    \label{eq:veff_incli}
\end{equation}

\subsection{Careful Design of Mass or Velocity Bins}
\label{sec:2tier}
Not all brute-force simulations are equally brute. While binning CO emitters according to mass or velocity and then applying Gaussian filters to each bin is the most straightforward and brute method of simulating line broadening that is still computationally feasible, carefully choosing the binning scheme should result in being able to maintain accuracy while saving computational cost.

In the case of our fiducial model, note that low-mass emitters (and thus narrow CO line profiles) will typically neither be resolvable in frequency space nor contribute significantly to the power spectrum at small scales where the effect of line broadening becomes marked. For instance, with CO(1--0) at $z\sim3$, the dominant contribution to shot noise will be from $M_h\sim10^{12}$ emitters with $v_\text{vir}\sim230$\,km\,s$^{-1}$. The COMAP science channelization of 15.6\,MHz quoted in~\cite{Ihle19} ($156$\,km\,s$^{-1}$ in velocity space) can resolve these profiles but not the profiles corresponding to emitters with $M_h\lesssim10^{11}\,M_\odot$ (for which $v_\text{vir}\lesssim100$\,km\,s$^{-1}$).

The native instrumental frequency resolutions of COMAP ($\sim2$\,MHz) and mmIME ($4$--$8$\,MHz) are signficantly finer than the $>10$\,MHz binning applied during data analyses, and in principle the spectrometers used are capable of resolving widths of $\lesssim100$\,km\,s$^{-1}$. However, in practice such line profiles would not likely be detectable due to the low line intensity associated. Even in aggregate, the contribution of these low-mass (narrow-width) emitters to the power spectrum---particularly the shot-noise component of the power spectrum---would remain subdominant.

Then consider a CO simulation at $z\sim3$ with halo masses ranging from $10^{10}\,M_\odot$ to $3\times10^{13}\,M_\odot$. This corresponds to $v_\text{vir}\in(50,700)$\,km\,s$^{-1}$, so a brute binning scheme might define equally spaced bins across this range. But one-sixth of these bins---not a majority of the bins, but not exactly a negligible fraction---will correspond to $M_h\in(10^{10},10^{11})\,M_\odot$. This population is about 100 times more numerous than the $M_h\gtrsim10^{11}\,M_\odot$ population, yet negligible for the purpose of simulating the line broadening effect for the reasons that we have described above.

Therefore, a less-brute two-tier scheme will bisect the halo population in our simulation around $M_h=10^{11}\,M_\odot$, applying no line broadening to the low-mass subset but binning and applying line broadening to the high-mass subset. This reduces the need to iterate repeatedly through the low-mass but far more numerous halo subset, and also reduces the number of bins used without significantly lowering accuracy, which we will demonstrate in~\autoref{sec:sim}.

\section{Preliminary Validation of Effective Line Width}
\label{sec:precheck}
While we will eventually simulate $z\sim3$ CO(1--0) based on lightcones from dark matter simulations, we undertake a sanity check in this section with numerical calculations based on the above analytic halo model, in order to set expectations for accuracy of using $v_\text{eff}$ relative to the full calculation of line broadening, both before and after inclination corrections. In the first two subsections we will check $v_\text{eff}$ across a subset of our fiducial model distribution. Then in the final subsection, we will actually make a diversion outside of $z\sim3$ CO(1--0) and examine line broadening of the CO lines at $z\sim1$--5 observed by mmIME to show that our effective line width is a reasonable description of the effect on the monopole $P_0(k)$ consistent with the calculations in Appendix A of~\cite{mmIME-ACA}.
\subsection{Fiducial Model Ensemble}
To check whether $v_\text{eff}$ from~\autoref{eq:veff_actual} is a good description of the ideal $v_\text{erf}$ of~\autoref{eq:veff_tofit2} (both before correcting for inclination effects), we take 1764 samples from the MCMC of~\autoref{sec:LMmodel} and fit for $v_\text{erf}$ (minimising the summed squared difference between the two sides of~\autoref{eq:veff_tofit2} across all $k$) as well as calculate $v_\text{eff}$ and the other ans\"{a}tze of $\avg{L^2v_\text{vir}}/\avg{L^2}$ and $\avg{L^2}/\avg{L^2v_\text{vir}^{-1}}$. We use the \texttt{lim} package\footnote{\url{https://github.com/pcbreysse/lim/tree/pcbreysse}} as our basis for all calculations.
\begin{figure}
    \centering
    \includegraphics[width=0.96\linewidth]{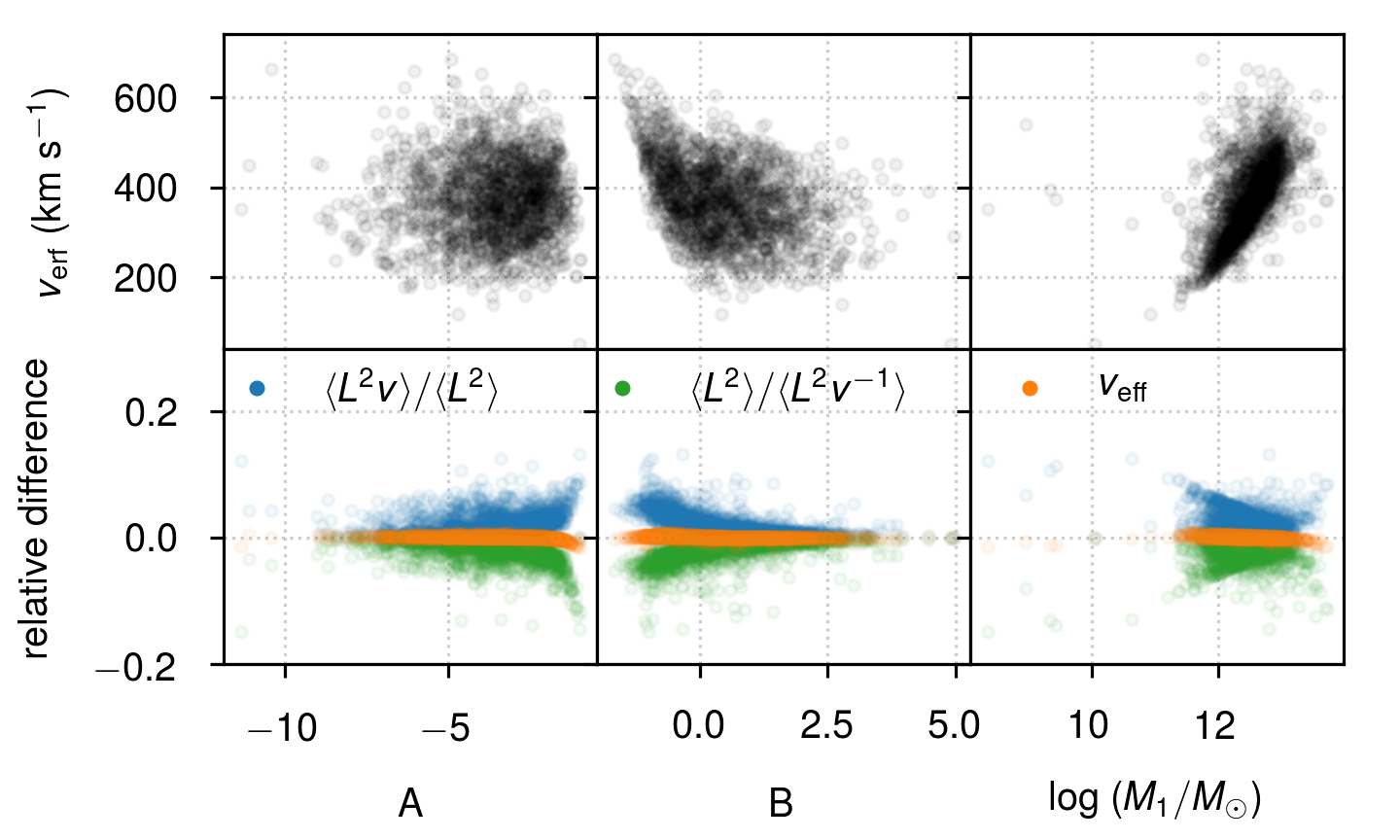}
    \caption{Comparison of effective line FWHM and ans\"{a}tze before accounting for inclination. \emph{Upper panels:} scatter plot of $v_\text{erf}$ for 1764 draws from our fiducial model posterior, against model parameters $A$ (\emph{left}), $B$ (\emph{middle}), and $M_1$ (\emph{right}), illustrating correlations or lack thereof as described in the main text. \emph{Lower panels:} scatter plot of the relative differences between our ans\"{a}tze against the same three model parameters as in the upper panels, once again showing correlations or lack thereof.}
    \label{fig:verf_compare}
\end{figure}

We show the results in~\autoref{fig:verf_compare}. First it is useful to note the range of $v_\text{erf}$, which falls predominantly in the 200--500\,km\,s$^{-1}$ range corresponding to the predominant range of $\log{(M_1/M_\odot)}$. However, while there is a very strong correlation indeed between $v_\text{erf}$ and $M_1$, it is not perfect in our data. This appears to be in part due to outliers where $M_1$ is so low---below our minimum emitter $M_h=10^{10}\,M_\odot$, in fact---that we have effectively ended up with a single power-law description of $L(M_h)$. There is however some additional scatter around the average correlation even at $M_1\gtrsim10^{12}\,M_\odot$, part of which seems to be from a weak anti-correlation between $B$ and $v_\text{erf}$. We can explain this based on the fact that if $B<0$, the double power-law $L(M_h)$ relation breaks into a shallower but still positive slope for $M_h>M_1$, and therefore the CO shot noise becomes dominated by very rare but very bright CO emitters.

Moving on to our ans\"{a}tze, we find that $v_\text{eff}$ is extremely close to $v_\text{erf}$---the relative difference between the two is below 0.5\% in the vast majority of cases. Meanwhile our other ans\"{a}tze differ by larger fractions from $v_\text{erf}$, but somewhat astonishingly in opposite directions by almost exactly opposite amounts. Our best qualitative explanation for this is that $\avg{L^2v}/\avg{L^2}$ better describes attenuation at intermediate scales ($k\sim1$\,Mpc$^{-1}$) while $\avg{L^2}/\avg{L^2v^{-1}}$ describes attenuation at small scales ($k\gtrsim1$\,Mpc$^{-1}$) almost exactly. The midpoint $v_\text{eff}$ is a compromise between the two and will thus match up with the fit $v_\text{erf}$\added{,} which has to make a similar compromise.

The difference increases with lower values of $B$, where compared to $v_\text{erf}$ we find that $\avg{L^2v}/\avg{L^2}$ is too high and $\avg{L^2}/\avg{L^2v^{-1}}$ too low by several percent. This is likely due to the larger range of mass scales that contribute to the shot noise as $L(M_h)$ monotonically increases rather than plateauing or declining. In these situations, $\avg{L^2}/\avg{L^2v^{-1}}$ actually probably still describes the attenuation at high $k$ more accurately, but the deviation at intermediate $k$ is likely greater and the relative difference from the fit $v_\text{erf}$ thus greater as well. We also find some correlation between relative differences and $M_1$ or $A$, although much of this may be driven by underlying weak correlations between the model parameters themselves.

In summary, we find that our $v_\text{eff}$ prescription is just as good as fitting $\operatorname{erf}{(k\sigma_\parallel)}/(k\sigma_\parallel)$ to the shot noise attenuation across ${k\in(10^{-2},10^1)}$~Mpc$^{-1}$ and solving for $\sigma_\parallel$. However, do note that our `failed' ans\"{a}tze are actually still reasonable descriptions within a few percent in most cases, and in particular we still wholly expect $\avg{L^2}/\avg{L^2v^{-1}}$ to be the better effective line width to use at very high $k$.

\begin{figure}
    \centering
    \includegraphics[width=0.96\linewidth]{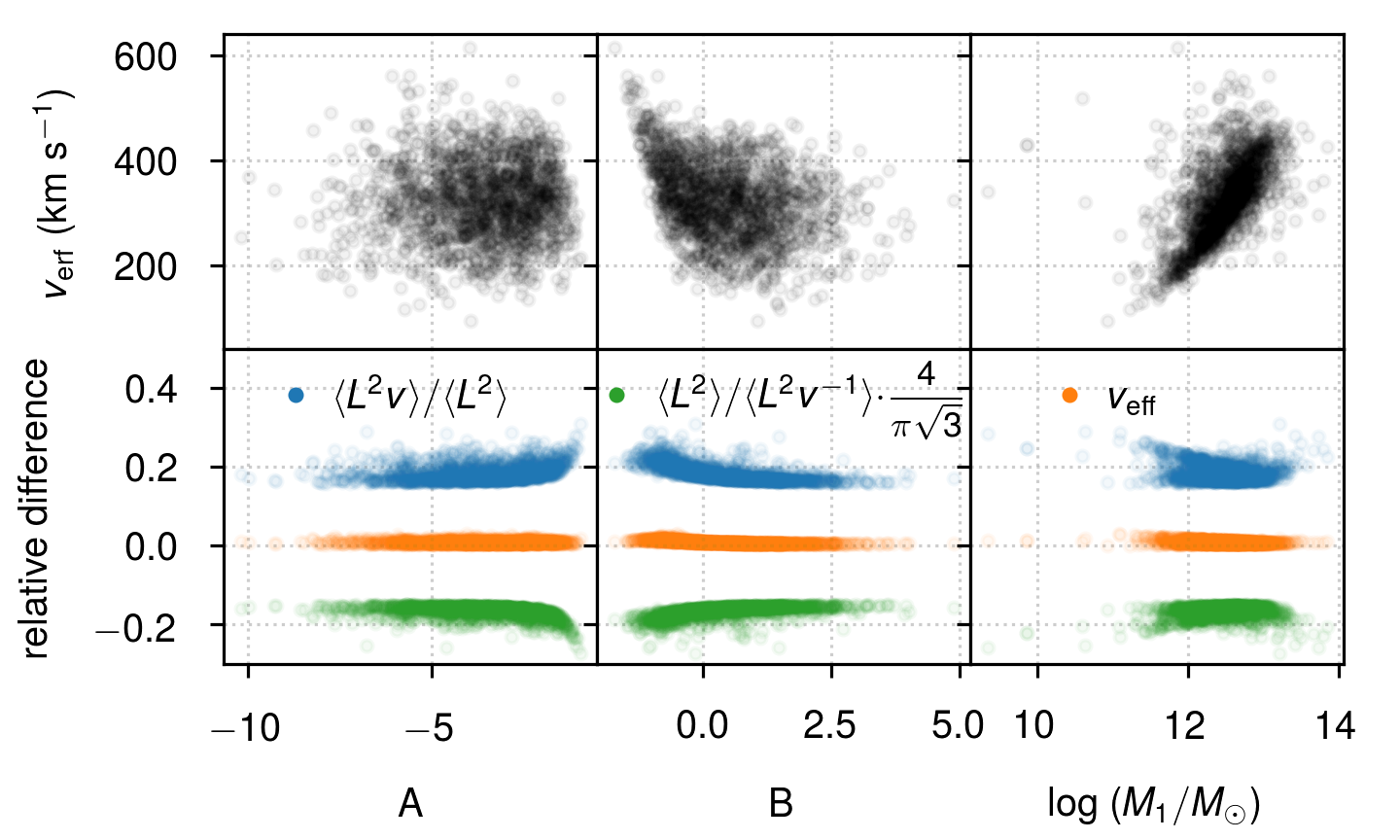}
    \caption{Same as~\autoref{fig:verf_compare} but after accounting for inclination.}
    \label{fig:verf_compare_incli}
\end{figure}

Note that two key caveats apply to these last couple of points. One is that these facts only hold as long as angular resolution effects are comparatively negligible, since these were not included in devising that ansatz. The other caveat is that the relative errors across the ans\"{a}tze will be much greater once we include corrections for inclination, as shown in~\autoref{fig:verf_compare_incli}. That said, the qualitative points about the different ans\"{a}tze still hold. This includes the point that $\avg{L^2}/\avg{L^2v^{-1}}$, or rather $4/(\pi\sqrt{3})\cdot\avg{L^2}/\avg{L^2v^{-1}}$, should be the best ansatz at very high $k$ in the absence of a sizeable angular beam. However, differences in functional form mean that the attenuation will asymptote less quickly to this description with inclination corrections than without them, meaning that errors may be so large for any scales relevant to actual power spectrum analysis that we may as well use the midpoint $v_\text{eff}$ instead. We will see this in~\autoref{sec:mmIME} in our consideration of power spectra observed by mmIME.

\subsection{A Closer Look at Attenuation for Specific Parameter Values}
\label{sec:extremedraw}
So far, we have looked at the shot noise attenuation in isolation. But it will be most instructive to examine the total $P_0(k)$ and $P_2(k)$, incorporating both clustering and shot-noise components and accounting for all redshift-space observational effects that we have discussed so far, including the effect of angular resolution in the context of a single-dish experiment like COMAP. This will come at somewhat increased computational cost, and we will be making equivalent calculations from N-body simulations later in this work. However, we will examine numerical realizations of our analytic model for two sets of parameter values\replaced{, mostly omitting accounting for inclination as our purposes here are mostly illustrative}{, accounting for inclination throughout}.

We first examine the effect of line broadening given the model parameter values given towards the end of~\autoref{sec:LMmodel} in Equations~\ref{eq:realparami} through~\ref{eq:realparamf}, which broadly represent the median expected $P(k)$ and luminosity function. This will give a sense of what our average expectation should be for attenuation of the total $P_0(k)$ and $P_2(k)$.
\begin{figure}
    \centering
    \includegraphics[width=0.96\linewidth]{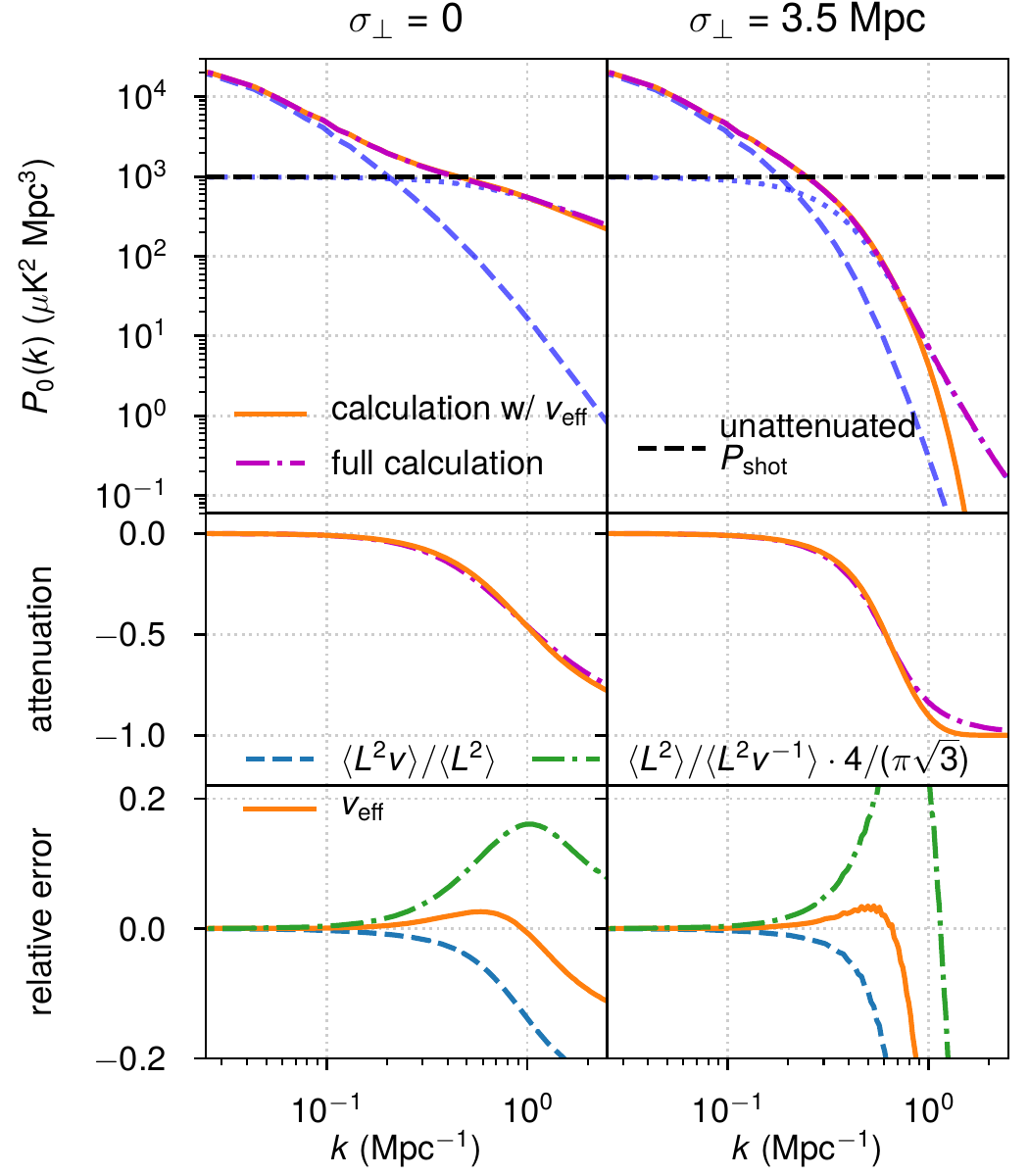}
    \caption{Illustration of the expected effect of line broadening on the $z\sim3$ CO(1--0) $P_0(k)$ predicted by the model parameter values of Equations~\ref{eq:realparami}--\ref{eq:realparamf}, both ignoring angular resolution (\emph{left panels}) and assuming smearing with a Gaussian beam of FWHM $4.5'$ or $\sigma_\perp=3.5\text{\,Mpc}$ (\emph{right panels}). \emph{Upper panels:} $P_0(k)$ calculated with the full formalism of~\autoref{sec:theory} (\replaced{but without corrections for inclination}{inclination-inclusive}; magenta dash-dotted) and with just $v_\text{eff}$ (orange). For the full calculation we also show the clustering (\replaced{magenta}{indigo} dashed) and shot-noise (\replaced{magenta}{indigo} dotted) contributions separately\replaced{. In addition, we show the full calculation including inclination corrections (cyan dotted). We}{, and} also show the unattenuated shot noise (black dashed) as a guide. \emph{Middle panels:} Attenuation due to the introduction of line broadening, calculated using the full formalism, both without inclination corrections (magenta solid) and with them (cyan dotted). \emph{Lower panels:} Error relative to the full calculation (\replaced{without}{with} inclination corrections) from using each of our three ans\"{a}tze for effective line width---$\avg{L^2v}/\avg{L^2}$ (blue dashed), $\avg{L^2}/\avg{L^2v^{-1}}\added{\cdot4/(\pi\sqrt{3})}$ (green dash-dotted), and the midpoint $v_\text{eff}$ (orange solid)\replaced{. We also show the error relative to the full calculation  including inclination corrections when using the corrected $v_\text{eff}$ of~\autoref{eq:veff_incli} (orange dotted)}{---all inclination-inclusive}.}
    \label{fig:veff_realistic}
\end{figure}

We show the effect on $P_0(k)$ in~\autoref{fig:veff_realistic}, first ignoring angular resolution and then accounting for the COMAP beam FWHM of $4.5'$ at 30 GHz, which corresponds to $\sigma_\perp\approx3.5$\,Mpc. The overall conclusion is that using a single $v_\text{eff}$ results in a very good approximation of the attenuation of $P_0(k)$ for scales relevant to COMAP, but it is worth noting that when incorporating a non-zero $\sigma_\perp$---which our definition of $v_\text{erf}$ in~\autoref{eq:veff_tofit2} does not---our approximation breaks down as we approach the comoving scales corresponding to the COMAP beam size ($\pi/\sigma_\perp\sim0.9$\,Mpc$^{-1}$). Indeed the high-$k$ ansatz of~\autoref{eq:veff_failed2} also fails because it too does not account for the presence of an angular beam.

The actual amount of $P_0(k)$ attenuation is also worth discussing briefly. While we defined $v_\text{erf}$ while ignoring angular resolution, the $P_0(k)$ attenuation from line broadening clearly must depend on beam smearing. Quantitatively,~\autoref{eq:Pclustv} and~\autoref{eq:Pshotv} show that the two effects are not separable into independent multipliers in front of the unattenuated power spectrum components. Qualitatively, the fact that beam smearing effectively discards angular modes means the spherical averaging of $P(k,\mu)$ into $P_0(k)$ must depend more heavily on line-of-sight modes, thus increasing the relative weight of line broadening in attenuation.

Nonetheless, these divergences between the cases of $\sigma_\perp=0$ and $\sigma_\perp\approx3.5$\,Mpc are somewhat esoteric in the context of the actual COMAP observation, where the very beam smearing that results in the breakdown of our approximation already introduces its own power loss in the same regime where this breakdown occurs. Given the transfer functions from both beam smearing and loss of large-scale modes due to filtering in the data pipeline prior to map-making---see Foss et al.~(in prep.)~for specific details---the COMAP $P_0(k)$ measurement will be most sensitive to $k\sim0.2$--0.3\,Mpc$^{-1}$. Therefore, in this case, line broadening should only introduce 7--8\% attenuation of $P_0(k)$ at scales relevant to COMAP.

\begin{figure}
    \centering
    \includegraphics[width=0.96\linewidth]{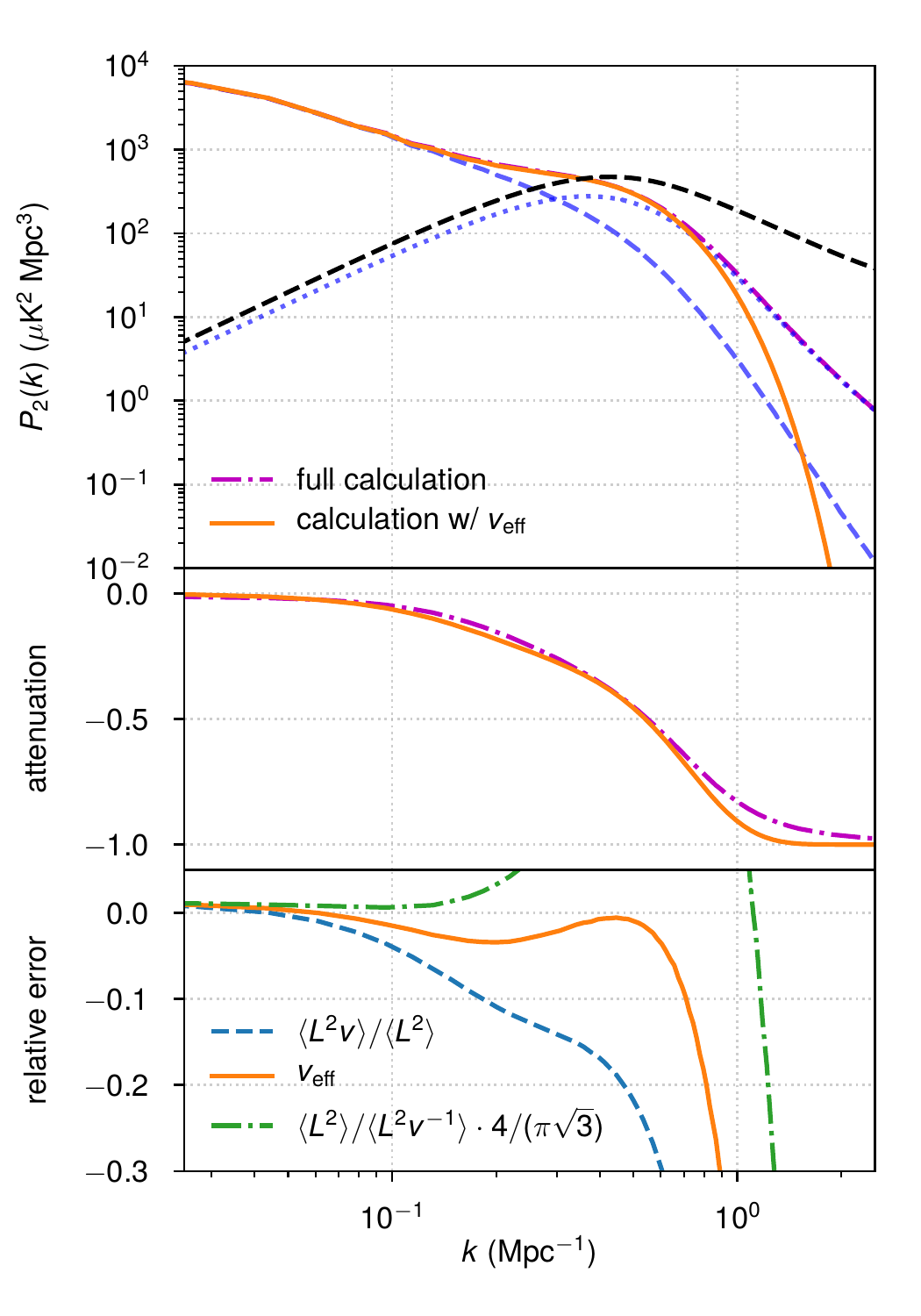}
    \caption{Similar to the right panels of~\autoref{fig:veff_realistic} (assuming smearing with a Gaussian beam of FWHM $4.5'$ or $\sigma_\perp=3.5\text{\,Mpc}$), but showing attenuation of $P_2(k)$. The unattenuated shot-noise contribution to the quadrupole (black dashed) shown in the top panel is calculated after beam smearing is taken into account but before line broadening is applied.}
    \label{fig:veff_realisticq}
\end{figure}
We also show the effect of line broadening on the quadrupole $P_2(k)$ in~\autoref{fig:veff_realisticq}. Here, we only show the case where we set ${\sigma_\perp\approx3.5}$~Mpc, as this is required (along with $\sigma_\perp>\sigma_\parallel$, which is the case here) for a positive shot-noise contribution to the quadrupole. The overall conclusions are similar to those for $P_0(k)$ in that the approximation breaks down near the COMAP beam scale but is otherwise acceptable. We do note however that the attenuation of the shot-noise component of the quadrupole due to line broadening is far more severe for even intermediate scales than it was for the monopole, at around \added{20--}30\% for $k\sim0.2$--0.3\,Mpc$^{-1}$.

We might naturally ask whether these conclusions still hold for a more extreme draw from our fiducial distribution with unusual values of $A$ or $B$. So we also show the same plots given the following parameter values:
\begin{align}
A &= -0.97,\label{eq:extremeparami}\\
B &= -0.35,\\
\log{C} &= 11.1,\\
\log{(M_1/M_\odot)} &= 12.8,\\
\sigma_L &= 0.5.\label{eq:extremeparamf}
\end{align}
\begin{figure}
    \centering
    \includegraphics[width=0.96\linewidth]{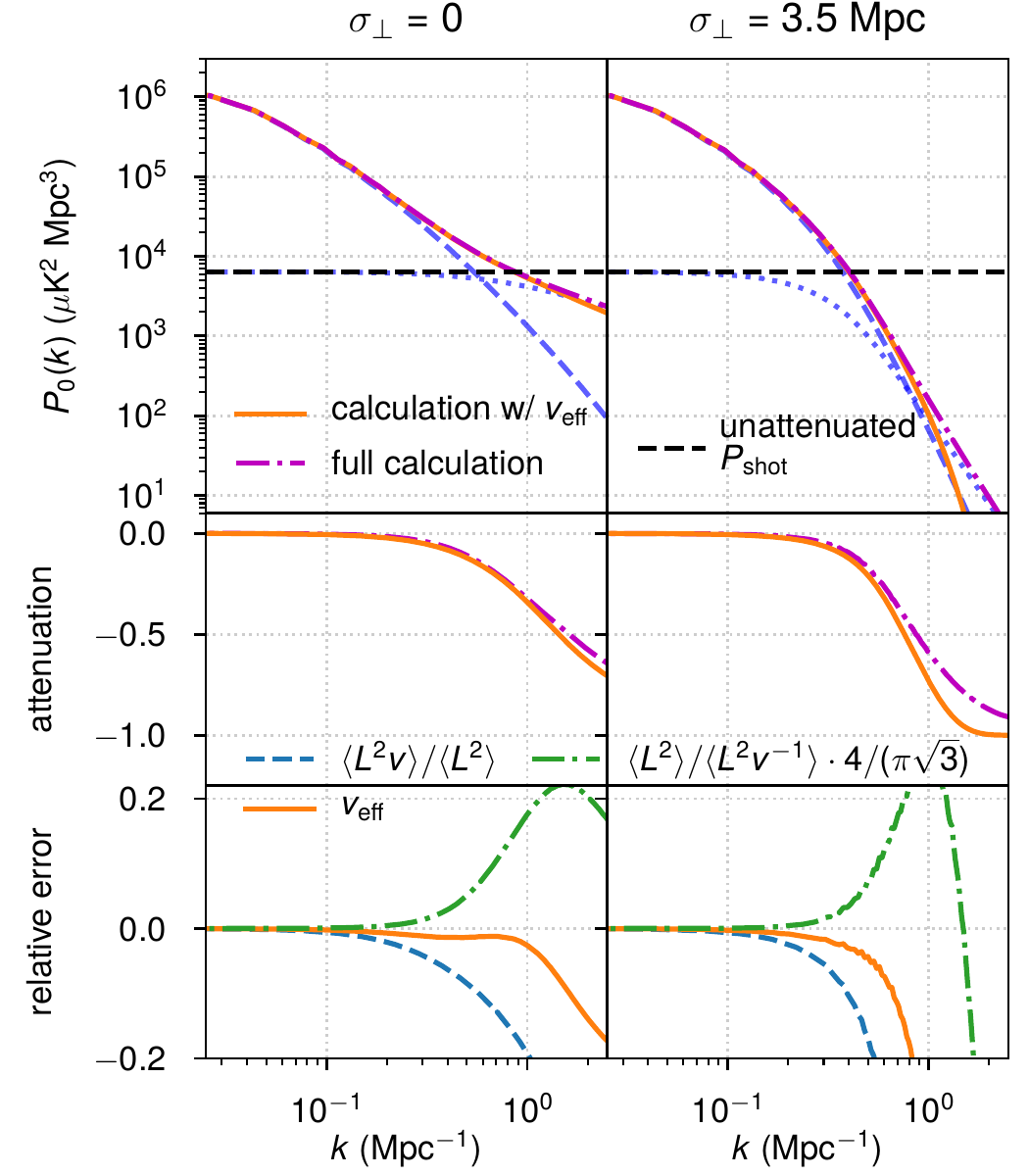}
    \caption{Same as~\autoref{fig:veff_realistic} but using the model parameter values of Equations~\ref{eq:extremeparami}--\ref{eq:extremeparamf}, representing a more extreme draw from our distribution with a high value of $A$.}
    \label{fig:veff_extremedraw}
\end{figure}
\begin{figure}
    \centering
    \includegraphics[width=0.96\linewidth]{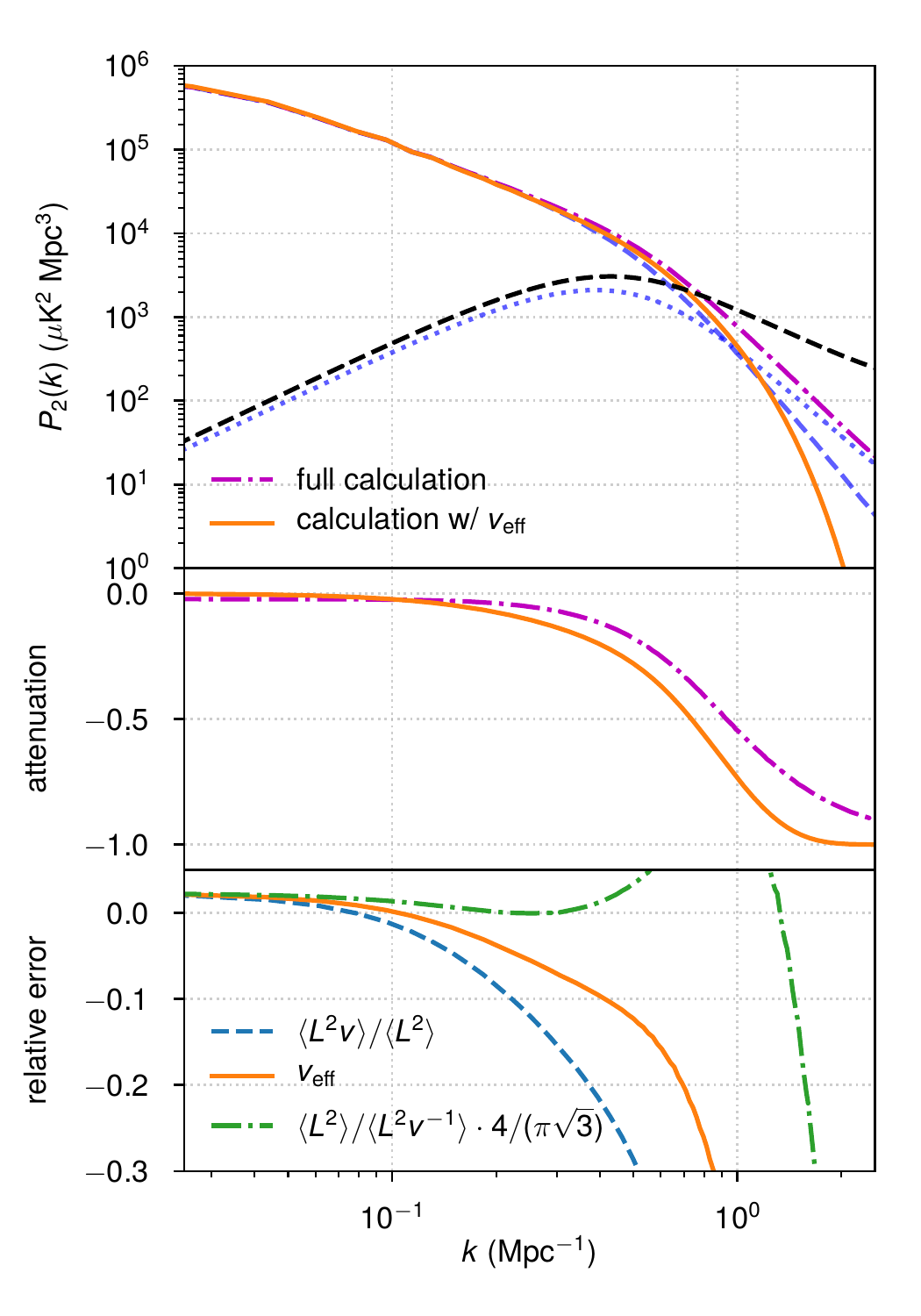}
    \caption{Same as~\autoref{fig:veff_realisticq} but using the model parameter values of Equations~\ref{eq:extremeparami}--\ref{eq:extremeparamf}.}
    \label{fig:veff_extremedrawq}
\end{figure}

We show $P_0(k)$ in~\autoref{fig:veff_extremedraw} and $P_2(k)$ in~\autoref{fig:veff_extremedrawq} for these parameter values. Importantly, the clustering component is much higher relative to the shot-noise component than in our more pedestrian parameter set, which is not necessarily surprising given the high value of $A$ and thus the shallow faint-end $L(M_h)$ relation. This has a series of implications for the accuracy of our $v_\text{eff}$ prescription because, as we noted when discussing~\autoref{eq:Pclustv}, the attenuation of the clustering component is weighted by $L(M_h)$ at each $M_h$ and thus should be less than the attenuation of the shot-noise component (weighted by $L^2$) for a monotonically increasing $v(M_h)$. Therefore, our error in estimating attenuation is greater than in~\autoref{fig:veff_realistic} and~\autoref{fig:veff_realisticq}, and we will always expect too much attenuation in these situations because our approximation is based on the shot-noise component.

Still, in scales relevant to COMAP analysis ($k\lesssim0.5$\,Mpc$^{-1}$), the relative error is typically within a few percent for the monopole (although greater for the quadrupole), and the amount of attenuation in the monopole is only around 3\%. This is much smaller than the amount of attenuation given our previous parameter set, precisely because the clustering component is so much more dominant. So our approximation behaves worse but the attenuation being approximated is smaller, both for the same reason.

Importantly, however, the fact that our approximation breaks down at small scales with the introduction of a non-negligible $\sigma_\perp$ does not bode well for its performance with respect to the VID. We will examine the VID explicitly in further simulations to follow in this work.

Note that while we have not discussed the effect of accounting for inclination in these cases, we do show it graphically in \replaced{Figures~\ref{fig:veff_realistic} through~\ref{fig:veff_extremedrawq}}{\autoref{sec:extremedraw_noincli}}. The effect on the attenuation is quite small for scales relevant to COMAP, although less negligible for $P_2(k)$. Furthermore, the accuracy of using $v_\text{eff}$ to describe the attenuation is largely the same after accounting for inclination.
\subsection{Beyond the Fiducial: Models for mmIME CO Observations at 100 GHz}
\label{sec:mmIME}
We have now mentioned the Millimetre-wave Intensity Mapping Experiment, or mmIME~\citep{mmIME-ACA}, on several occasions. While mmIME will be using data from several community interferometers across a wide frequency range, the first analysis work of~\cite{mmIME-ACA} looks at data from the Atacama Large Millimetre-wave Array (ALMA) in compact configurations observing at 100 GHz. Being aware of the line broadening effect but not having a detailed model of it, the key step taken is avoidance by excluding modes above a certain $k_\parallel$ from the analysis. Specifically, using the lag coordinate $\eta$ (written implicitly in inverse frequency units of $\nu_\text{obs}^{-1}$), the excluded modes correspond to $\eta>500$, and since $\eta$ and $k_\parallel$ are related by
\begin{equation}
    k_\parallel = 2\pi\eta\cdot\frac{H(z)\nu_\text{obs}}{c(1+z)},
\end{equation}
the threshold of $\eta=500$ is equivalent to $k_\parallel\approx0.75$\,Mpc$^{-1}$ for $z=2.5$, corresponding to the frame for CO(3--2) observations at 100 GHz. The cut is designed to exclude all modes of redshift-space comoving wavelength $8.4$\,Mpc and below, which corresponds to 600\,km\,s$^{-1}$ and below in velocity space. Since observations discussed in~\autoref{sec:LWmodel} have found line widths below this are typical, the cut is fairly conservative. At the same time, as~\cite{mmIME-ACA} note, the cut will not entirely eliminate suppression of power from line broadening. Line profiles are finite in extent and are not perfect periodic modes, so line profiles that are $\sim300$\,km\,s$^{-1}$ wide will still lead to some attenuation of $\eta<500$ modes.

\cite{mmIME-ACA} do not explicitly correct for this attenuation, but do note that if the shot noise is dominated by CO emitters with line widths of $\sim300$\,km\,s$^{-1}$, the measurement is likely attenuated by $\sim25\%$ and the necessary correction thus an upward shift by one-third. Here we will examine whether we are able to derive similar corrections with our own (inclination-inclusive) model.

First, we review the emission models for the CO lines observed by mmIME. In essence, the models use the basic flow of~\cite{Li16}, relating halo mass to star-formation rate via~\cite{Behroozi13a,Behroozi13b}, star-formation rate to IR luminosity via a simple scaling of $10^{10}\,L_\odot\,M_\odot^{-1}$ yr, and then IR luminosity to CO luminosity. Since~\cite{Li16} model only CO(1--0) emission at $z\sim3$, the models of~\cite{mmIME-ACA} link IR and CO luminosities via fits found in~\cite{Kamenetzky16} from a broad sample of $z<1$ galaxies observed with {\it Herschel}. These models inform how much each line should contribute to the total measurement, and are scaled up uniformly in luminosity to match the mmIME data. We will consider the models without this final scaling as it should not affect any results concerning attenuation from line broadening.

For this section only we will change our cosmological parameters to match those of~\cite{mmIME-ACA}, namely $\Omega_m=0.27$ instead of 0.286. However, as~\cite{mmIME-ACA} do not note all parameters that may affect predictions (the baryonic matter density fraction, for instance), we do not expect to perfectly reproduce the predictions of~\cite{mmIME-ACA} for line intensity. Nonetheless, we are able to recreate the $L(M_h)$ for the various CO lines at 100 GHz, and are able to reproduce the $P_\text{shot}$ values for each line to within 10\% (except for CO(5--4) at $z\sim4.8$, where we fall within 20\%).
\begin{deluxetable}{ccccc}
\tablecolumns{5}
\tablecaption{CO lines observed by~\protect\cite{mmIME-ACA} at 100 GHz as part of mmIME, with redshifts, predictions for shot noise power, and our calculated (inclination-inclusive) $v_\text{eff}$ for each line.\label{tab:mmIME_veffs}}
\tablehead{\colhead{Line}&\colhead{$z$}&\colhead{$P_\text{shot}$ from}&\colhead{$P_\text{shot}$ from}&\colhead{$v_\text{eff}$} \\
         & & \colhead{\cite{mmIME-ACA}} & \colhead{this work} & \\
         & & \colhead{($\mu$K$^2$\,Mpc$^3$)} & \colhead{($\mu$K$^2$\,Mpc$^3$)} & \colhead{(km\,s$^{-1}$)}}
    \startdata
         CO(2--1) & 1.3 & $100h^{-3}=292$ & 315 & 168 \\
         CO(3--2) & 2.5 & $160h^{-3}=466$ & 519 & 213 \\
         CO(4--3) & 3.6 & $\hphantom{1}80h^{-3}=233$ & 209 & 220 \\
         CO(5--4) & 4.8 & $\hphantom{1}20h^{-3}=58\hphantom{3}$ & \phn46 & 199
    \enddata
\end{deluxetable}

For $v(M_h)$, we will actually use the exact same prescription as for CO(1--0) at $z\sim3$, which is to set the line FWHM equal to the virial velocity. Partly this is because devising $v(M_h)$ for each line at each redshift is well beyond the scope of this paper, but partly we also expect the same prescription to be a reasonable one for other high-redshift CO lines, at least in the absence of high information. Looking at the three sources in~\cite{VLASPECS} robustly detected in CO(1--0), we find that their CO(1--0) line widths are consistent with their CO(3--2) line widths from~\cite{ASPECS-LP}. Although we broadly expect higher-$J$ CO lines to have steeper gas density profiles due to the higher gas temperatures required for excitation of these lines, we do not expect this to be an overwhelmingly large effect for the lines observed by mmIME.

Thus, using essentially the same $v(M_h)$ as in~\autoref{sec:LWmodel} (albeit with appropriate modifications for redshift and cosmology) but swapping out the $L(M_h)$ models to match~\cite{mmIME-ACA}, we can find $v_\text{eff}$ for each line. We show these values in~\autoref{tab:mmIME_veffs} alongside our reproduced $P_\text{shot}$ values. Note that $v_\text{eff}$ tends to be lower at lower redshift---despite the continued growth of halo masses, the decline in star-formation activity after $z\sim3$ means that the halo mass scales that dominate the CO shot noise are smaller for lower redshift.

While our $v_\text{eff}$ values are somewhat lower than the 300\,km\,s$^{-1}$ expectation of~\cite{mmIME-ACA}, the comparison is not exactly even because of our choice of profile shape. In Appendix A, \cite{mmIME-ACA} consider a simple Gaussian profile, a double-Gaussian profile, and a top-hat profile, and find that the simple Gaussian profile results in the most attenuation. The difference in attenuation is at a $\sim20$\% level, but so is our difference in line widths. So our prediction of attenuation really should be broadly consistent with the $\approx25$\% expectation of~\cite{mmIME-ACA}.

\begin{figure*}
    \centering
    \includegraphics[width=0.96\linewidth]{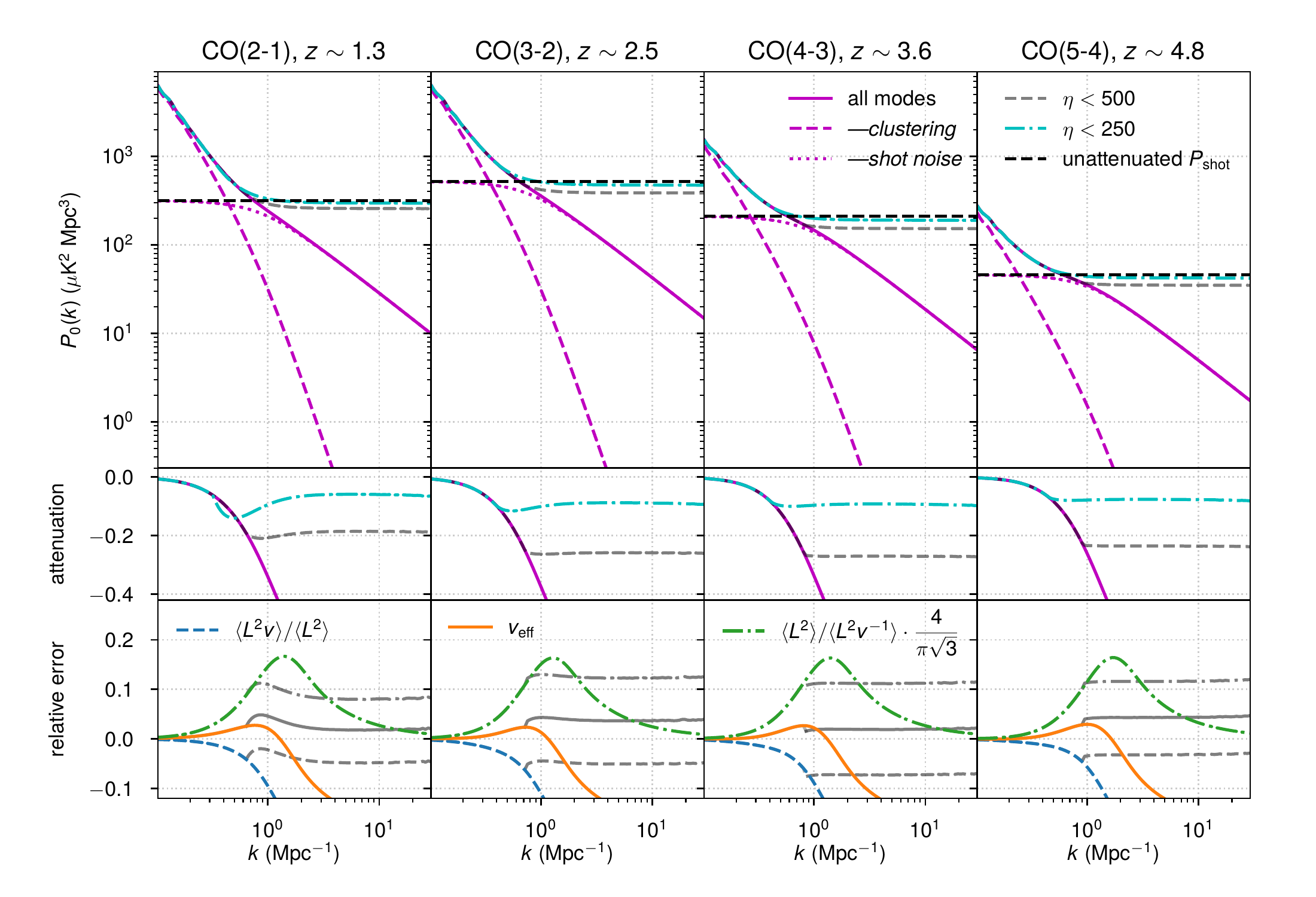}
    \caption{Illustration of the expected effect of line broadening on the individual lines observed by mmIME at 100 GHz (indicated above each column of panels), using the $L(M_h)$ model reproduced from~\protect\cite{mmIME-ACA} and our $v(M_h)$ model. \emph{Upper panels:} $P_0(k)$ calculated with the full formalism of~\autoref{sec:theory} (magenta solid), including the effect of random inclinations. For the full calculation we also show the clustering (magenta dashed) and shot-noise (magenta dotted) contributions separately. We also show the unattenuated shot-noise contribution to the quadrupole (black dashed), as well as the expected $P_0(k)$ when restricting calculations to modes with $\eta<500$ (grey dashed) or $\eta<250$ (cyan dash-dotted). \emph{Middle panel:} Attenuation due to the introduction of line broadening, calculated using the full formalism (magenta solid). We again show the altered attenuation when restricting calculations to modes with $\eta<500$ (grey dashed) or $\eta<250$ (cyan dash-dotted). \emph{Lower panel:} Error relative to the full calculation from using each of our three ans\"{a}tze for effective line width---$\avg{L^2v}/\avg{L^2}$ (blue dashed), $4/(\pi\sqrt{3})\cdot\avg{L^2}/\avg{L^2v^{-1}}$ (green dash-dotted), and the midpoint $v_\text{eff}$ (orange solid). We also show what the error would be against the $\eta<500$ calculation if we froze the attenuation calculation at the corresponding $k$ value for each ansatz (respective line styles, in grey).}
    \label{fig:mmIME_pspecs}
\end{figure*}

We calculate the expected effect of line broadening for each of the lines individually, and show this in~\autoref{fig:mmIME_pspecs}. Note that we set $\sigma_\perp=0$, since angular resolution does not have the same relevance in (visibility-space) interferometric power spectrum measurements that it does in (image-space) single-dish measurements like COMAP.

First, as we have previously discussed, $\avg{L^2}/\avg{L^2v^{-1}}$ is actually a better approximation than $v_\text{eff}$ in situations where $k$ is high and $P_0(k)$ is predominantly shot noise. However, the issue is that the approximation only converges for $k\gtrsim10$\,Mpc$^{-1}$, by which point the attenuation of the raw $P_0(k)$ is extremely large (having already exceeded 30\% by $k\sim1$\,Mpc$^{-1}$). That said, although we do not show the approximate $P_0(k)$ using any of these ans\"{a}tze in the topmost panels of~\autoref{fig:mmIME_pspecs}, the lowermost panels show that the approximate calculation using $v_\text{eff}$ is still within a few percent of the full calculation up to $k\sim1$\,Mpc$^{-1}$.

Second, we simulate not only restricting $\eta<500$ in the power spectrum analysis, but also a more stringent $\eta<250$ cut as briefly discussed in both Section 4.1 and Appendix A of~\cite{mmIME-ACA}. Recalling that for $z\sim2.5$ these correspond to roughly $k_\parallel<0.75$\,Mpc$^{-1}$ and $k_\parallel<0.38$\,Mpc$^{-1}$, it should not be terribly surprising that for all lines, the cuts arrest attenuation of the shot-noise component of the signal at the corresponding $k$. The level of attenuation differs slightly between each line, and in particular the CO(2--1) line at $z\sim1.3$ in our model shows the least attenuation, which makes sense for the same reasons we discussed for $v_\text{eff}$ being markedly lower at this redshift. However, they are broadly consistent with the $\sim25$\% and $\sim10$\% predictions from~\cite{mmIME-ACA} for $\eta<500$ and $\eta<250$ given dominant line widths of $\sim300$\,km\,s$^{-1}$ (again keeping in mind the minor differences that arise from the choice of profile shape even for the same line width).

For the approximate attenuations calculated using our different ans\"{a}tze, we could consider taking the value at $k\approx k_\parallel(\eta=500)$ and comparing this to the full calculation of attenuation with the $\eta<500$ cut. The results are shown in the lowermost panels of~\autoref{fig:mmIME_pspecs}, and suggest that using the approximate calculation from $v_\text{eff}$ at the relevant $k$-value is accurate to within a few percent. The other ans\"{a}tze do not yield calculations nearly as accurate---note in particular that while $4/(\pi\sqrt{3})\cdot\avg{L^2}/\avg{L^2v^{-1}}$ ultimately converges to the full calculation, the fact that it is significantly deviant at $k\sim1$\,Mpc$^{-1}$ results in $\sim10\%$ errors in estimated high-$k$ attenuation for $\eta<500$.

We note incidentally that the slightly greater attenuation of the \emph{total} $P_0(k)$ with $\eta$ cuts at $k\lesssim1$\,Mpc$^{-1}$ is perhaps counterintuitive but not necessarily an unexpected effect of these cuts. At these scales the clustering component is non-negligible, and the redshift-space enhancement in $P_\text{clust}$ increases with $\mu$, albeit polynomially and not exponentially. So for values of $k$ low enough for this enhancement to grow faster with $\mu$ than the line-broadening suppression, restricting calculations to lower values of $\eta$ and thus $k_\parallel=k\mu$ would mildly suppress this enhancement, thus suppressing clustering and (to a somewhat lesser extent) the total $P_0(k)$. The consideration is largely immaterial for mmIME, which measures $P_0(k)$ well above the $k$-range where this suppression would be relevant. It is also likely to be esoteric in general, as surveys specifically looking to measure $P_0(k)$ at these intermediate scales would probably simply access these modes with the intrinsic attenuation rather than apply any data cuts---recall that mmIME applies these cuts in $\eta$ to arrest attenuation at a level one would otherwise only expect at much lower $k$ than the values central to mmIME.

Finally, we can project all of these to the comoving frame used for CO(3--2) at $z\sim2.5$---see~\autoref{sec:projection} for a discussion of how this is done---and consider the attenuation of the total $P_0(k)$. Again we see the attenuation arrested at $k_\parallel\approx0.75$\,Mpc$^{-1}$ with $\eta<500$ and $k_\parallel\approx0.38$\,Mpc$^{-1}$ with $\eta<250$. At $k\sim10^1$\,Mpc$^{-1}$, corresponding to the typical scales relevant for the new observations presented by~\cite{mmIME-ACA}, the total attenuation is 23\% for $\eta<500$ (or 8\% for $\eta<250$), within a couple of percentage points of the predictions given by~\cite{mmIME-ACA}.

We therefore suggest that an upward correction of the total spectral shot power by roughly one-third---31\% if we believe the above calculation of 23\% attenuation---is entirely justified. However, this should not necessarily translate to an identical upward correction of the estimated shot noise levels for \emph{individual} CO lines. Taking our models at face value, we would apply somewhat smaller corrections closer to 23\% for CO(2--1) at $z\sim1.3$ and somewhat larger upward corrections closer to $35$\% for CO(3--2) at $z\sim2.5$ and $37$\% for CO(4--3) at $z\sim3.6$. In the case of CO(3--2) at $z\sim2.5$, using the fiducial conversion from~\cite{mmIME-ACA} of $r_{31}=0.42$ to convert the mmIME result into an estimate of \mbox{CO(1--0)} shot noise, applying this correction would change the best estimate from $(1140^{+870}_{-500})h^{-3}\,\mu\text{K}^2\text{\,Mpc}^3=(3.3^{+2.5}_{-1.5})\times10^3\,\mu\text{K}^2\text{\,Mpc}^3$ to $(1540^{+1170}_{-680})h^{-3}\,\mu\text{K}^2\text{\,Mpc}^3=(4.5^{+3.4}_{-2.0})\times10^3\,\mu\text{K}^2\text{\,Mpc}^3$.

Note that this upward correction would appear to reduce tension against the COPSS result from~\cite{COPSS} of $(3.0\pm1.3)h^{-3}\,\mu\text{K}^2\text{\,Mpc}^3=(8.7\pm3.8)\times10^3\,\mu\text{K}^2\text{\,Mpc}^3$. The difference between the two measurements would change from 1.5--$2\sigma$ to just over $1\sigma$, without the need to allow for $T^2b^2\gg10$\,$\mu$K$^2$ as in the re-analysis of the COPSS result in Section 5.3 of~\cite{mmIME-ACA}. However, the COPSS result itself may require its own upward correction, potentially by a similar fraction, depending on the relative contribution of the clustering and shot-noise components to that measurement. Furthermore, the conversion from CO(3--2) intensity to CO(1--0) intensity is highly uncertain at these redshifts, meaning that the tension may be overestimated in the first place from omitting this uncertainty.
\begin{figure}
    \centering
    \includegraphics[width=0.96\linewidth]{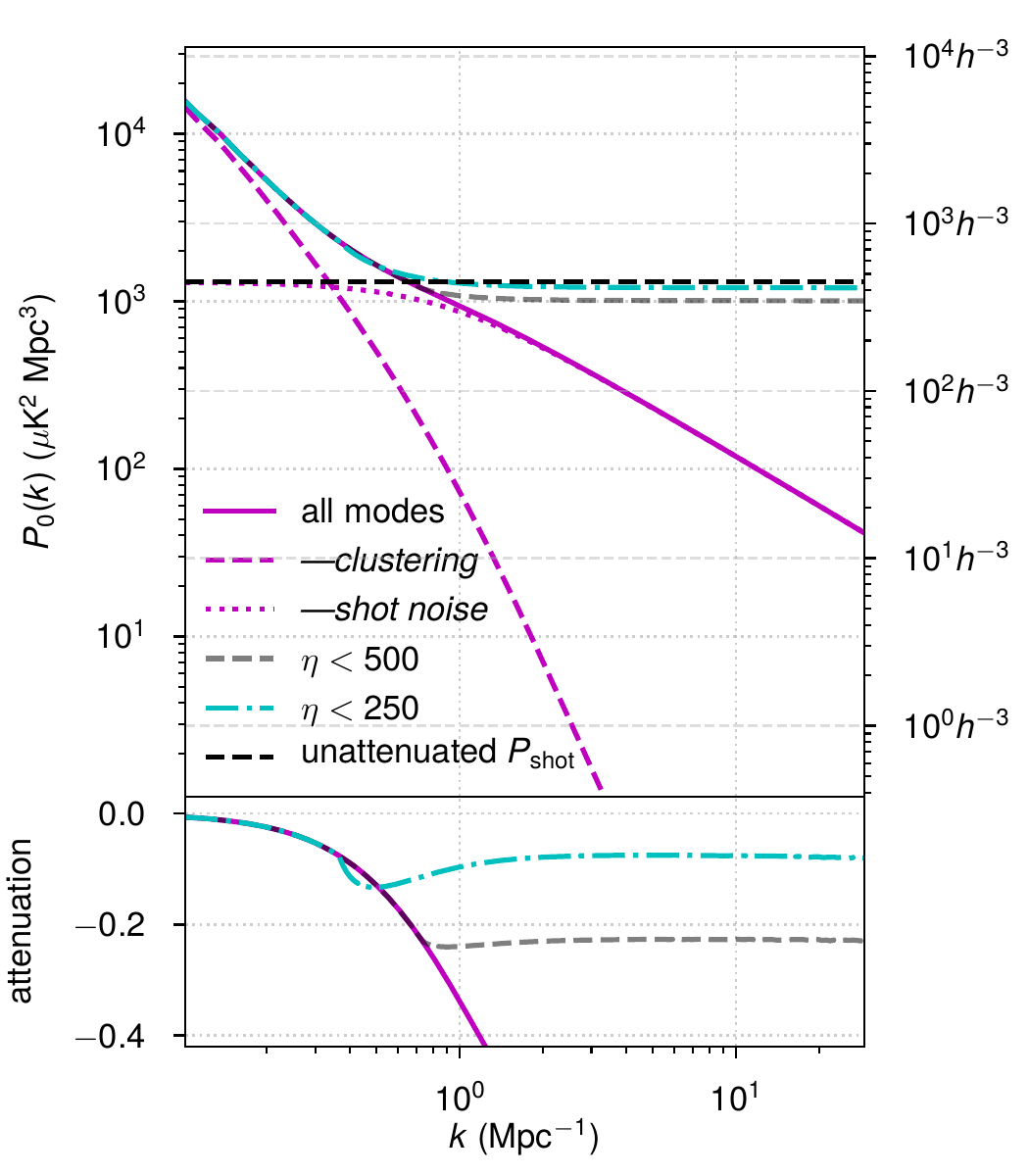}
    \caption{Illustration of the expected effect of line broadening on the total signal observed by mmIME at 100 GHz (indicated above each column of panels), with power spectra for all lines projected to the CO(3--2) frame at $z\sim2.5$ and summed together. \emph{Upper panel:} $P_0(k)$ calculated with the full formalism of~\autoref{sec:theory} (magenta solid). For the full calculation we also show the clustering (magenta dashed) and shot-noise (magenta dotted) contributions separately. We also show the unattenuated total shot noise (black dashed), as well as the expected $P_0(k)$ when restricting calculations to modes with $\eta<500$ (grey dashed) or $\eta<250$ (cyan dash-dotted). \emph{Lower panel:} Attenuation due to the introduction of line broadening, calculated using the full formalism (magenta solid). We again show the altered attenuation when restricting calculations to modes with $\eta<500$ (grey dashed) or $\eta<250$ (cyan dash-dotted).}
    \label{fig:mmIME_pspec_total}
\end{figure}
\section{Detailed Simulations}
\label{sec:sim}
Returning to the context of COMAP and our fiducial model of CO(1--0) at $z\sim3$, we find there is a need for explicit simulations of the CO intensity map across our fiducial distribution. The use of lightcones from a N-body cosmological simulation to generate these intensity maps mirrors the actual analysis expected for COMAP, and will also allow us to simulate the VID (with and without line broadening) without making the simplifying assumption of a log-normal galaxy count distribution as in~\cite{Breysse17}.
\subsection{Methods}
The cosmological simulation used is the {\tt c400-2048} box from the Chinchilla suite. \cite{Li16} used the same simulation and provide implementation details of the simulation and subsequent halo identification. The simulation spans $400h^{-1}$\,Mpc on each side, and has a dark matter particle mass of $5.9\times10^8h^{-1}\,M_\odot$. Since we only include the dark matter halo population with $M_h\geq10^{10}\,M_\odot$ in our analysis, and although the box size is not quite large enough to encompass the full comoving line-of-sight span of the COMAP observation, we are mainly concerned with $k\gtrsim0.1$\,Mpc$^{-1}$ in this work. All in all, the size and resolution of this simulation should be sufficient for the purposes of this work.

We generate 100 lightcones spanning $z=1.5$--3.5 and a flat-sky area of $100'\times100'$, each with its associated dark matter halo catalogue. As information relevant to the spatial orientation of the halo was not included at time of catalogue generation, we assign a random inclination angle $i$ to each halo. We then use \texttt{limlam\_mocker}\footnote{\url{https://github.com/georgestein/limlam_mocker}} to assign a line luminosity to each halo according to the model outlined in~\autoref{sec:LMmodel}. We generate 1280 CO(1--0) realizations, iterating through a random draw of 1280 samples of parameter values from our fiducial distribution, each assigned to one lightcone.

With the line luminosities fixed for each halo in each realization, we simulate a (noiseless) line-intensity cube spanning the full solid angle of the lightcone and 26--34 GHz in observing frequency (or $z=2.4$--3.4), with each voxel spanning $0.4'\times0.4'\times15.625$\,MHz. For each realization, we generate cubes both with and without the COMAP beam of $4.5'$ applied, and with the following five variations on simulating line broadening.
\begin{enumerate}
    \item No line widths are specified---i.e., $v(M_h)=0$ and thus $\sigma_\parallel(M_h)=0$ for all $M_h$.
    \item \texttt{vmax64}: Each halo has a calculated $v_\text{max}$, so we calculate the CO FWHM for each halo as $v_\text{max}\sin{i}/0.866$ and then bin the halo population by this FWHM in 64 linearly spaced bins. \replaced{The CO cube is generated}{We generate a CO cube} for each velocity bin \replaced{individually}{separately}, and \replaced{a Gaussian filter applied}{apply a Gaussian filter} to each cube along the line of sight, with the width of the Gaussian given by the median FWHM in each bin. We then sum the cubes across all bins to give the total CO temperature field. The maps generated using this approach serve as our ground truth for line broadening.
    \item \texttt{vvir64}: We take the same approach as in \texttt{vmax64} but using $v_\text{vir}(M_h)$ instead of $v_\text{max}$, which (as previously discussed) peak-patch or other approximate N-body simulations may not provide.
    \item \texttt{veff}: We apply a single Gaussian filter to the total CO temperature field along the line of sight, with the width of the Gaussian determined by $v_\text{eff}$ from~\autoref{eq:veff_incli}, calculated from the analytical halo model.
    \item \texttt{2tier}: This is the two-tier approach described in~\autoref{sec:2tier}. We generate a CO cube from the $M_h<10^{11}\,M_\odot$ halo population with no line broadening, and bin the $M_h>10^{11}\,M_\odot$ population in 16 linearly spaced bins of $v_\text{vir}\sin{i}/0.866$. The appropriate Gaussian filter is applied along the line of sight to the CO cube for each FWHM bin, and we sum the CO cubes for all FWHM bins plus the $M_h<10^{11}\,M_\odot$ halo subset to give the total CO temperature field.
\end{enumerate}
Whenever line broadening is applied, the CO cubes are initially generated with voxels that are four times finer in frequency space than the final voxels, so as to accurately model how the line intensity from each source is distributed across frequency channels. This combined with the Gaussian filter does increase the time required to compute the CO cube by almost a factor of 8, but as the calculation of various statistics of the cube takes significantly more time, there is not nearly as much impact on the time required to complete all calculations around each realization. Simulating line broadening across 16 bins (as in \texttt{2tier}) or 64 bins (as in \texttt{vmax64}), this total time does increase approximately by factors of 3 to 10, but such overhead would be acceptable for COMAP analysis.

We calculate $P_0(k)$ for all cases. We also calculate $P_2(k)$ and the VID for all line broadening variations, but only when the COMAP beam is applied. For the VID, the CO cube is first coarsened to voxels of $4'\times4'\times15.625$\,MHz to match the beam size, and then we calculate voxel counts $B_i$ across 75 log-spaced bins of $T_i\in(10^0,10^3)\,\mu$K. (Note that the mean temperature has been subtracted from the cube at this point.)
\subsection{Results}
\begin{figure}
    \centering
    \includegraphics[width=0.96\linewidth]{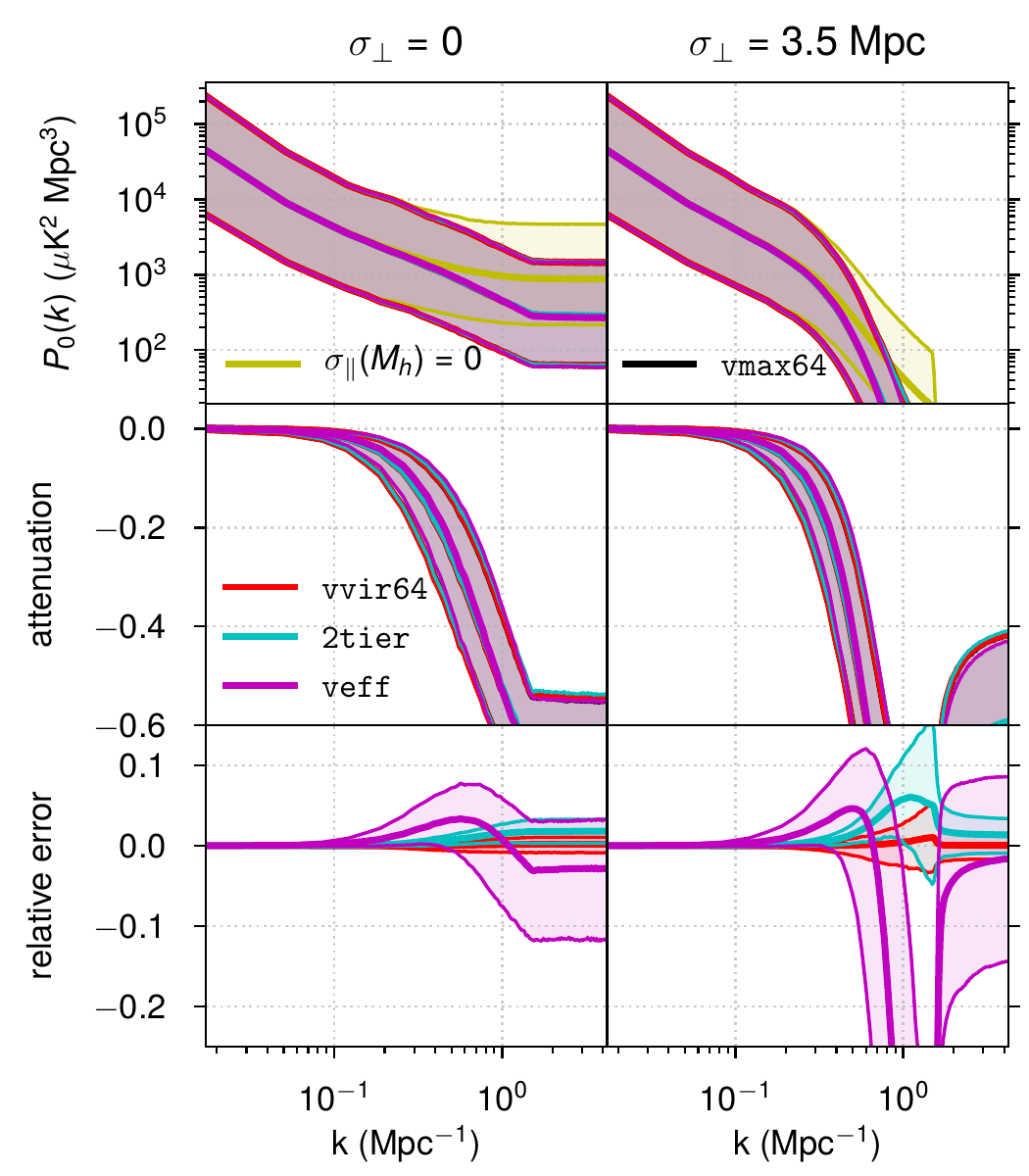}
    \caption{Summary of simulated $z\sim3$ CO(1--0) $P_0(k)$ across 1280 draws from our model posterior, using \texttt{c400-2048} lightcones. We show medians and 90\% intervals for each calculation, both ignoring angular resolution (\emph{left panels}) and assuming smearing with a Gaussian beam of FWHM $4.5'$ or $\sigma_\perp=3.5\text{\,Mpc}$ (\emph{right panels}). \emph{Upper panels:} $P_0(k)$ calculated without line broadening (yellow), using \texttt{vmax64} (black) to calculate ground truth with line broadening, and three other variations: \texttt{vvir64} (red), \texttt{2tier} (cyan), and \texttt{veff} (magenta). \emph{Middle panels:} Attenuation due to the introduction of line broadening, for each of the four variations prescribing non-zero line widths. \emph{Lower panels:} Error from using \texttt{vvir64}, \texttt{2tier}, or \texttt{veff}, relative to the ground truth \replaced{established with}{(}\texttt{vmax64}\added{)}.}
    \label{fig:test2}
\end{figure}

We show $P_0(k)$ from our simulations in~\autoref{fig:test2}, both with and without beam smearing, and with all variations on line broadening. Note that the stalling of attenuation above $k\sim1$\,Mpc$^{-1}$ is an artefact of the limited line-of-sight resolution (we expect $k_{\parallel,\text{max}}\approx0.9\text{\,Mpc}^{-1}$, corresponding to the 15.625\,MHz voxel width).

Overall the results are not very surprising. Without beam smearing, all our approximations of line broadening perform very well, landing within several percent of ground truth (\texttt{vmax64}). while there is some breakdown with beam smearing introduced, it is confined to $k\gtrsim0.5\text{\,Mpc}^{-1}$ where $P_0(k)$ is already significantly attenuated. The overall expected effect on a COMAP $P_0(k)$ detection is $\approx7$\% attenuation\replaced{, }{---consistent with the preliminary calculations from~\autoref{sec:extremedraw} using the representative parameters of Equations~\ref{eq:realparami} through~\ref{eq:realparamf}---}with the 90\% interval spanning 3\% to 14\% if we take possible variations in the $L(M_h)$ model into account.\added{ This range neglects possible variation due to other factors like source inclinations or the $v(M_h)$ model (which, if allowed to vary, may contribute almost equally to uncertainty in expected attenuation); we estimate the impact of these factors in~\autoref{sec:morevariations}.}

\begin{figure}
    \centering
    \includegraphics[width=0.96\linewidth]{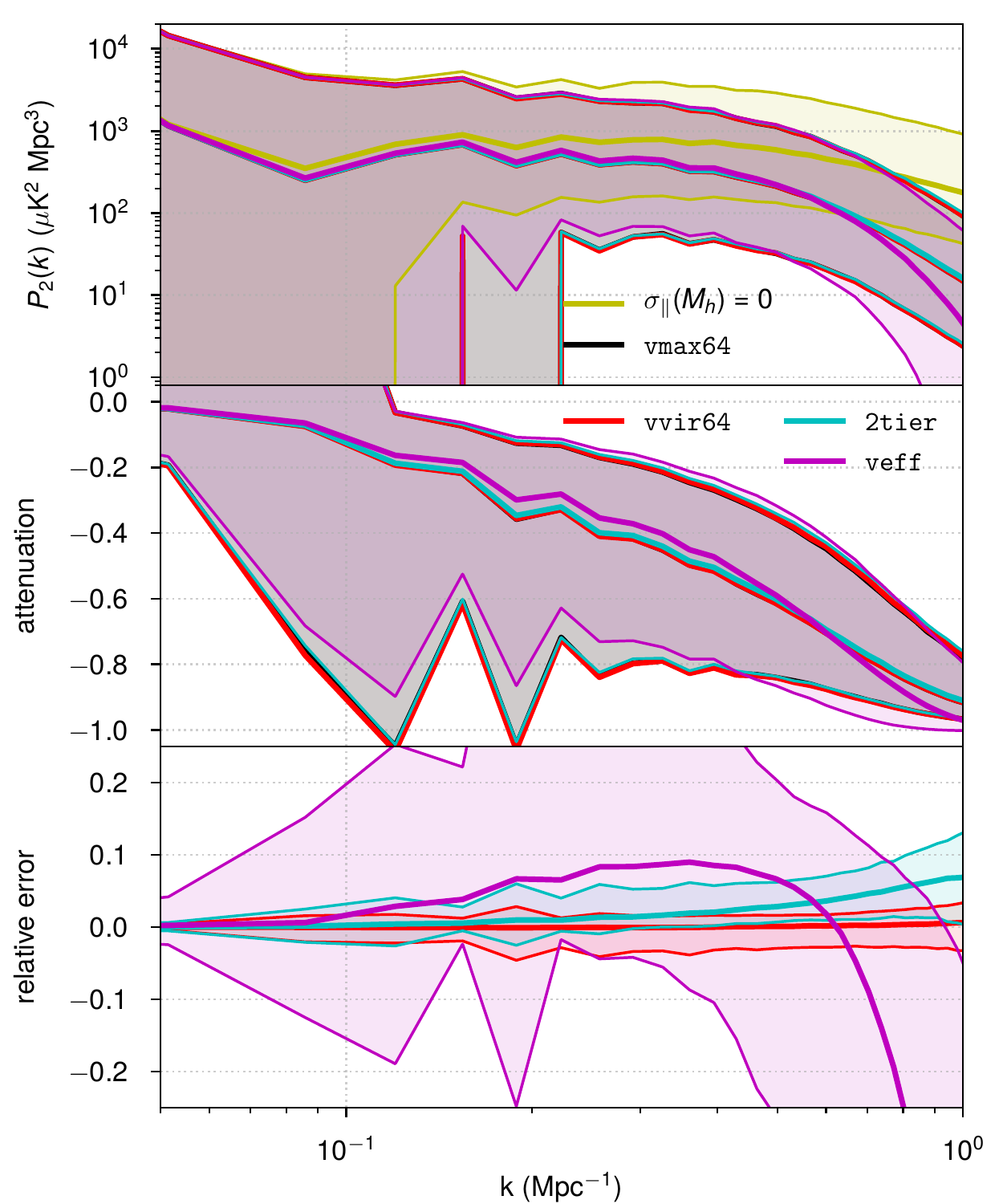}
    \caption{Similar to the right panels of~\autoref{fig:test2} (assuming smearing with a Gaussian beam of FWHM $4.5'$ or $\sigma_\perp=3.5\text{\,Mpc}$), but summarising $P_2(k)$.}
    \label{fig:test2q}
\end{figure}

We also show $P_2(k)$ from our simulations (only in the presence of beam smearing) in~\autoref{fig:test2q}, although plotting only over a $k$-range of $(0.05,1)$\,Mpc$^{-1}$ to redact both effects of finite box size and effects of limited spectral resolution in the simulated CO cube\added{ (although some numerical effects remain, leading to some visible ringing in the simulated $P_2(k)$ values)}. The attenuation is much greater across $k$ than in $P_0(k)$, exceeding 30\% by $k\sim0.2$\,Mpc$^{-1}$ in a majority of model draws. We also see that while the \emph{median} relative error from ground truth\added{ (in the lowermost panel of~\autoref{fig:test2q})} is within 10\% up to $k\sim0.6$\,Mpc$^{-1}$, the 90\% interval in relative error from our allowed variations in $L(M_h)$ can be much greater. The \texttt{2tier} approach actually still results in acceptable accuracy here, even with the relatively low number of velocity bins. The distribution of relative error at higher $k$ values shows a negative skew, particularly for \texttt{veff}. This makes sense if we recall the discussion from~\autoref{sec:extremedraw}. I\replaced{n situations where}{f} the shot noise is much lower than the clustering component, using $v_\text{eff}$ will tend to over-attenuate, whereas the approximation will not tend to under-attenuate if the shot noise is dominant.

\begin{figure}
    \centering
    \includegraphics[width=0.96\linewidth]{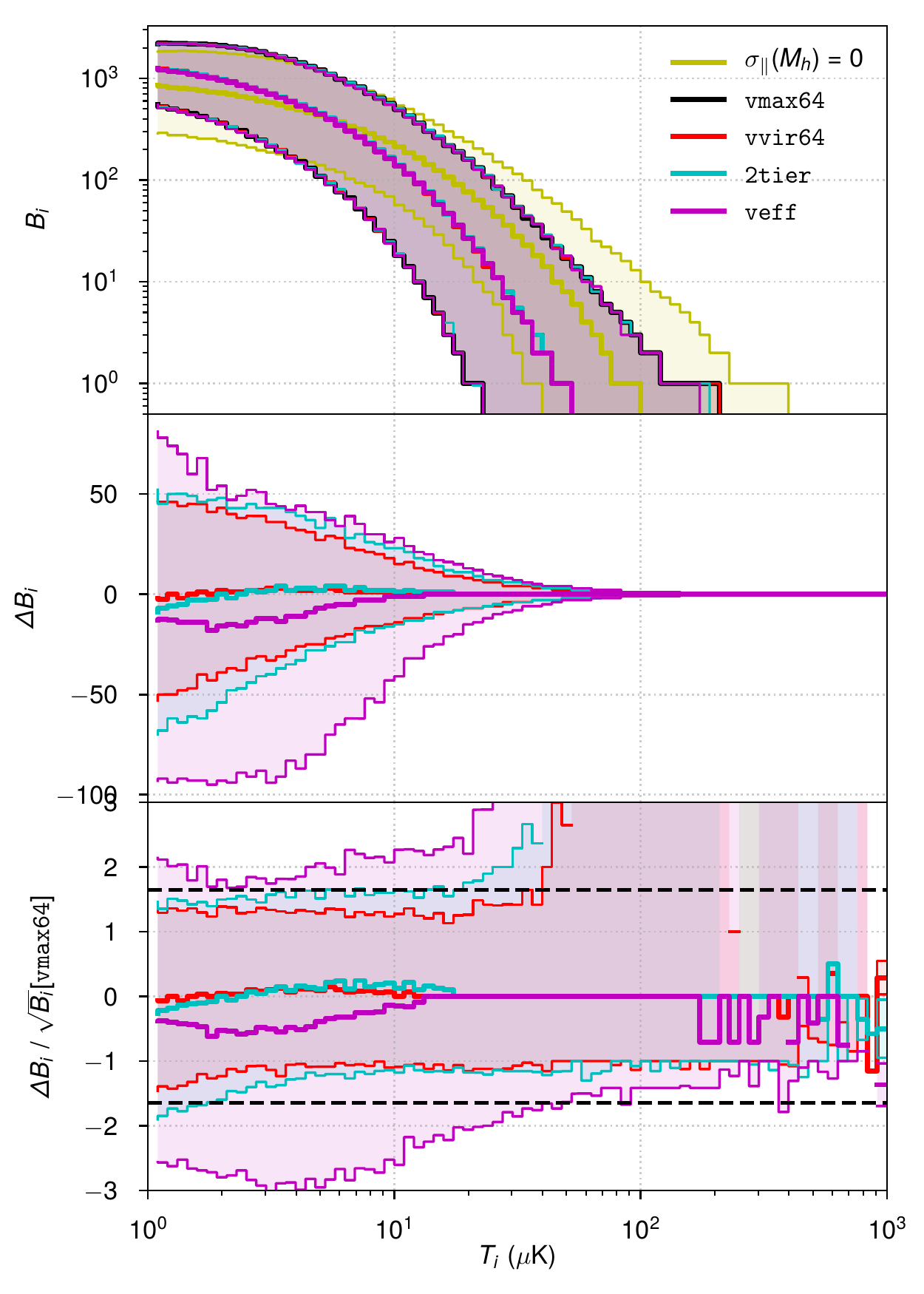}
    \caption{Summary of the simulated $z\sim3$ CO(1--0) VID across 1280 draws from our model posterior, using \texttt{c400-2048} lightcones. We show medians and 90\% intervals for each calculation, both ignoring angular resolution (\emph{left panels}) and assuming smearing with a Gaussian beam of FWHM $4.5'$ or $\sigma_\perp=3.5\text{\,Mpc}$ (\emph{right panels}). \emph{Upper panels:} Bin counts $B_i$ calculated without line broadening (yellow), using \texttt{vmax64} (black) to calculate ground truth with line broadening, and three other variations: \texttt{vvir64} (red), \texttt{2tier} (cyan), and \texttt{veff} (magenta). \emph{Middle panels:} Absolute difference from ground truth (using \texttt{vmax64}) in $B_i$ for each of \texttt{vvir64}, \texttt{2tier}, and \texttt{veff}. \emph{Lower panels:} Difference from ground truth divided by the square root of the ground truth $B_i$, representing the relative difference in units of expected Poisson error. \replaced{We show b}{B}lack dashed lines at $\Delta B_i/\sqrt{B_i}=\pm1.645$ \replaced{to r}{r}epresent the 90\% interval \replaced{we would expect if we saw}{expected with} perfect Poisson errors across all bins.}
    \label{fig:test3}
\end{figure}
Finally, in~\autoref{fig:test3} we consider the VID for the first time. Broadly, all line broadening simulation variations result in the same effect, shifting voxel counts from high $T_i$ to low $T_i$. But while the resulting $B_i$ are indistinguishable on a log-log plot, the deficiencies of the \texttt{veff} approach become clear when looking at a linear plot of the absolute error $\Delta B_i$ relative to ground truth. While using \texttt{veff} at least resulted in a reasonable estimate of $P_2(k)$ on average, here we see a clear systematic error at low $T_i$. When we consider $\Delta B_i/B_i^{1/2}$---that is, the ratio of this error to the Poisson error expected from the `true' $B_i$---we find that \texttt{veff} is the only approach where the relative error from the approximation \added{is} clearly \replaced{exceeds}{in excess of} Poisson error.
\section{Discussion}
\label{sec:discussion}
\subsection{The Inevitable Limitations of a Description of Line Broadening Using a Single Parameter}
In the present state of line-intensity mapping, the primary observable for autocorrelation experiments similar to COMAP and mmIME remains the spherically-averaged monopole power spectrum $P_0(k)$, for which we have shown that the above $v_\text{eff}$ approximation works very well as a replacement for a numerically calculated transfer function for the shot-noise component. The results of the following subsection will also demonstrate that most of the time, this approximation allows us to match ground truth to within 10--20\% at scales relevant to COMAP, a significant improvement over the systematic errors that would arise from neglecting line broadening. Therefore, for forecasting or analysis of the line-intensity $P_0(k)$, using a single $v_\text{eff}$ is \emph{adequate}.

However, clear shortcomings to this approach exist. One shortcoming, which is perhaps not overly relevant in application to analysis, is that the approximation breaks down at sufficiently high $k$ in the presence of an angular beam of sufficiently large comoving size. In the case of COMAP, an angular size of 4.5 arcminutes on sky corresponds to $\sigma_\perp\approx3.5$\,Mpc at the central COMAP redshift, compared to typical values of $\sigma_{\parallel,\text{eff}}\sim1.5$--2\,Mpc that we would expect from $v_\text{eff}\approx250$--350\,km\,s$^{-1}$. Yet we have not accounted for the angular beam in our approximation above.

We might contemplate accounting for $\sigma_\perp$ explicitly, \replaced{in which case the requirement for an effective $\sigma_\parallel$ changes from~\autoref{eq:veff_tofit2} to this:
\begin{equation}\frac{\operatorname{erf}{\left[k(\sigma_{\parallel,\text{erf},\sigma_\perp}^2-\sigma_{\perp}^2)^{\frac{1}{2}}\right]}}{k(\sigma_{\parallel,\text{erf},\sigma_\perp}^2-\sigma_{\perp}^2)^{\frac{1}{2}}}\approx\frac{\avg{L^2\frac{\text{erf}{\left[k(\sigma_v(M_h)^2-\sigma_{\perp}^2)^{\frac{1}{2}}\right]}}{k(\sigma_v(M_h)^2-\sigma_{\perp}^2)^{\frac{1}{2}}}}}{\avg{L^2}}.\label{eq:veff_tofit3}\end{equation}
(Note that due to the properties of the error function, this is a well-defined real-valued function even for $\sigma_\perp>\sigma_\parallel$.) For brevity, we forgo showing results from fitting for $\sigma_{\parallel,\text{eff}}$ using this alternate prescription. But importantly, replacing $\sigma_{\parallel,\text{erf}}$ with $\sigma_{\parallel,\text{erf},\sigma_\perp}$ does not actually result in appreciable improvement in relative error.

In essence, no matter what exact value we use for the effective line width, the shape of the left- and right-hand sides of~\autoref{eq:veff_tofit3} differ too much across $k$ when the angular smearing of the signal in comoving space is comparable to or exceeds the line-of-sight smearing from line broadening. We thus}{via appropriately tweaking~\autoref{eq:veff_tofit2}. However, calculations in~\autoref{sec:madansatz} show that doing so does not actually improve relative error in any appreciable fashion. The fundamental truth is that we} fully expect these kinds of approximations to fail at sufficiently high $k$\replaced{---in practice we see the median relative error in $P_0(k)$ exceed}{. We note that, looking at the lower right panel of~\autoref{fig:test2} for the median relative error in $P_0(k)$ versus ground truth when using $v_\text{eff}$, this relative error only exceeds} 10\% for $k\gtrsim0.9\text{\,Mpc}^{-1}\approx\pi/\sigma_\perp$\added{ for our COMAP CO(1--0) simulations}. \replaced{T}{So in practice, t}his concern at least should not be too relevant to analysis, where the $P_0(k)$ loss around this $k$-range is already sufficiently high ($\gtrsim80$\%) that data here are likely to be discarded.

However, our approximation remains wanting in analysis due to concerns that are extremely relevant in the context of analyses like~\cite{Ihle19} that deal with the VID rather than just $P(k)$. The central motivation behind using the VID jointly with $P(k)$, as explained in previous works like~\cite{Breysse17}, is that the CO intensity field observed by COMAP is highly non-Gaussian, so that the VID contains substantial information beyond $P(k)$. Just as $P(k)$ therefore does not fully determine the VID, the $P_0(k)$ transfer function does not fully determine the VID transfer function. Therefore, a single parameter describing the $P_0(k)$ transfer function from line broadening---which summarizes our approach with a single $v_\text{eff}$---would never be able to fully describe the effect of line broadening on the VID.

Even without thinking about non-Gaussianities explicitly, we can consider an extreme scenario. Suppose we have 1000 emitters with 999 of them at $L\sim10^4\,L_\odot$ and negligible line broadening, and only one at $L\sim10^6\,L_\odot$ but line width of $v\sim300$\,km\,s$^{-1}$, so that
    \begin{equation}\frac{dn}{dL}\propto999\delta_D(L-10^4\,L_\odot)+\delta_D(L-10^6\,L_\odot),\end{equation}
where we use $\delta_D$ to denote the (one-dimensional) Dirac delta function. \replaced{With this distribution of luminosities}{Given this $dn/dL$}, clearly the fainter but more numerous emitters dominate the clustering component and occupy the overwhelming majority of voxels, but the single bright emitter will dominate any $L^2$-weighted statistics by itself, including shot noise and our \replaced{approximate $v_\text{eff}$. In particular,}{effective line width:}
    \begin{align}
        v_\text{eff}&=\frac{\int dL\,\frac{dn}{dL}\,L^2v(L)}{2\int dL\,\frac{dn}{dL}\,L^2}+\frac{\int dL\,\frac{dn}{dL}\,L^2}{2\int dL\,\frac{dn}{dL}\,L^2/v(L)}\\&=\frac{(10^6\,L_\odot)^2\cdot300\text{\,km\,s}^{-1}}{999\cdot(10^4\,L_\odot)^2+(10^6\,L_\odot)^2}\approx270\text{\,km\,s}^{-1}\replaced{,}{.}
    \end{align}
\replaced{s}{S}o if we use a single Gaussian filter across the entire CO cube as in the previous subsection's treatment, then we will broaden all line profiles by almost the full line width of the single bright emitter. This may not affect $P_0(k)$ too much---the full 3D $P(\mathbf{k})$ may be somewhat affected but the effect on the averaged $P_0(k)$ will be small at smaller scales (where the bright emitter dominates the shot noise power spectrum anyway) and even less at larger scales. However, this $v_\text{eff}$ prescription will distort the VID significantly, as we see in~\autoref{fig:test3}. Voxels around the locations of fainter emitters, which \emph{should} be in low-temperature bins of $T_i\sim10^0$--$10^1\,\mu\text{K}$ in a proper simulation, instead end up with \emph{overly} low temperatures of $\lesssim10^0\,\mu\text{K}$ due to excessive line broadening. No procedure can avoid this effect if it will apply the same line width to a halo subset spanning orders of magnitude in various properties, which is to say that a very coarse binning of the halo catalogue in a simulation into two or three subsets is unlikely to substantially correct for VID distortion.\footnote{This does not exclude the possible devising of a method to correct the VID, but we consider this to be beyond the scope of the present work.}

In summary, for any analyses that involve $P_0(k)$ detection by itself, our $v_\text{eff}$ prescription is valid. But we need a more thorough line broadening simulation procedure for anything more advanced---not just the VID, but other statistics like higher-order moments of the full 3D power spectrum that we could consider as part of future COMAP science. What distinguishes line-intensity mapping from line candidate scans or targeted galaxy surveys is the measurement of aggregate line emission from both bright and faint galaxies. We therefore neglect faint line emitters at our own peril, not only in models of the signal but also in models of systematic effects.
\subsection{Challenges for Interpretation of Single-dish and Interferometric LIM Surveys}
Qualitatively, line broadening poses the same challenges for single-dish surveys like COMAP as for interferometric surveys like COPSS and mmIME. The primary challenge is the attenuation of the power spectrum to begin with, which \replaced{means}{somewhat raises} sensitivity requirements to achieve a given signal-to-noise ratio \replaced{are somewhat higher than previously thought}{for $P(k)$}. The secondary challenge---just as important---is in interpretation of a detection, which requires accurate correction for the attenuation.

The work here lets us gauge the former, but not necessarily the latter. We have seen above that, in the context of an initial detection with a signal-to-noise ratio of 3--5, the attenuation of the power spectrum due to line broadening would be small or can be reduced with appropriate redaction of the data to a level that is subdominant to uncertainties from noise. However, once LIM surveys reach higher detection significance, we will no longer be able to ignore either this correction or the uncertainties around the various assumptions around line profile width and shape that underlie the correction.

Quantitatively, speaking \emph{very specifically} about the problem of line broadening as it relates to detection of $P_0(k)$, single-dish surveys are at an advantage over interferometric surveys. The attenuation expected for COMAP at $k\sim0.2$--0.3\,Mpc$^{-1}$ is around 10\%, whereas for mmIME the attenuation expected at $k\sim10$\,Mpc$^{-1}$ is around 25\% even with the calculations restricted to $\eta<500$ (which discards modes above a certain $k_\parallel$ and thus is part of a sensitivity trade-off between number of Fourier modes and signal recovery). This means COMAP requires only a 11\% correction, versus 33\% for mmIME. If model uncertainties add relative error to this correction, interferometric surveys have the larger correction to begin with and thus will bear more of a challenge.

This, however, is only part of the story. If COMAP analyses make use of statistics that put more weight on small-scale data, such as the quadrupole $P_2(k)$ (with its shot-noise component from anisotropic smearing) or the VID, these analyses will rely heavily on models of line broadening. Even without undertaking rigorous simulations, we can consider a more extreme scenario where the true line profile is described by either a top-hat profile or a Gaussian profile. These two shapes are clearly going to result in very different kernels for the VID, and if \replaced{we think these are equally likely shapes}{the two are equally likely} then we will need to split the difference and indicate appropriate uncertainties.

In the case of high-redshift CO, we think the Gaussian assumption is well-justified as it has served well in the context of untargeted high-redshift CO line searches like ASPECS, where the Gaussian shape is a reasonable fit for most (although not all) of the detections by~\cite{ASPECS-LP}. However, if we did want to seriously consider other possible line shapes like double-Gaussian or double-horned profiles, we must account in inferences for the uncertainty from \replaced{our inability to decide the true line shape}{such variations}. We leave this problem to future work.
\section{Conclusions}
\label{sec:conclusion}
The scope of this work, simply put, was to answer three questions posited in the Introduction. As we conclude, we find an answer to each question (partly reproduced):
\begin{enumerate}
\item \emph{What is the level of signal attenuation that we can expect for experiments like COMAP and mmIME due to line broadening?} Our median expectation for COMAP is a $\lesssim10\%$ attenuation of the line-intensity $P_0(k)$ at relevant scales, with the 90\% interval from our fiducial model ensemble being 3--14\%. Our prediction of the effect for mmIME is larger, broadly agreeing with the expectation of~\cite{mmIME-ACA} of 25\% attenuation after data cuts.
\item \emph{Is it sufficient to describe the effect of line broadening using a single parameter, such as an effective global line width?} We have described a way to calculate an effective global line width that \emph{does}, in fact, allow for a reasonable approximation of the effect of line broadening on the monopole $P_0(k)$.
\item \emph{How does this simplification fail?} The approximation has greater errors for the quadrupole $P_2(k)$, and results in significant systematic error in the context of the VID. These \replaced{are fundamental features coming from}{results demonstrate} the overdetermined nature of attempting to define a single effective global line width to describe non-Gaussian statistics.
\end{enumerate}

The effect of line broadening is thus less critical for initial upper limits or detections (where uncertainties from noise will likely dominate over any uncertainty in corrections for line broadening), but has serious implications for more advanced analyses\added{, especially those} hoping to make use of other statistics like $P_2(k)$ or the VID. The strong attenuation of $P_2(k)$ \replaced{would}{may} have \replaced{serious}{non-negligible} effects on the signal-to-noise required for cosmological analyses of the kind proposed by~\cite{Bernal19b,Bernal19a}\added{, and work is already underway to quantify impacts on such analyses. While LIM cosmology may be somewhat insulated from line broadening due to the focus on larger scales, we expect the effects will be more severe for astrophysical analyses of the kind discussed by~\cite{SchaanWhite21}. Their proposal to disentangle any 1-halo clustering contribution from shot noise at intermediate to small scales requires measurements of $P_\ell(k)$ at the kinds of scales where we predict significant attenuation of $P_\ell(k)$}. Although the challenge is not insurmountable\replaced{, }{ in theory, }it will necessitate building more sensitive instruments or operating longer surveys than we may have previously thought.

We make no claim that our $v(M_h)$ model for CO(1--0) at $z\sim3$ is entirely accurate, but it is consistent with the information we have at the time of writing. It is however somewhat tenuous as tying the line width directly to the halo circular velocity carries assumptions about gas density profiles and dynamics that we discussed briefly in~\autoref{sec:LWmodel}. Any future work that directly contradicts these assumptions would automatically demand revisions of our model.

It will be important going forward to study in great detail the line profile shape and width of both faint and bright high-redshift galaxies in various lines, and how these correlate with properties that we may be able to connect to halo properties in dark matter simulations\added{, or at least properties of galaxies in semi-analytic models or baryon-inclusive cosmological simulations}. This will be particularly important in studying the luminosity function and non-Gaussianities via the VID, where the line shape may perhaps be even more important than the width.

Paradoxically it may be line-intensity mapping that may be best suited to study some of these properties in certain regimes. If we find, for instance, that LIM observables like $P_\ell(k)$ and the VID are not well described using models of line broadening devised with data from untargeted line searches like ASPECS, then it may indicate something about the line profiles of faint emitters that differs from the line profiles of bright emitters as studied by line searches (or at least the extrapolation from bright emitters to faint emitters that would have been used).

Overall, line broadening is a systematic obstacle to detection and interpretation of LIM signals, but by no means a catastrophic one. We trust that future work---not only in simulation but also in observation---will continue to shed light on the problem and offer fresh approaches.

\begin{acknowledgements}

DTC is supported by a CITA/Dunlap Institute postdoctoral fellowship. The Dunlap Institute is funded through an endowment established by the David Dunlap family and the University of Toronto. KC acknowledges support from the National Science Foundation under Grant Nos.\ 1518282 and 1910999. Work at the University of Oslo is supported by the Research Council of Norway through grant 251328. HP acknowledges support from the Swiss National Science Foundation through Ambizione Grant PZ00P2\_179934. JOG acknowledges support from the Keck Institute for Space Studies, NSF AST-1517108, and the University of Miami.\added{ LCK was supported by the European 
Union’s Horizon 2020 research and innovation programme under the Marie 
Skłodowska-Curie grant agreement No.~885990.} Thanks to Sarah Church and Karto Keating for early discussions that motivated and informed the present work, and to Tim Pearson and other members of the COMAP collaboration for helpful comments on the manuscript. We thank Matthew Becker and Risa Wechsler for access to the Chinchilla cosmological simulation (\texttt{c400-2048}) used in this work. We thank Riccardo Pavesi for access to the COLDz ABC posterior sample used in this work. COMAP is supported by the National Science Foundation under Grant No.\ 1910999. This research made use of NASA's Astrophysics Data System Bibliographic Services. Some of the computing for this project was performed on the Sherlock cluster. DTC would like to thank Stanford University and the Stanford Research Computing Center for providing computational resources and support that contributed to these research results.\added{ Part of this work was shown pre-publication at the 2021 Line Intensity Mapping Workshop hosted by the University of Chicago and the Kavli Institute for Cosmological Physics, whose (virtual) hospitality we gratefully acknowledge. Finally, we thank an anonymous referee whose comments helped improve the manuscript.}
\end{acknowledgements}

\software{\texttt{hmf}~\citep{hmf}; Matplotlib~\citep{matplotlib}; Astropy, a community-developed core Python package for astronomy~\citep{astropy}.}

\appendix

\section{Explicit Analytic Corrections for Inclination}
\label{sec:incli}
Under the specific but reasonable assumption of randomly oriented, rotation-dominated emitters, we may introduce an analytic correction to the expressions of~\autoref{sec:theory}. Instead of $\sigma_v(M_h)$, the applicable width is $\sigma_v$ times a random $\sin{i}$---or rather $\sin{i}/0.866$, if we assume that the mean $\sigma_v$ value corresponds to the median value of $\sin{i}=\sqrt{3}/2$. For random orientation, the distribution of $\cos{i}$ is uniform such that the average attenuation of the shot noise at fixed $k$ and $\mu$ may be calculated as follows (partly via Mathematica\footnote{Wolfram Research, Inc.; Version 12.1, 2020.}):
\begin{align}
    P_{\text{shot,v},i}(k,\mu)&=C_{LT}^2\int dM_h\int_0^1d{(\cos{i})}\,\frac{dn}{dM_h}\,L^2(M_h)\exp{\left[-k^2\sigma_\perp^2(1-\mu^2)-k^2\sigma_v^2(M_h)\mu^2\cdot\frac{1-\cos^2{i}}{3/4}\right]}\nonumber\\
    &=C_{LT}^2\int dM_h\,\frac{dn}{dM_h}\,L^2(M_h)\exp{\left[-k^2\sigma_\perp^2(1-\mu^2)\right]}\left[\frac{2k\sigma_v(M_h)\mu}{\sqrt{3}}\right]^{-1}F{\left(\frac{2k\sigma_v(M_h)\mu}{\sqrt{3}}\right)},\label{eq:Pshotvi}
\end{align}
where $F(x)$ is Dawson's integral. (For our purposes, we consider $F(x)/x$ to evaluate to $1$ at $x=0$.) With similar replacements, we can calculate the clustering component as follows:
\begin{align}P_{\text{clust},v,i}(k,\mu)
&=C_{LT}^2P_m(k)\left\{\int dM_h\,\frac{dn}{dM_h}\,L(M_h)\left[b(M_h)+\Omega_m(z)^{0.55}\mu^2\right]\times\right.\nonumber\\&\qquad\left.\exp{\left[-\frac{k^2\sigma_\perp^2(1-\mu^2)}{2}\right]}\left[\frac{\sqrt{2}k\sigma_v(M_h)\mu}{\sqrt{3}}\right]^{-1}F{\left(\frac{\sqrt{2}k\sigma_v(M_h)\mu}{\sqrt{3}}\right)}\right\}^2.\label{eq:Pclustvi}\end{align}

We have inadvertently assumed that perfectly face-on galaxies will have an observed line FWHM of zero, which cannot possibly\added{ be} true for several reasons. The foremost astrophysical reason is that the CO gas will have its own velocity dispersion independent of the overall rotation. For instance,~\cite{deBlokFabian14} consider dispersions of 10 or 50\,km\,s$^{-1}$ for low- and high-redshift galaxies. Furthermore, finite frequency resolution means we would never observe even point sources as point sources. Gas dispersions of 10--50\,km\,s$^{-1}$ are certainly subdominant to the $\approx16$\,MHz channelization used both in this work in~\autoref{sec:sim} and in~\cite{mmIME-ACA} for mmIME data. On a more pragmatic level, the randomly drawn inclinations will rarely hit very close to face-on, and certainly never exactly face-on---less than 0.5\% of simulated emitters would have $\sin{i}<0.1$---so we believe this oversight may be overlooked.

We can get an approximate idea of the level of correction this introduces for the shot noise by calculating the ratio of the integrands in~\autoref{eq:Pshotv} and~\autoref{eq:Pshotvi}. For $k\sigma_v\mu\sim0.1$ the ratio is within a few percent of unity, but for $k\sigma_v\mu\gtrsim1$ the relative difference begins to exceed 10\%. Therefore, as the effect is greater at smaller scales, it will clearly affect simulations of the VID.

We turn to implications for our effective line widths. The main task is to adjust our high-$k$ ansatz, as the smaller scales are most affected. The integral of~\autoref{eq:Pshotvi} in $\mu\in(0,1)$ (setting $\sigma_\perp=0$) is unfortunately somewhat less pleasant than before:
\begin{equation}
    P_{0,\text{shot},v,i} = C_{LT}^2\avg{L^2\,{}_2F_2\left(\frac{1}{2},1;\frac{3}{2},\frac{3}{2};-\frac{4k^2\sigma_v^2}{3}\right)},
\end{equation}
where ${}_2F_2$ denotes the generalized hypergeometric function. Looking at $k\sigma_v\to\infty$, the leading-order behaviour is such that
\begin{equation}
    {}_2F_2\left(\frac{1}{2},1;\frac{3}{2},\frac{3}{2};-\frac{4k^2\sigma_v^2}{3}\right)\sim\frac{\pi^{3/2}}{4}\left(\frac{2}{\sqrt{3}}k\sigma_v\right)^{-1}.
\end{equation}
Compare to~\autoref{eq:S0} and~\autoref{eq:veff_tofit}, where the prefactor for $\operatorname{erf}{(k\sigma)}/(k\sigma)\sim(k\sigma)^{-1}$ is $\pi^{1/2}/2$. Thus we should multiply the high-$k$ ansatz by the ratio of the two prefactors, which is $4/(\pi\sqrt{3})\approx0.735$. (Note that if we were assuming $\sigma_v(M_h)$ were the value for face-on galaxies as opposed to $\cos{i}=0.5$, the correction would be greater, requiring multiplication by $2/\pi\approx0.637$.) Thus we derive the inclination-corrected $v_\text{eff}$ of~\autoref{eq:veff_incli} as the midpoint between the inclination-corrected high-$k$ ansatz and $\avg{L^2v}/\avg{L^2}$ as is, as the latter approximates the line broadening effect at values of $k$ where incorporating inclination does not yield appreciable changes.
\section{Approximate Analytic Validation of Effective Line Width Ansatz}
\label{sec:ansatzcheck}
An exact analytic validation of our ansatz for the effective line width is challenging on multiple levels. The error function is a decidedly non-elementary function, and $dn/dM_h$ also takes a somewhat complex form, specified for instance by~\cite{MRP18} by a generalized Schechter function (i.e., a power law with exponential cutoff of variable sharpness). We can however show for a highly simplified model that it really is a reasonable ansatz through a basic polynomial approximation to the error function.

As the model of~\cite{MRP18} suggests, $dn/dM_h\sim M_h^{-2}$ up to the exponential cutoff, although some variation in this power-law slope exists. Here we simplify the picture drastically so that $dn/dM_h$ is such a power law up to some maximum $M$ and equal to zero above it. If we suppose $L(M_h)$ follows a power law up to that cutoff, we broadly expect that power-law slope to be at least positive if not super-linear. We could leave $v(M)$ as a generic power law leaving the slope unspecified, but for illustrative purposes and given physical expectations, we will fix $v(M)\sim M^{1/3}$.

Up to various constants and coefficients that we will omit, our ansatz is
\begin{align}
    \frac{\avg{L^2v}}{\avg{L^2}}+\frac{\avg{L^2}}{\avg{L^2v^{-1}}}&=\frac{1}{2}\left(\frac{\int dM_h\,M_h^{a+1/3}}{\int dM_h\,M_h^a}+\frac{\int dM_h\,M_h^a}{\int dM_h\,M_h^{a-1/3}}\right)\nonumber\\&= \frac{M^{1/3}}{2}\left(\frac{a+1}{a+4/3}+\frac{a+2/3}{a+1}\right).
\end{align}

Now we consider solving for $\sigma_\parallel$ in~\autoref{eq:veff_tofit2} using different approximations in different regimes. We will however replace $k\sigma_{\parallel,\text{eff}}$ with $M_\text{eff}^{1/3}$ and $k\sigma_\parallel(M_h)$ with $M_h^{1/3}$ to simplify our calculations. It will not affect our conclusions around the relative error of different approximations in different regimes.

We can broadly think of three different regimes: $k\sigma\ll1$, $k\sigma\lesssim1$, and $k\sigma\gg1$. In the first regime of very large scales and/or very small masses or velocities, attenuation is minimal and~\autoref{eq:veff_tofit2} is almost tautological. In the last regime of very small scales and/or very large masses or velocities, $\operatorname{erf}{(x)}/x\approx1/x$ and so we actually approach the second component of our ansatz, $\avg{L^2}/\avg{L^2v^{-1}}$.

In the middle regime, we use a second-order Taylor series approximation: $\operatorname{erf}(x)/x\approx 2\pi^{-1/2}(1-x^2/3)$ (within 5\% up to $x=0.8$). Since the prefactor will cancel anyway when calculating $M_\text{eff}$, we consider the $L^2$-weighted average $1-M_h^{2/3}/3$:
\begin{align}
    1-\frac{M_\text{eff}^{2/3}}{3} = \frac{\int dM_h\,M_h^a(1-M_h^{2/3}/3)}{\int dM_h\,M_h^a} = 1-\frac{a+1}{3a+5}M^{2/3},
\end{align}
which is to say that
\begin{equation}
    M_\text{eff}^{1/3}=\left(\frac{a+1}{a+5/3}\right)^{1/2}M^{1/3}.
\end{equation}
So now we can compare the coefficients in front of $M^{1/3}$ for each approximation---the $\avg{L^2v}/\avg{L^2}$ ansatz; the $\avg{L^2}/\avg{L^2v^{-1}}$ ansatz which the truth will approach as $k\sigma\to\infty$; the $k\sigma\lesssim1$ approximation; and the midpoint ansatz:
\begin{align}
    C_{v} &= \frac{a+1}{a+4/3}\\
    C_{v^{-1}} &= \frac{a+2/3}{a+1}\\
    C_{k\sigma\lesssim1} &= \left(\frac{a+1}{a+5/3}\right)^{1/2}\\
    C_\text{eff} &= \frac{C_v+C_{v^{-1}}}{2} = \frac{1}{2}\left(\frac{a+1}{a+4/3}+\frac{a+2/3}{a+1}\right)
\end{align}
If $dn/dM_h\sim M_h^{-2}$, then $a=0$ would correspond to a roughly linear $L(M_h)$, and values above this to models with super-linear $L(M_h)$ like our fiducial model ultimately is at low mass. Therefore we can heuristically consider how our different approximations compare for different faint-end $L(M_h)$ slopes.

It is clear by inspection that $C_v > C_{v^{-1}}$ and that for high $a$ (i.e., for a sufficiently steep $L(M_h)$ power law) these approximations become more similar. The latter point is not necessarily worth much thought when we return to more realistic models, as the approximations may still diverge depending on the model behaviour at high mass (where we have not specified the exponential cutoff and bright-end power law at all). Even with the former point, we do note that for values of $a$ around 2 or 3 (which we might expect for $dn/dM_h\sim M^{-2}$ and a faint-end $L(M)\sim M_h^2$ or $M_h^{2.5}$), the difference is not too great at around 10\%. So while $C_v$ clearly differs from $C_{v^{-1}}$ it is still surprisingly close for a blind guess.

Perhaps even more surprising is the closeness between $C_v$ and $C_{k\sigma\lesssim1}$, the latter being the best we can do analytically within reason at intermediate scales. It is perhaps clearer if we explicitly express the ratio between the two,
\begin{equation}
    \frac{C_v}{C_{k\sigma\lesssim1}} = \frac{[(a+1)(a+5/3)]^{1/2}}{a+4/3}.
\end{equation}
We note first that the ratio is quite close to unity---$C_v$ is within 5\% of $C_{k\sigma\lesssim1}$ for values of $a$ around 2 or 3. But we also note that this ratio is one between an arithmetic mean and a geometric mean. Since the arithmetic mean of two non-negative real numbers is always greater than their geometric mean, it is always true that $C_v < C_{k\sigma\lesssim1}$ (as long as a real value of $C_{k\sigma\lesssim1}$ exists for comparison).

We thus have $C_{v^{-1}}<C_v<C_{k\sigma\lesssim1}$, with $C_v$ being close to truth at intermediate scales  and $C_{v^{-1}}$ being close to truth at $k\sigma\gg1$ but necessarily further away from truth at intermediate scales compared to $C_v$. The midpoint ansatz represented by $C_\text{eff}$ is a clear compromise---not the optimal choice in any single regime but able to reduce the maximum error across all regimes.
\added{\section{A Closer Look at Attenuation for Specific Parameter Values, But Without Analytic Corrections for Inclination}
\label{sec:extremedraw_noincli}
In~\autoref{fig:veff_noincli} we show the same plots as in~\autoref{sec:extremedraw}, but without any accounting of random source inclinations. We also show the results with inclination for reference, but the high-$k$ ansatz does not include the correction for inclination previously described. We note that the effect is relatively subtle for most $k$, and that the expected accuracy of the appropriate $v_\text{eff}$ is quite similar with or without inclination accounting.

\begin{figure}
    \centering
    
        \includegraphics[width=0.42\linewidth]{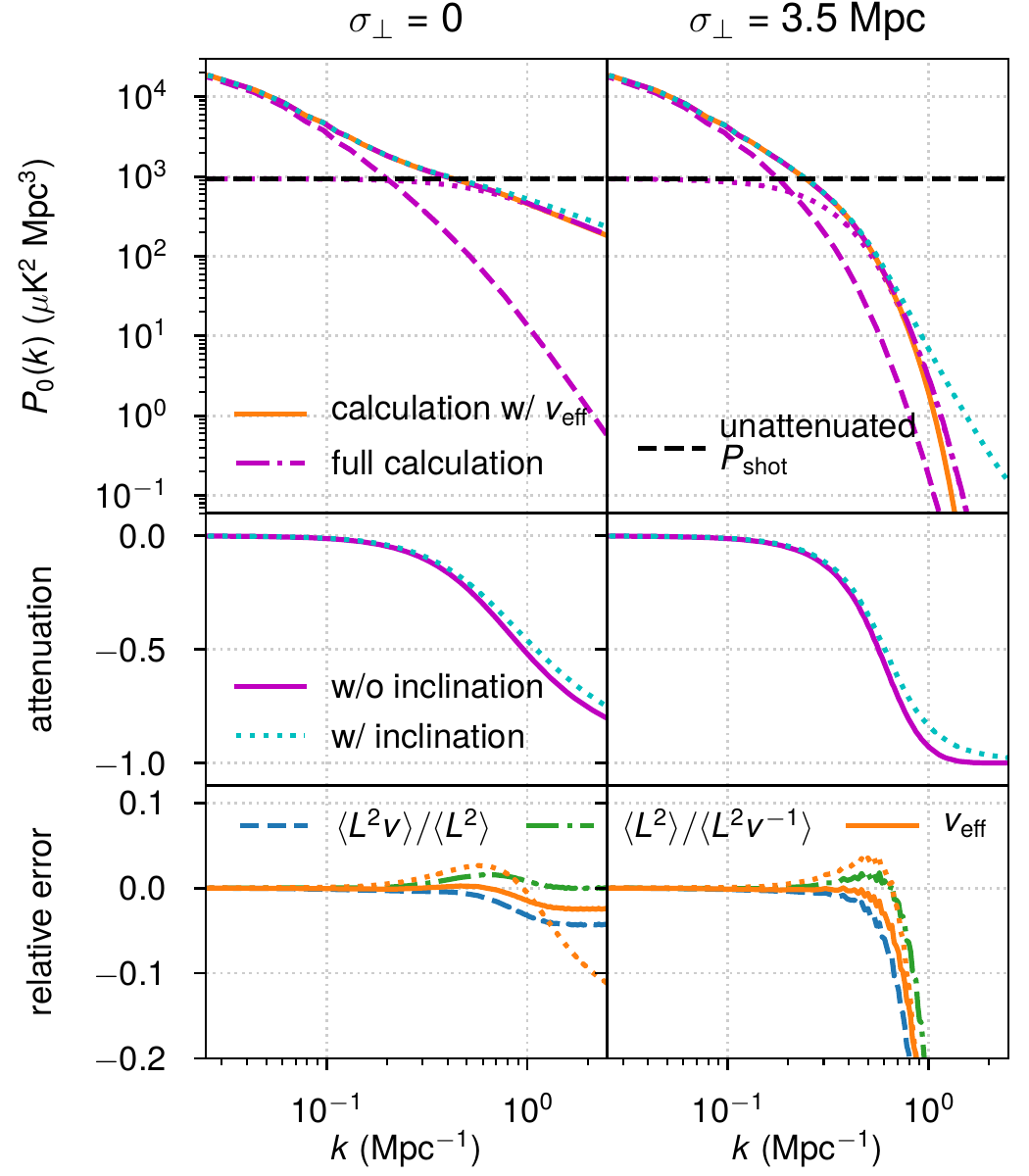}\includegraphics[width=0.42\linewidth]{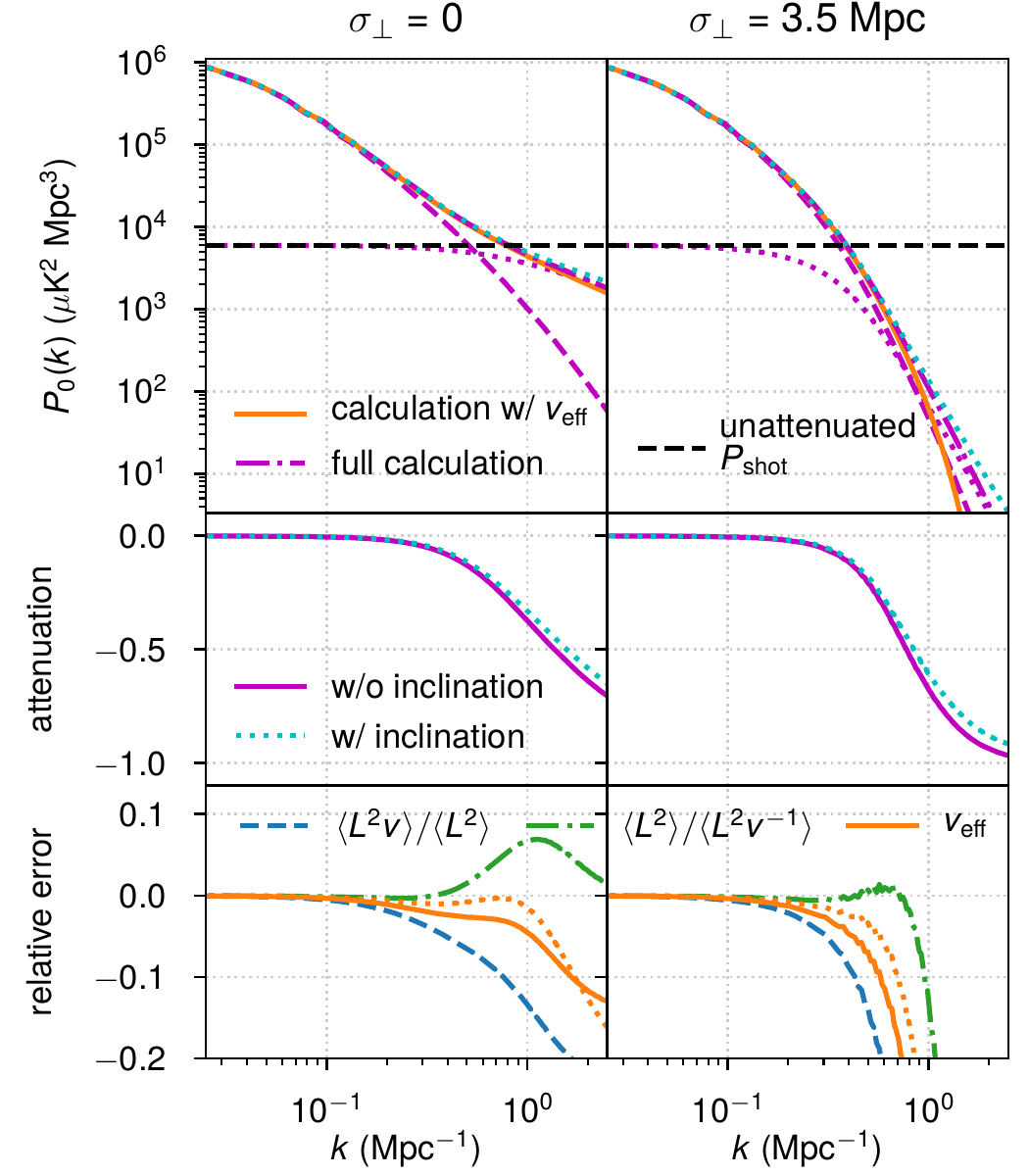}
        
        \includegraphics[width=0.42\linewidth]{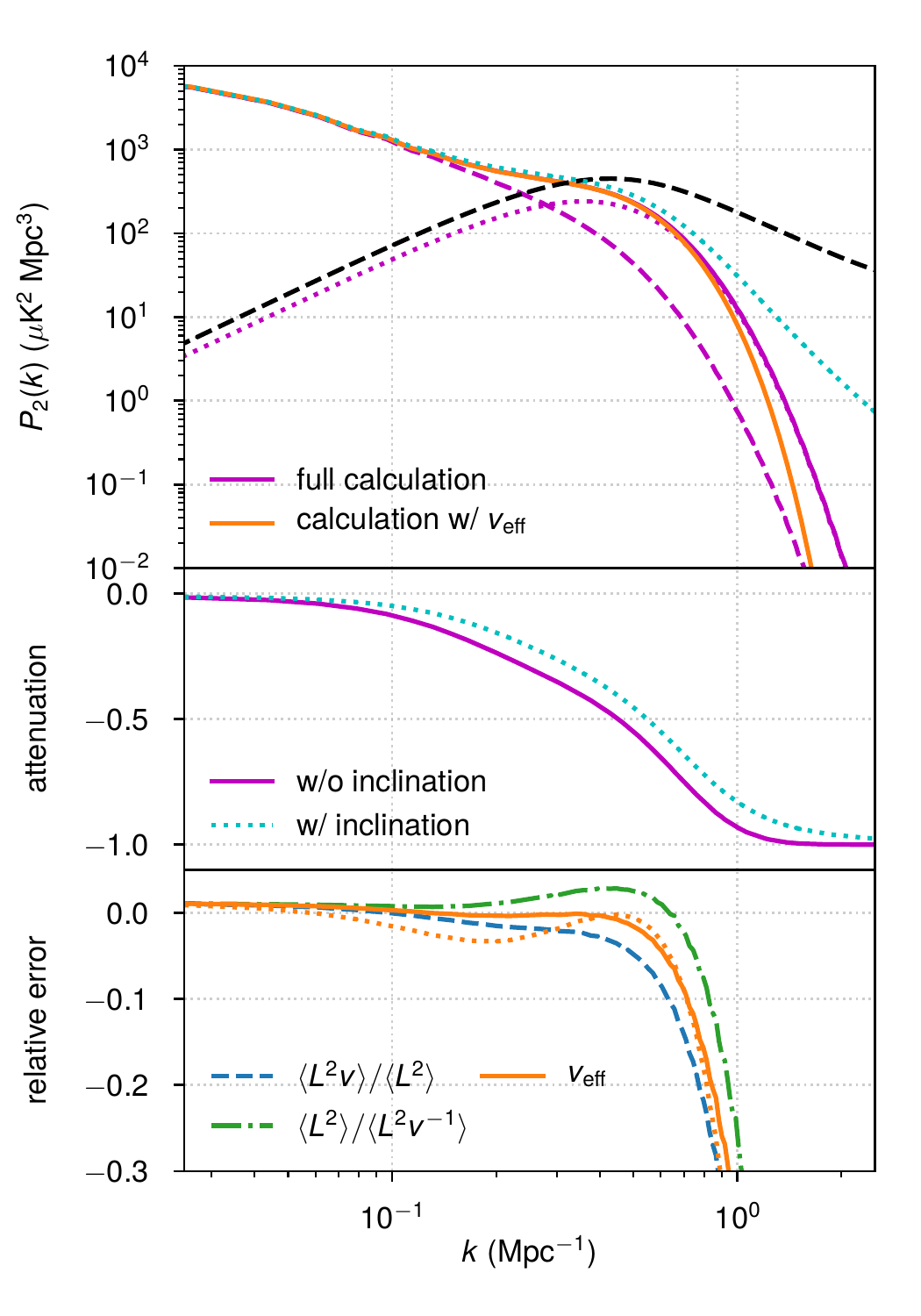}\includegraphics[width=0.42\linewidth]{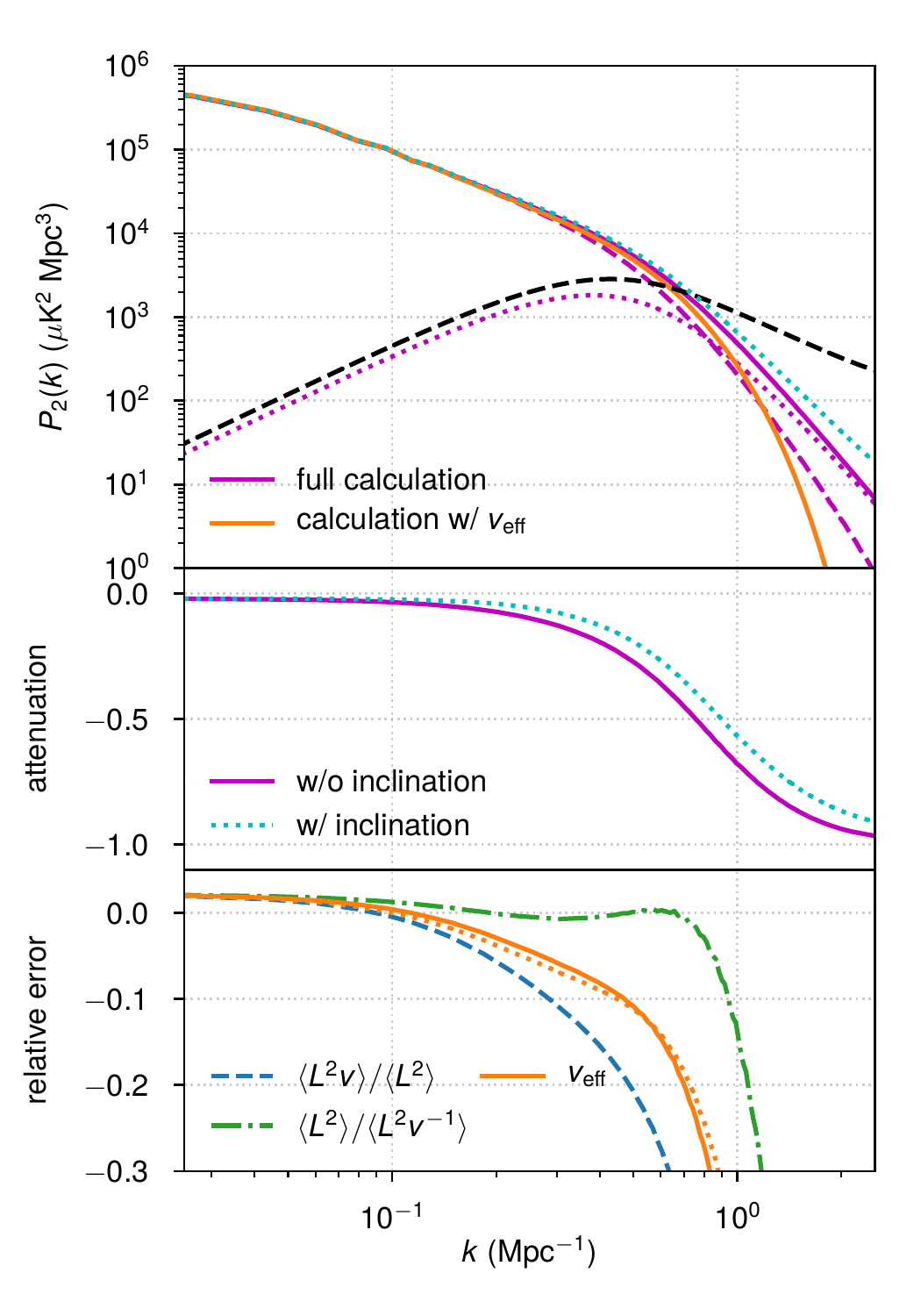}
    \caption{Same as~\autoref{fig:veff_realistic} (upper left subfigure),~\autoref{fig:veff_realisticq} (lower left subfigure),~\autoref{fig:veff_extremedraw} (upper right subfigure), and~\autoref{fig:veff_extremedrawq} (lower right subfigure), but the calculations neglect the corrections for inclination described in~\autoref{sec:incli}. The exceptions are the cyan dotted lines in the upper two rows of each subfigure, which show $P_\ell(k)$ or relative attenuation with inclination, and the orange dotted line in the lowermost row of each subfigure, which shows the relative error of the inclination-inclusive $v_\text{eff}$ of~\autoref{eq:veff_incli} compared to the inclination-inclusive full calculation. Note also that the clustering/shot-noise breakdown for the full calculation is shown in magenta instead of indigo.}
    \label{fig:veff_noincli}
\end{figure}
}
\section{Projection of mmIME Power Spectra}
\label{sec:projection}
We refer to~\cite{LidzTaylor16} for the calculations required to project each of the power spectra in~\autoref{fig:mmIME_pspecs} to a common frame---namely the $z\sim2.5$ frame applicable for CO(3--2)---and sum these together for the result shown in~\autoref{fig:mmIME_pspec_total}.

To map the wavevector $\mathbf{k}_J$ for the CO($J\to J-1$) line at redshift $z_J$ to the apparent wavevector $\mathbf{k}$ in the CO(3--2) frame at redshift $z=z_{J=3}$, we need only consider how the parallel and perpendicular components $k_{\perp,J}$ and $k_{\parallel,J}$ map to $k_\perp$ and $k_\parallel$ in the CO(3--2) frame. Equations 5 and 6 of~\cite{LidzTaylor16} give these mappings (notation altered to match our own):
\begin{equation}
    k_\parallel=\underbrace{\frac{H(z)}{H(z_J)}\frac{1+z_J}{1+z}}_{\alpha_{\parallel,J}}k_{\parallel,J};\qquad
    k_\perp=\underbrace{\frac{R(z_J)}{R(z)}}_{\alpha_{\perp,J}}k_{\perp,J}.
\end{equation}
Here we have used our assumption of a flat universe to replace the comoving angular diameter distance used in Equation 6 of~\cite{LidzTaylor16} with simply the comoving distance $R(z)$.

Then Equation 7 of~\cite{LidzTaylor16} gives the anisotropic power spectrum as a function of $k_\parallel$ and $k_\perp$, so that we can describe the sum of the anisotropic power spectra in the common CO(3--2) frame,
\begin{equation}
    P_\text{tot}(k_\parallel,k_\perp) = \sum_J\frac{P_J(k_\parallel/\alpha_{\parallel,J},k_\perp/\alpha_{\perp,J})}{\alpha_{\parallel,J}\alpha_{\perp,J}^2}.\label{eq:Ptot1}
\end{equation}
In this work, we have always dealt with power spectra as functions of $k$ and $\mu$. Recall that $\mu$ is the dot product between $\hat{\mathbf{k}}$ and the line-of-sight unit vector, so that
\begin{equation}
k_\parallel=k\mu;\qquad k_\perp=k(1-\mu^2)^{1/2}.
\end{equation}
Then we can map the apparent $k$ and $\mu$ in the CO(3--2) frame to the true $k_J$ and $\mu_J$ corresponding to the arguments to $P_J$ in~\autoref{eq:Ptot1}:
\begin{equation}
    k_J=k\left(\frac{\mu^2}{\alpha_{\parallel,J}^2}+\frac{1-\mu^2}{\alpha_{\perp,J}^2}\right)^{1/2};\qquad\mu_J=\frac{k\mu}{\alpha_{\parallel,J}k_J}.
\end{equation}
So given the above relations, we can write
\begin{equation}
    P_\text{tot}(k,\mu) = \sum_J\frac{P_J[k_J(k,\mu),\mu_J(k,\mu)]}{\alpha_{\parallel,J}\alpha_{\perp,J}^2}.\label{eq:Ptot2}
\end{equation}
We use this rewritten expression to sum the anisotropic power spectra for $J=2$ through $J=5$ in the main text and thus evaluate the total $P_0(k)$ in the CO(3--2) frame as shown in~\autoref{fig:mmIME_pspec_total}.\added{

\section{Bracketing Additional Variations Possible in Line Broadening-Induced Attenuation of the COMAP Monopole Power Spectrum Measurement}
\label{sec:morevariations}

In~\autoref{sec:sim} we discussed the fact that given the possible range of $L(M_h)$ parameters, our expectation for attenuation of $P_0(k)$ as observed by COMAP also spans a certain range. The 90\% interval from the possible variation in $L(M_h)$ can be expressed as $7^{+7}_{-4}\%$. However, the discussion of our line model in~\autoref{sec:linemodel} shows that this is far from the only possible aspect of our model with room for uncertainty. In particular we will discuss two potential additional aspects of our model that can be varied: the $v(M_h)$ relation and its scaling with source inclinations.
\subsection{Varying the Halo Mass--Line FWHM Relation}
We can consider two ways to vary our $v(M_h)$ model---the shape or functional form of the model, and the amplitude or general range of velocities predicted overall by the model. We propose that in bracketing uncertainties in $P_0(k)$ attenuation, the shape is less important than the amplitude. We will demonstrate this in~\autoref{sec:shapevary} before moving to consider the impact of varying the amplitude in~\autoref{sec:ampvary}.
\subsubsection{The Relative Unimportance of Shape}
\label{sec:shapevary}
The shape of the $v(M_h)$ relation is highly degenerate with respect to the level of $P_0(k)$ attenuation, at least for scales relevant to COMAP. This should be quite clear based on the demonstration of the adequacy of our $v_\text{eff}$ prescription for $P_0(k)$. If setting $v(M_h)$ to a constant is almost as good as using $v_\text{vir}(M_h)$, clearly the shape cannot be so important.

In fact, we will consider yet another form that yields similar attenuation for the representative median-like $L(M_h)$ given by the parameters of Equations~\ref{eq:realparami} through~\ref{eq:realparamf}. In this alternative functional form, we will simply use the power-law fit between line luminosity and line FWHM described in~\autoref{eq:vLpowerlaw}, such that
\begin{equation}
    \log{\frac{v(M_h)}{\text{km\,s}^{-1}}}=\delta_0+\delta_1\log{\frac{L'(M_h)}{10^{10}\text{ K\,km\,s}^{-1}\text{ pc}^2}}=\delta_0+\delta_1\log{\frac{10^{-10}\,C}{(M_h/M_1)^A+(M_h/M_1)^B}}.
\end{equation}
It is incredibly easy to devise values of $\delta_0$ and $\delta_1$ to mimic the attenuation expected from setting $v(M_h)=v_\text{vir}(M_h)$, as we show in~\autoref{fig:varyvofM}. A `flat' $v(M_h)$ model setting $\delta_0=2.50$ and $\delta_1=0$ matches the attenuation within a few percent, which we expect since the resulting $v(M_h)$ is itself within 10 percent of the $v_\text{eff}$ value of 283 km s$^{-1}$ predicted for this particular $L(M_h)$ model. However, a `steep' $v(M_h)$ model setting $\delta_0=2.35$ and $\delta_1=0.5$ works just as well. Note incidentally that this `steep' model also happens to predict $v(M_h)$ exactly at the (log-)midpoint between the `spherical' and `disk' models from \cite{ASPECS-LPISM} discussed briefly in~\autoref{sec:LWchara}.

\begin{figure}
    \centering
    \includegraphics[width=0.6\linewidth]{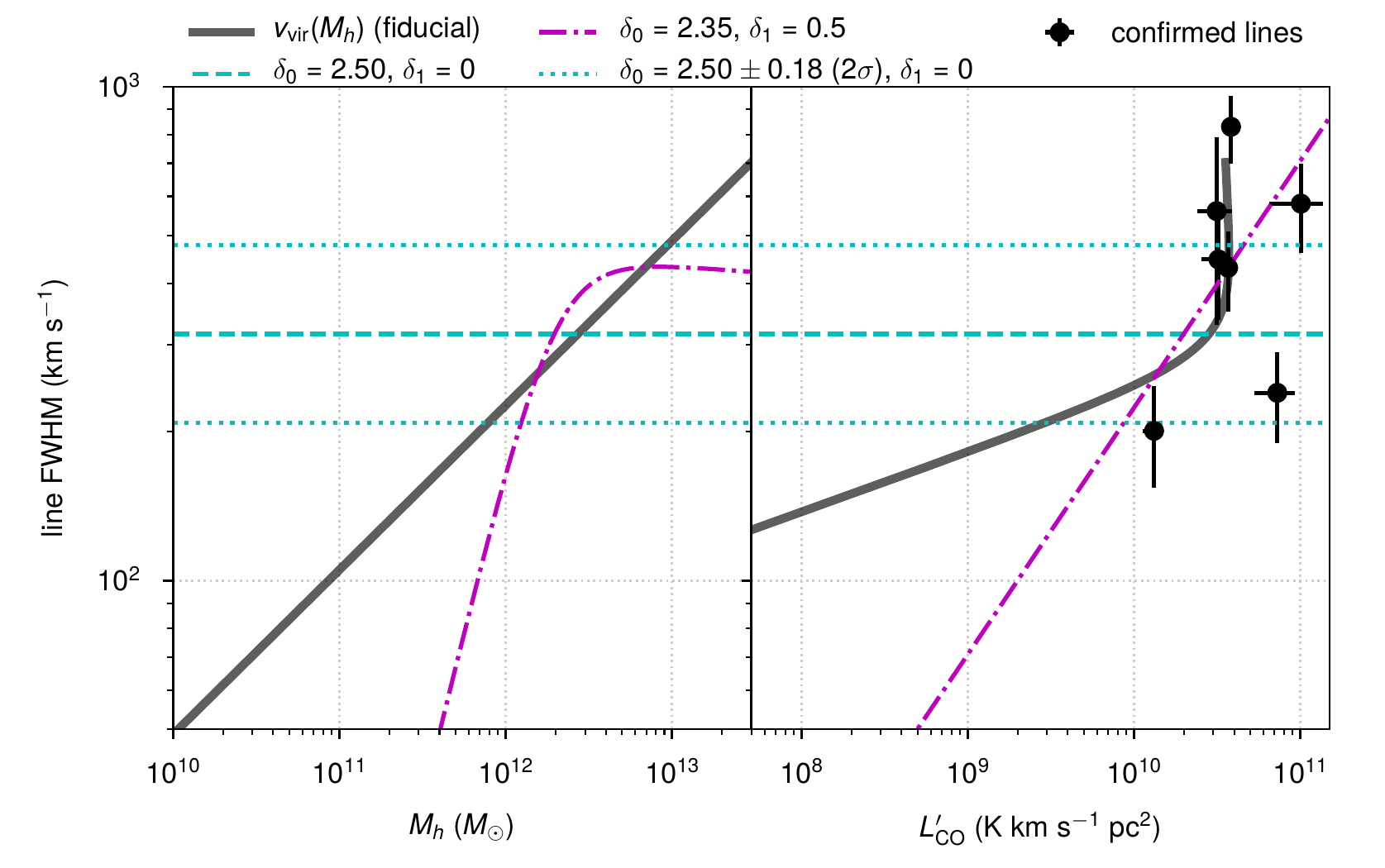}\includegraphics[width=0.375\linewidth]{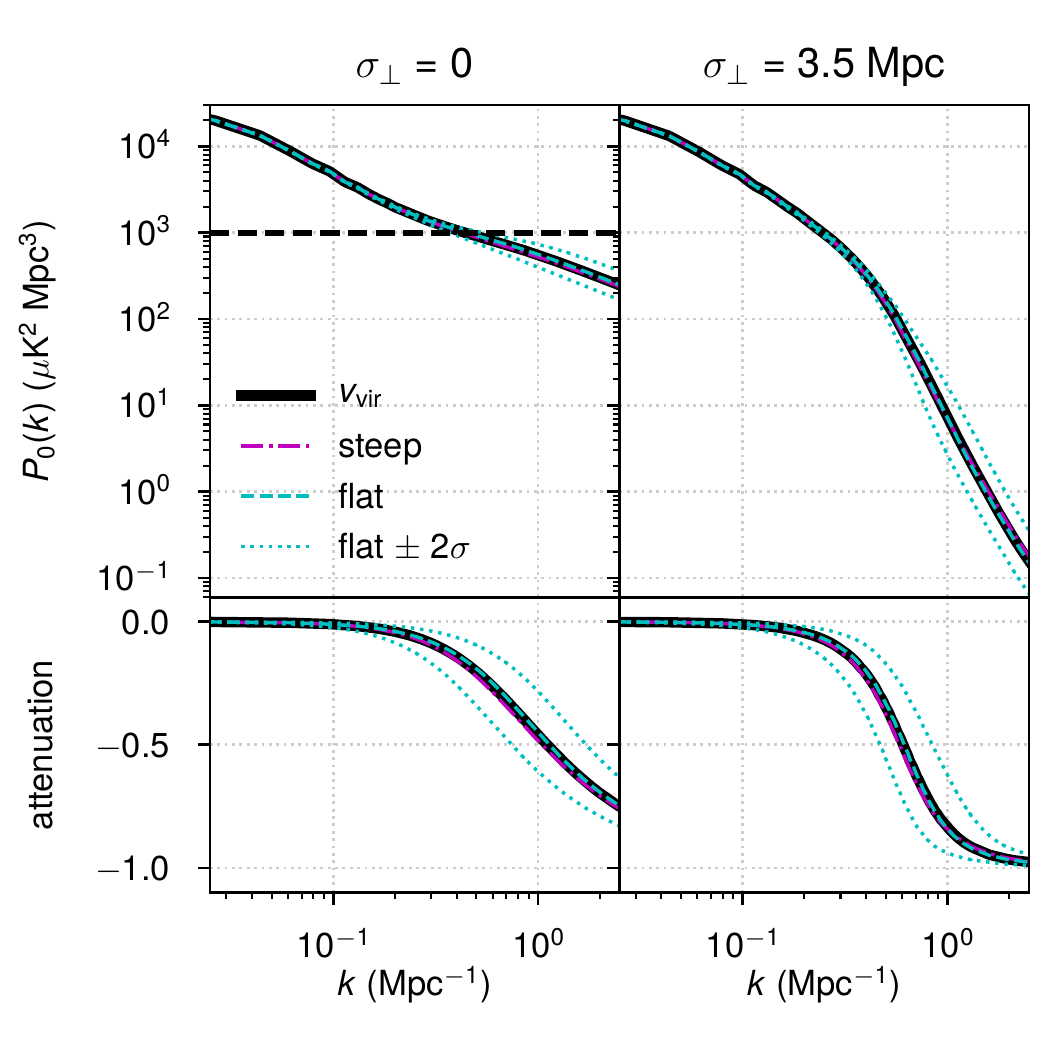}
    \caption{Illustration of possible variations in the $v(M_h)$ relation. \emph{Left subfigure:} model line FWHM as a function of halo mass (left panel) and $L'_\text{CO}$ (right panel). The models shown are $v_\text{vir}(M_h)$ (black thick curves; our fiducial expectation for preliminary calculations in~\autoref{sec:ansatzcheck}) and the variations on the power-law $v(L')$ model described in the main text that mimic the same attenuation. We also show confirmed COLDz and VLASPECS sources as we did before in~\autoref{fig:linewidths}. \emph{Right subfigure}: attenuated $P_0(k)$ (upper panels) and the attenuation relative to $P_0(k)$ calculated without line broadening (lower panels), both without transverse smoothing (left panels) and with transverse smoothing corresponding to the COMAP angular beam size (right panels). The `steep' and `flat' models show very similar results to the $v_\text{vir}$ model, bracketed by the $\pm2\sigma$ variations on the `flat' model.}
    \label{fig:varyvofM}
\end{figure}
\subsubsection{Varying the Amplitude}
\label{sec:ampvary}
We previously discussed that a fit to confirmed CO-selected CO(1--0) line emitters suggested $\delta_0=2.395\pm0.208$ and $\delta_1=0.193\pm0.302$. These parameters have significant covariance, and fixing $\delta_1=0$ (noting that the data do not indicate any statistically significant $L'$--FWHM correlation), the same fitting procedure finds $\delta_0=2.50\pm0.09$. The central value matches our `flat' mimic model considered above. Assuming the $\delta_0$ likelihood is described by a $t$-distribution with 6 degrees of freedom (one parameter fit to seven confirmed line candidates), a range of $\pm2\sigma$ around the central value, or $2.50\pm0.18$, gives the 90\% confidence interval for $v(M_h)$ variation.

Compared to the 7\% overall attenuation of the COMAP measurement expected from either the $v(M_h)=v_\text{vir}(M_h)$ model or its mimic counterparts, the edges of our 90\% interval of $\delta_0=2.50\pm0.18$ (for fixed $\delta_1=0$) yield estimates of 3\% and 13\% attenuation (lower and upper edges respectively). This interval is similar to the 3--14\% interval spanned by 90\% of simulated $L(M_h)$ variations for fixed $v(M_h)$. Therefore, we can say that the uncertainties associated the line width model are similar to those associated with the line luminosity model.
\subsection{Turning Source Inclination Multipliers On or Off}
We showed in~\autoref{sec:incli} the impact of going from assuming $v(M_h)$ describes all line widths to assuming that $v(M_h)$ describes the line widths for the median value of the sine of a randomly determined inclination angle, effectively switching from assuming \emph{no} sources are rotation-dominated to assuming \emph{all} sources are rotation-dominated. The upshot was that the high-$k$ ansatz for $v_\text{eff}$ was adjusted down by a little over 26\%, so we should expect a similar order-of-magnitude effect to be present in $P_0(k)$ attenuation.

Indeed, when running the simulations of~\autoref{sec:sim} without accounting for inclination, we find the expected attenuation is $10^{+8}_{-6}\%$ (90\% interval), instead of $7^{+7}_{-4}\%$. In other words, assuming \emph{no} sources are rotation-dominated yields 30--40\% greater attenuation compared to assuming \emph{all} sources are rotation-dominated, with less binary assumptions presumably landing somewhere in between. Compared to the sizes of the 90\% interval from variations in $L(M_h)$ alone or $v(M_h)$ alone, this is a subdominant effect.
\subsection{Summary of Effects}
If we account for the range of allowable line width models in addition to the range of allowable line luminosity models, our best estimate for the expected overall fractional attenuation due specifically to line broadening is
\begin{equation}
    \frac{\Delta P_{0,\text{COMAP}}}{P_{0,\text{COMAP}}}=-0.07\ _{-\ 0.07}^{+\ 0.04}\text{ (90\% lum.) }_{-\ 0.06}^{+\ 0.04}\text{ (90\% wid.)}, 
\end{equation}
for a COMAP measurement of $P_0$, with a possible systematic effect of up to 40\% relative increase (meaning to $-0.10$, not to $-0.37$) from assuming that part or all of the emitter population is not rotation-dominated and thus the inclinations of those sources do not narrow line widths relative to the median expectation.

We also recall that angular beam widths will impact the attenuation, and this will vary from experiment to experiment. However, the estimated overall $P_0(k)$ attenuation for scales relevant to COMAP changed only by a few percent when weighted by preliminary sensitivities at each $k$. (The change is greater when looking at a specific wavenumber like $k=0.25$\,Mpc$^{-1}$, but $\lesssim20$\% even so.) The effect of the angular beam on measurements is thus ultimately highly subdominant to other uncertainties (including modelling uncertainties) at present, and is an esoteric consideration for the current experimental and theoretical LIM landscape.

\section{The Ineffectiveness of Altering the Error Function-based Fit for Global Effective Line Width to Account Explicitly for the Presence of An Angular Beam}
\label{sec:madansatz}
Consider devising an alternate $v_\text{eff}$ that explicitly accounts for beam width, altering the requirement for an effective $\sigma_\parallel$ from~\autoref{eq:veff_tofit2} to this:
\begin{equation}\frac{\operatorname{erf}{\left[k(\sigma_{\parallel,\text{erf},\sigma_\perp}^2-\sigma_{\perp}^2)^{\frac{1}{2}}\right]}}{k(\sigma_{\parallel,\text{erf},\sigma_\perp}^2-\sigma_{\perp}^2)^{\frac{1}{2}}}\approx\frac{\avg{L^2\frac{\text{erf}{\left[k(\sigma_v(M_h)^2-\sigma_{\perp}^2)^{\frac{1}{2}}\right]}}{k(\sigma_v(M_h)^2-\sigma_{\perp}^2)^{\frac{1}{2}}}}}{\avg{L^2}}.\label{eq:veff_tofit3}\end{equation}
(Note that due to the properties of the error function, this is a well-defined real-valued function even for $\sigma_\perp>\sigma_\parallel$.) We omit consideration of source inclination angles in this section, as these are not relevant to the particular comparison at hand. The key result here is that replacing $\sigma_{\parallel,\text{erf}}$ with $\sigma_{\parallel,\text{erf},\sigma_\perp}$ does not actually result in appreciable improvement in relative error.

\begin{figure}
    \centering
    \includegraphics[width=0.96\linewidth]{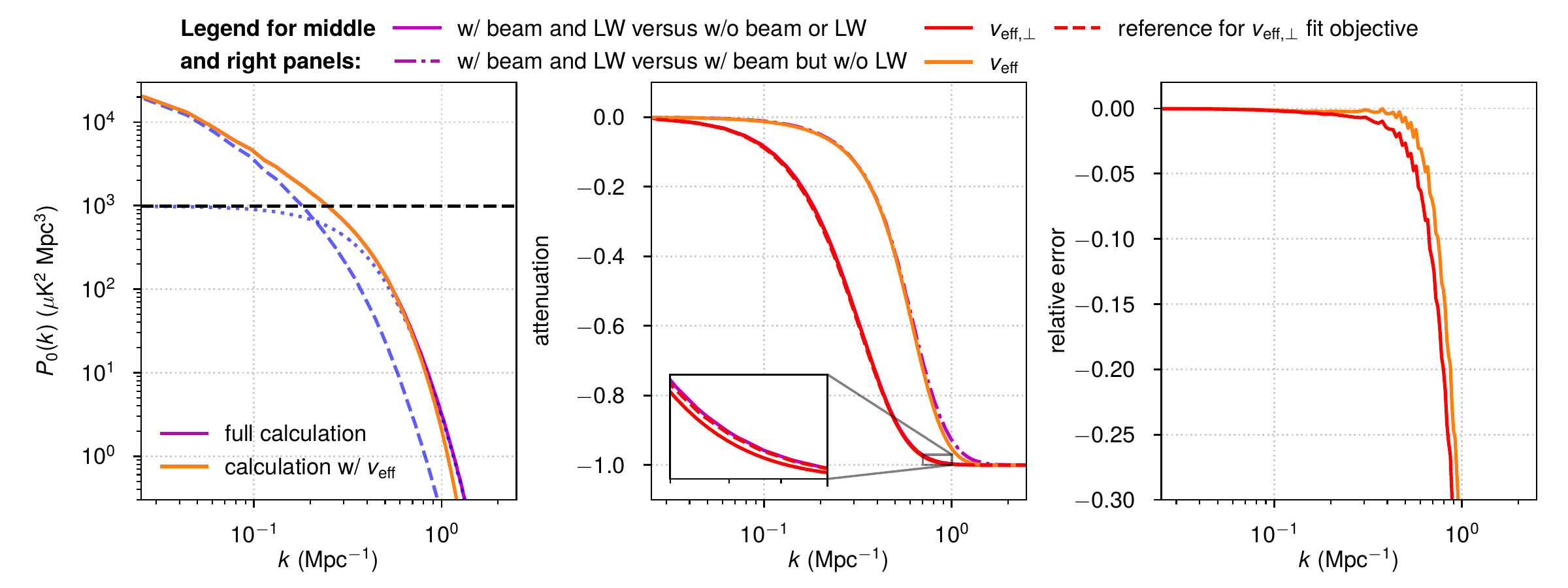}
    \caption{Comparison of $v_{\text{eff},\perp}$ against $v_\text{eff}$. \emph{Left panel:} Attenuated $P_0(k)$ as previously shown in the uppermost right panel of the upper left subfigure of~\autoref{fig:veff_noincli}. \emph{Middle panel:} Attenuation relative to $P_0(k)$ calculated without line broadening. To visually separate the attenuation described by $v_\text{eff}$ (orange) and $v_{\text{eff},\perp}$ (red solid), we show the attenuation in relation to $P_0(k)$ calculated without line widths but with the beam for $v_\text{eff}$ and the same for the full mass-dependent calculation (magenta dash-dotted), and show the attenuation in relation to $P_0(k)$ calculated without either line widths or beam smearing for $v_{\text{eff},\perp}$ and again for the mass-dependent calculation (magenta solid). We also show the estimated shot-noise attenuation from the combination of beam smearing and line widths (red dashed), which corresponds to the right-hand side of~\autoref{eq:veff_tofit3} and thus informs the objective function for the fit that produces $v_{\text{eff},\perp}$. \emph{Right panel:} error in the approximate attenuated $P_0(k)$ based on either $v_\text{eff}$ (orange) or $v_{\text{eff},\perp}$ (red).}
    \label{fig:veffalt}
\end{figure}

\autoref{fig:veffalt} illustrates this by showing calculations using the criteria from both~\autoref{eq:veff_tofit2} and~\autoref{eq:veff_tofit3}; we will dub the velocity associated with the former $v_\text{eff}$ and the velocity associated with the latter $v_{\text{eff},\perp}$. The $L(M_h)$ model used is again given by the parameters of Equations~\ref{eq:realparami} through~\ref{eq:realparamf}.

Somewhat paradoxically, the value of $v_{\text{eff},\perp}=333$\,km\,s$^{-1}$ is somewhat higher than the value of $v_\text{eff}=325$\,km\,s$^{-1}$, even if only by a few percent. (Note again we neglect source inclinations in this section, so the values are higher than the inclination-inclusive $v_\text{eff}=283$\,km\,s$^{-1}$ briefly mentioned in~\autoref{sec:shapevary}.) We have verified that this is not a numerical fluke but a real difference between the optima of the fit objective functions (the sum across $k$ of the square of the difference between the two sides of either~\autoref{eq:veff_tofit2} or~\autoref{eq:veff_tofit3}). The relative error is thus similar, and actually a little larger, when using $v_{\text{eff},\perp}$ compared to when using $v_\text{eff}$.

This is not a shortcoming of the design of the fit---the estimated shot-noise attenuation being fit to, shown in~\autoref{fig:veffalt} alongside the actual and approximated attenuation curves, ties more closely to the full calculation of the attenuation than to the approximate calculation using $v_{\text{eff},\perp}$. At first glance, both the actual and approximate attenuation curves follow an inverted sigmoid function. But the error function-based fit, corresponding to the left-hand side of~\autoref{eq:veff_tofit3}, asymptotes at high $k$ differently (more quickly to 100\% attenuation, to be specific) compared to even the attenuation of the shot-noise component in isolation, corresponding to the right-hand side of~\autoref{eq:veff_tofit3}. No matter what exact value we use for the effective line width, the shape of the left- and right-hand sides of~\autoref{eq:veff_tofit3} differ too much across $k$ when the angular smearing of the signal in comoving space is comparable to or exceeds the line-of-sight smearing from line broadening.

Ultimately, as we note in~\autoref{sec:discussion}, the point is largely esoteric for real-world analyses as they would likely not be sensitive to (and thus not make significant use of) such high-$k$ modes in the presence of such an angular beam. But it is another illustration of the shortcomings of attempting to describe the attenuation of $P_0(k)$ with only one parameter.
}

\bibliography{linebroad,keycites,software,morecites}{}
\bibliographystyle{aasjournal}



\end{document}